\documentclass[aps,prd,showpacs,nofootinbib,sort&compress,10pt]{revtex4-1}
\usepackage{blindtext}
\usepackage{amsmath}
\usepackage{amssymb}
\usepackage{mathtools}
\expandafter\ifx\csname package@font\endcsname\relax\else
\expandafter\expandafter
 \expandafter\usepackage
 \expandafter\expandafter
 \expandafter{\csname package@font\endcsname}%
\fi
\usepackage[utf8]{inputenc}
\usepackage{bm}
\usepackage{url}
\usepackage{color}
\usepackage{times}
\usepackage{graphicx}
\usepackage{textcomp}
\usepackage[update,prepend]{epstopdf}
\usepackage{rotating}
\usepackage{txfonts}
\providecommand{\url}[1]{{#1}}

\expandafter\ifx\csname urlstyle\endcsname\relax
  \else
  \fi

\def\be{\begin{equation}}
\def\ee{\end{equation}}
\def\ba{\begin{eqnarray}}
\def\ea{\end{eqnarray}}

\def\a{\alpha}
\def\a {\alpha}
\def\b{\beta}
\def\g{\gamma}     
\def\d{\delta}     \def\D{\Delta}

\def\vk{\varkappa}
\def\e{\epsilon}

\def\o{\omega}   \def\O{\Omega}
\def\p{\phi}     

\def\th{\theta}

\def\r{\rho}

\def\s{\sigma}

\def\la{\label}

\def\lt{\left}
\def\rt{\right}

\newcommand{\arccot}{{\rm arccot~}}

\begin{document}
\title{Normal gravity field in relativistic geodesy}
\author{Sergei Kopeikin}
\email{kopeikins@missouri.edu}
\affiliation{Department of Physics and Astronomy, University of Missouri, Columbia, Missouri
65211, USA \\and\\ Shanghai Astronomical Observatory, 80 Nandan Road, Shanghai, 200030, China }
%\altaffiliation[Also at:]{ Siberian State University of Geosystems and Technologies, Plakhotny Street 10,
%Novosibirsk 630108, Russia}
\author{Igor Vlasov}
\affiliation{Sternberg Astronomical Institute, Lomonosov Moscow State University,\\ Universitetsky Prospect 13, Moscow 119992, Russia}
\author{Wen-Biao Han}
\affiliation{Shanghai Astronomical Observatory, 80 Nandan Road, Shanghai, 200030, China\\and\\
School of Astronomy and Space Science, University of Chinese Academy of Sciences, Beijing 100049, China }
%\begin{keywords}
%general relativity -- relativistic geodesy -- reference ellipsoid -- gauge freedom -- -- geoid -- Pizzetti's
%theorem -- Clairaut's theorem.
%\end{keywords}

\date{\today}
\begin{abstract}
Modern geodesy is subject to a dramatic change from the Newtonian paradigm to Einstein's theory of general relativity.  This is motivated by the ongoing advance in development of quantum sensors for applications in geodesy including quantum gravimeters and gradientometers, atomic clocks and fiber optics for making ultra-precise measurements of the geoid and multipolar structure of the Earth's gravitational field. At the same time, Very Long Baseline Interferometry, Satellite Laser Ranging and Global Navigation Satellite System have achieved an unprecedented level of accuracy in measuring 3-d coordinates of the reference points of the International Terrestrial Reference Frame and the world height system. The main geodetic reference standard to which gravimetric measurements of the of Earth's gravitational field are referred, is a normal gravity field represented in the Newtonian gravity by the field of a uniformly rotating, homogeneous Maclaurin ellipsoid which mass and quadrupole momentum are equal to the total mass and (tide-free) quadrupole moment of the Earth gravitational field. The present paper extends the concept of the normal gravity field from the Newtonian theory to the realm of general relativity. We focus our attention on the calculation of the post-Newtonian approximation of the normal field that is sufficient for current and near-future practical applications. We show that in general relativity the level surface of homogeneous and uniformly rotating fluid is no longer described by the Maclaurin ellipsoid in the most general case but represents an axisymmetric spheroid of the fourth order with respect to the geodetic Cartesian coordinates. At the same time, admitting  a post-Newtonian inhomogeneity of the mass density in the form of concentric elliptical shells  allows to preserve the level surface of the fluid as an exact ellipsoid of rotation. We parametrize the mass density distribution and the level surface with two parameters which are intrinsically connected to the existence of the residual gauge freedom, and derive the post-Newtonian normal gravity field of the rotating spheroid both inside and outside of the rotating fluid body. The normal gravity field is given, similarly to the Newtonian gravity, in a closed form by a finite number of the ellipsoidal harmonics. We employ transformation from the ellipsoidal to spherical coordinates to deduce a more conventional post-Newtonian multipolar expansion of scalar and vector gravitational potentials of the rotating spheroid.  We compare these expansions with that of the normal gravity field generated by the Kerr metric and demonstrate that the Kerr metric has a fairly limited application in relativistic geodesy as it does not match the normal gravity field of the Maclaurin ellipsoid already in the Newtonian limit.  We derive the post-Newtonian generalization of the {\it Somigliana} formula for the normal gravity field measured on the surface of the rotating spheroid and employed in practical work for measuring the Earth gravitational field anomalies. Finally, we discuss the possible choice of the gauge-dependent parameters of the normal gravity field model for practical applications and compare it with the existing EGM2008 model of gravitational field. 
\end{abstract}
\maketitle
\tableofcontents
\newpage
\section{Introduction}\la{sec1}

\subsection{Earth's gravity field in the Newtonian theory}

Gravitational field of the Earth has a complicated spatial structure that is also subject to short and long temporal variations \citep{Vanicek_1986_book,Moritz_2006_book,Torge_2012_book}. Studying this structure and its time evolution is a primary goal of many scientific disciplines such as fundamental astronomy, celestial mechanics, geodesy, gravimetry, etc. The principal component of the Earth's gravity field is well approximated by radially-isotropic field that can be thought as being generated by either a point-like massive particle located at the geocenter or by a massive sphere (or a shell) having a spherically-symmetric distribution of mass inside it. According to the Newtonian gravity law the spheres (shells) of different size and/or of different spherically-symmetric stratifications of the mass density generate the same radially-isotropic gravitational field under condition that the masses of the spheres (shells) are equal. The same statement holds in general relativity where it is known under the name of Birkhoff's theorem \citep{LL_book2}. The radially-isotropic component of the Earth's gravity field is often called a monopole as it is characterized by a single parameter - the Earth's mass, $M$. Generally speaking, the total mass $M$ of the Earth is not constant because of the loss of hydrogen and helium from atmosphere, gradual cooling of the Earth's core and mantle, energy loss due to tidal friction, the dust accretion from an outer space, etc. Nonetheless, the temporal change of the Earth's overall mass is minuscule, $\dot M/M\le 10^{-15}$ [\textcolor{blue}{\url{https://en.wikipedia.org/wiki/Earth_mass}}], and can be neglected in  most cases. Thus, in the present paper we consider the Earth's mass, $M$, as constant. The time variability of mass does not affect the radial isotropy of the monopole field. It only changes its strength.

Monopole approximation is a good textbook example for discussion in undergraduate physics courses but it is insufficient for real scientific applications because Earth’s figure is not spherically-symmetric causing noticeable deviations from the radial isotropy of the gravity field. These deviations are taken into account by applying the next approximation in the multipolar expansion of the gravity field called ellipsoidal \citep{Moritz_1967,Vanicek_1986_book}. This is achieved by modelling the distribution of Earth's matter as a rotating bi-axial ellipsoid with its center of mass located at the geocenter and the minor axis coinciding with the Earth's polar principal axis of inertia. Moreover, gravitational potential on the surface of the rotating ellipsoid is equated to the value of the gravitational potential, $W_0$, measured on geoid that is the equipotential surface which coincides with the undisturbed level of the world ocean \citep{Torge_2012_book,Vanicek_1986_book}. It is the ellipsoidal approximation which is called the normal gravity field and the corresponding ellipsoid of revolution is known as a (global) reference ellipsoid \citep[chapter 4.2.1]{Torge_2012_book}. 

In the Newtonian theory of gravity the normal gravity field is uniquely specified in the exterior domain to ellipsoid by four parameters: the geocentric gravitational constant, $GM$, the semi-major axis of the reference ellipsoid, $a$, its flattening, $f$, and the nominal value of the Earth's rotational velocity, $\omega$ which are considered as fundamental geodetic constants \citep[Table 1.1]{petit_2010}. The normal field is used as a reference in description of the actual gravity field potential, $W$, of the Earth which can be represented as a linear superposition of the normal gravity field potential, $U$ and a disturbing potential $T$, that is \citep{Moritz_1967,Torge_2012_book}
\be
W=U+T\;.
\ee 
Notice that on the surface of rotating Earth there is also a centrifugal force besides the force of gravity. The potential $\Phi$ of the centrifugal force is considered as a well-defined quantity which can be easily calculated at each point of space. Therefore, although the potential $\Phi$ is small compared with $U$, it is not included to the perturbation $T$ but considered as a part of the normal gravity field potential $U$ that consists of the gravitational potential $V$ of a non-rotating Earth, and the centrifugal potential $\Phi$,
\be
U=V+\Phi\;.
\ee
The disturbing potential, $T$, includes all high-order harmonics in the multipolar expansion of the gravitational field of the Earth associated with the, so-called, anomalies in the distribution of the mass density. The multipolar harmonics are functions of spatial coordinates (and time) that can be expressed in various mathematical forms. For example, the multipolar harmonics expressed in spherical coordinates are known as the gravity potential coefficients \citep{Vanicek_1986_book,Torge_2012_book} representing the, so-called, gravity disturbances or anomalies. 

The gravity anomalies of the disturbing potential $T$ have been consistently measured for a long time with a gradually growing accuracy. Originally, their measurements were limited to rather small, local regions of the Earth's surface and were conducted by means of gravimeters giving access to the absolute value of the gravity force and the deflection of vertical (plumb line) at the measuring point. The gravimetric ground-based measurements are indispensable for regional studies of the gravity field anomalies but they remain sparse and insufficient to build a comprehensive model of the global gravity field which is the primary task of geodesy. Advancement in constructing the global model of the gravity field was achieved with the help of the dedicated geodetic space missions like LAGEOS and satellite laser ranging techniques \citep{lagslr}. More recently, further progress have been spawned with the advent of space gravity gradiometers -- GRACE and GOCE \citep{bender_2003SSRv}. 
The overall set of measurements of the gravity field anomalies has been processed and summarized in 2008 in the form of the Earth Gravitational Model (EGM2008) that has been built and publicly released by the National Geospatial-Intelligence Agency (NGA) EGM Development Team in 2012 \citep{egm2008,egm2008corr}. The disturbing gravitational potential $T$ of this model is complete up to all spherical harmonics of the degree and order 2159, and contains some additional potential coefficients up to degree 2190. Full access to the model's coefficients is provided on website of NGA
[\textcolor{blue}{\url{http://earth-info.nga.mil/GandG/wgs84/gravitymod/egm2008/index.html}}]

\subsection{Why do we need general relativity in geodesy?}

%Indeed, it is well known that one and the same gravitational field potential $V$ in the exterior domain of a massive extended body can be generated by different distributions of density inside the body. The mathematical reason for this is that the mass density, $\rho({\bm x}$, appears in the form of a kernel of an integral equation
%\be
%V({\bm x})=\int_{\cal V}\frac{\rho({\bm x}')d^3x'}{|{\bm x}-{\bm x}'|}\;,
%\ee  
%which is a formal solution of the Poisson equation, representing a functional relationship between the density $\rho({\bm x})$ and the gravity potential $V({\bm x})$. Solution $\rho({\bm x})$ of the integral equation for the given value of $V({\bm x})$ is not unique, and should be chosen by resorting to additional assumptions and measurements stemming from geophysical studies (for example, propagation of seismic waves).  
Positions of reference points (geodetic stations) on the Earth’s surface can now be determined with precision at the level of few millimeters and their variation over time at the level of 1 mm/year, or even better \citep{FuLL_2016}. Continuous geodetic observations become more and more fundamental for many Earth-science applications at the global and local levels like large scale and local Earth-crust deformation; global tectonic motion; redistribution of geophysical fluids on or near Earth’s surface including ocean, atmosphere, cryosphere, and the terrestrial hydrosphere; monitoring of the mean sea level and its variability for evaluation its impact on global warming, and many others \citep{Torge_2012_book}. All these important applications depend fundamentally on the availability and accuracy of the global International Terrestrial Reference System (ITRS). In addition to the above-mentioned geoscience applications, the ITRS -- through its realization by an International Terrestrial Reference Frame (ITRF), is an indispensable reference needed to ensure the integrity of Global Navigation Satellite System (GNSS), such as GPS, GLONASS, Galileo, and clock's synchronization \citep{petit_2010}. 

It is believed that the requirements of geoscience to measurement precision, including the most stringent one -- the mean sea level variability, are to reach the availability of the reference frame that will be reliable, stable and accessible at the positional accuracy down to 1 mm, and stability of 0.1 mm/year \citep{Pearlman_2009book}. It is crucial to reach this accuracy from both scientific and practical points of view as economy and safety of modern society is extremely vulnerable to even small changes in sea level \citep{fu_2013AdSpF}. Stability of the reference frame means that no discontinuity or drift should occur in its time evolution, especially for its defining physical parameters, namely the origin and the scale. Unfortunately, the current level of reference frame accuracy (based on the latest ITRF realization) is about ten times worse than the science requirement \citep{altamimi_2016}. New technological methods and theoretical models are required to fill up this gap. 

Relativistic effects of gravitational field of Earth have a fractional order of 1 ppb or about 1 cm on a geospatial scale. Albeit small, they depend globally on the geographic position of observer and produce a systematic bias in the height measurements, unless properly taken into account. It is, therefore, mandatory to switch from the Newtonian paradigm to general theory of relativity in order to thoroughly accommodate relativistic effects to geodesy. Nowadays, it is commonly accepted that a network of high precision clocks and their comparison will be able to significantly contribute to the solution of the problem of stability and accuracy of a new generation ITRF through very precise determination of height differences and relative velocities of clocks participating in the network \citep{issi_2017arXiv}. In addition to the highly precise geometrical coordinates of the ITRF (such as ellipsoidal heights), clock measurements will help to consistently provide physical heights at the reference points of observatories \citep{Bondarescu_2012GeoJI,Mai_2013ZGGL,Mai_2014}. Since clocks, according to general relativity, directly measure the difference of gravitational potential this opens up a fundamentally new conceptual basis for physical geodesy, that is, an unambiguous geoid determination and realization of a new global dynamic reference system. 

For thorough theoretical description of clock's behaviour in gravitational field, one has to take into account all special and general relativistic effects like gravitational red shift, Doppler effect, gravitational (Shapiro) time delay, Sagnac effect, and even Lense-Thirring effect which appears as gravitomagnetic clock effect \citep{hackmann_2014,cohen_1993PhLA}. All these effects depend on clock’s relative motion and strength of gravitational field. Based on this we can solve the inverse problem to model the mass density and height variations affecting the clock measurement, e.g., related to the solid Earth tides. Gravitational effects associated with Earth's rotation and tides limit the metrological network of  ground-based atomic clocks at fractional level $10^{-16}$ \citep{fateev_2015}, and must be accurately calculated and subtracted from clock' readings.

This section would be incomplete without mentioning the other branches of modern geodesy which are tightly connected with the experimental gravitational physics and fully based on the mathematical apparatus of general relativity. This includes Very Long Baseline Interferometry (VLBI) that is used as a main tool of the International Earth Rotation Service (IERS) for monitoring precession, nutation and wobble (polar motion) as well as for producing the International Celestial and Terrestrial Reference Frames (ICRF and ITRF respectively). VLBI requires taking into account a stunning number of relativistic effects which are outlined in corresponding papers and recorded in IAU resolutions (see, e.g. \citep{Soffel_2003AJ,petit_2010,Soffel_Kopeikin2017}). Motion of geodetic satellites must take into account a significant number of relativistic effects as well, like geodetic precession, Lense-Thirring effect, relativistic quadrupole, relativistic tidal effects, etc. General-relativistic model of relativistic effects in the orbital motion of geodetic and navigation spacecraft has been worked out by Brumberg \& Kopeikin \citep{BK_1989KFNT,BK_1989NCimento} (c.f. \citep{DSX_1994PhRvD_4}). It was numerically analysed in a number of recent papers \citep{SanMiguel_2009CeMDA,GNSS_emission2015,Roh_2014JASS,Roh_2016AdSpR} studying a feasibility of observing various relativistic effects. A particular attention has been recently paid to experimental measurement of the Lense-Thirring effect in the orbital motion of LAGEOS and LARES satellites \citep{Ciufolini_2004Natur,Ciufolini_2007Natur,Ciufolini_2012NewA,Ciufolini_2017IJMPD}. This experimental study is especially important for relativistic astrophysics because the Lense-Thirring effect is considered as a main driving mechanism for the enormous release of energy in quasars and active galactic nuclei caused by accretion of matter on a central, supermassive Kerr black hole \citep{blandford_thorne_1982IAUS}. 

One of the important relativistic problem in geodesy which has not yet been explored in theoretical papers are relativistic effects in the normal gravity field of the Earth associated with the international reference ellipsoid that is used as a reference for geodetic measurements. The goal of the present paper is to give solution of this problem.

\subsection{The normal gravity field in classic geodesy and in general relativity}

In classic geodesy the normal gravity field of the Earth is generated by a rigidly rotating bi-axial ellipsoid which is made of a perfect (non-viscous) fluid of uniform density, $\rho_{\rm c}$ which value is determined from the known total mass and volume of the Earth. In the Newtonian theory this is the only possible distribution of mass density because any other mass distribution of rotating fluid yields the shape of the body being different from the bi-axial ellipsoid \citep{Moritz_1967,chandra_book}. Relativistic geodesy is an advanced branch of physical geodesy that is based on Einstein's general relativity which supersedes the Newtonian theory of gravity. General relativistic approach requires reconsidering the concept of the normal gravity field by taking into account the curvature of spacetime manifold and other post-Newtonian effects caused by Earth's mass. 

General theory of relativity replaces a single gravitational potential, $V$, with ten potentials which are components of the metric tensor $g_{\a\b}$, where, here and anywhere else, the Greek indexes $\a,\b,\g,\ldots$ take values from the set $\{0,1,2,3\}$. General relativity modifies gravitational field equations of the Newtonian theory correspondingly. More specifically, instead of a single Poisson equation for the scalar potential $V$, general relativity introduces ten partial differential equations of the second order for the metric tensor components. These equations are known as Einstein's equations \citep{Kopeikin_2011_book}
\be\la{eineq6}
R_{\a\b}-\frac12 g_{\a\b} R=\frac{8\pi G}{c^4}T_{\a\b}\;,
\ee
where $R_{\a\b}$ is the Ricci tensor, $R=g^{\a\b}R_{\a\b}$ is the Ricci scalar, $T_{\a\b}$ is the tensor of energy-momentum of matter which is the source of gravitational field, $G$ is the universal gravitational constant, and $c$ is the fundamental speed of the Minkowski spacetime that is equal to the speed of light in vacuum or to the speed of propagation of weak gravitational waves. 

The left part of \eqref{eineq6} is called the Einstein tensor which is a hyperbolic differential operator of the second order in partial derivatives applied to the metric tensor $g_{\a\b}$ \citep{avr}. The Einstein tensor is non-linear and, for this reason, Einstein's equations \eqref{eineq6} cannot be solved exactly in the most physical situations of practical importance. In order to circumvent this difficulty researchers resort to iterative approximations to solve the Einstein equations. One of the most elaborated iterative schemes is called the post-Newtonian approximations (PNA) which basics are discussed in section \ref{sec2} in more detail. In many astrophysical applications (especially in gravitational-wave astronomy) one needs to make several post-Newtonian approximations for calculating observable effects \citep{blanchet_2014llr}. For the purposes of relativistic geodesy and celestial mechanics of the solar system the first post-Newtonian approximation is usually sufficient \citep{Kopeikin_2011_book} though there are indications that one may soon need a second PN approximation \citep{Soffel_2016_JGeod} and exact, axially-symmetric solutions of general relativity \citep{kop_2015PhLA,Philipp_2017PhRvD,Oltean_2016}.

The problem of determination of a figure of rotating fluid body is formidably difficult already in the Newtonian theory \citep{chandra_book}. It becomes even more complicated in general relativity because of non-linearity of the Einstein equations. Geophysics is interested in finding distribution of mass density inside the Earth to understand better its thermal behaviour and seismological response. The interior structure of the Earth is also important for the International Earth Rotation Service (IERS) to account properly for free core nutation (FCN) in calculation of polar wobble and tidal variations of the Earth's rotational velocity affected by the elasticity of the Earth's interior \citep{petit_2010}. Geodesy does not require precise distribution of mass density inside the Earth as it basically needs to know the surface of equal geopotential (geoid) and the gravity anomalies in the domain being exterior to geoid. Geoid's reconstruction from the gravity anomalies utilizes the normal gravity field for solving the integral equations of the Stokes -Molodensky problem \citep{Torge_2012_book}. As a rule, the most simple, homogeneous distribution of mass density inside the Earth is used to model the normal gravity field. Attempts to operate with more realistic distributions of mass inside the Earth led to the models of the normal gravity field which turned out to be too complicated for practical computations and were abandoned.  

We emphasize that the internal density distribution and the surface of the rotating fluid body taken for modelling the normal gravity field must be consistent with the laws of the theory of gravitation. In the Newtonian theory the surface of the uniformly rotating homogeneous fluid is a bi-axial ellipsoid of revolution - the Maclaurin ellipsoid \citep{chandra_book}. More realistic, non-homogeneous distribution of mass of the rotating fluid does not allow it to be the ellipsoid of revolution yielding more complicated figure having a spheroidal surface \citep{Moritz_1967,Torge_2012_book}. Such models have less practical significance in geodesy because of more entangled structure of the normal gravity field.

One would think that modelling the normal gravity field in relativistic geodesy could be achieved by finding an exact solution of the Einstein equations which Newtonian limit corresponds to the homogeneous Maclaurin ellipsoid. Unfortunately, the exact solutions of general relativity describing gravity field of a single body consisting of homogeneous, incompressible fluid are currently known only for spherically-symmetric, non-rotating configurations \citep{Krori_1975JPhA,ponce_1986JMP,Lindblom_1988JMP}. There is a certain progress in understanding the general relativistic structure of rotating fluid configurations \citep{Islam_1985_book,Gourgoulhon_2010arXiv,friedmann_2013_book} but whether a rigidly rotating fluid body can be made of incompressible, homogeneous fluid, is not yet known. In the post-Newtonian approximation of general relativity it was found that a rigidly-rotating body consisting of a perfect fluid with homogeneous distribution of mass density inside it, can exist but it is not a bi-axial ellipsoid \citep{Chandr_1965ApJ142_1513,Chandr_1967ApJ147_334,pyragas_1974,Petroff_2003PhRvD}. On the other hand, by assuming that the distribution of mass density has a post-Newtonian ellipsoidal component in addition to the constant density $\rho_c$, we can chose the parameters of the density distribution such that the figure of the rigidly rotating fluid will remain exactly ellipsoidal in the first (and higher-order) post-Newtonian approximations \citep{kop_2016}. Thus, we have to make a decision what type of the post-Newtonian distribution of mass and the figure of the rotating fluid are to be used in relativistic geodesy. Depending on the choice we shall have slightly different relativistic descriptions of the normal gravitational field outside rotating spheroid. We will proceed by assuming that the surface of the rotating fluid has a small post-Newtonian spheroidal deviation and that the distribution of the fluid density is almost homogeneous with a small post-Newtonian correction taken in the form of a homeomorphic ellipsoidal distribution. 

Thus, the normal gravity field in relativistic geodesy is a solution of the Einstein field equations with matter consisting of a uniformly rotating, perfect fluid of nearly constant density occupying a spheroidal volume. We notice that a number of researchers solved the Einstein equations to find out gravitational field of uniformly rotating bi-axial ellipsoid of constant density \citep{clark_1948,tessand_1978PhRvD,cheng_2007ChAA}. Their solutions are not directly applicable in relativistic geodesy for the shape of a uniformly rotating and incompressible homogeneous fluid is not an ellipsoid. Moreover, the authors of the papers \citep{clark_1948,tessand_1978PhRvD,cheng_2007ChAA}  were mostly interested in astrophysical applications 
and never approached the problem from the geodetic point of view. 

The present paper intends to overcome the shortcomings of the previous articles, and is organized as follows. Section \ref{sec2} explains the post-Newtonian approximations in general relativity. We pay a special attention to various coordinate systems used in relativistic geodesy and transformations between them as well as to the Green functions used for solving the Einstein equations. Section \ref{sec3} discusses gravitational potentials and the model of matter distribution used as a source of gravitational field. Section \ref{sec4} is devoted to a comprehensive calculation of the Newtonian gravitational potential. Section \ref{sec5} presents details of the calculation of the post-Newtonian scalar and vector gravitational potentials. In section \ref{sec6} we discuss relativistic multipole expansion of gravitational field of rotating spheroid which includes both mass and spin multipole moments. Section \ref{sec7} gives a full description of the normal gravity field of rotating spheroid in relativistic geodesy including definitions of equipotential surfaces, the gravity field potential, the figure of equilibrium of the rigidly rotating fluid, and the Somigliana formula for the normal gravity force on the surface of the rotating spheroid. Section \ref{sec8} calculates the normal gravity field of the Kerr metric and compares it with that of a rigidly rotating spheroid made out of the ideal fluid. It proves that the Kerr metric is unsuitable for purposes of relativistic geodesy due to the peculiar structure of its multipole expansion.     

\subsection{Mathematical symbols and notations}
The following notations are used throughout the paper:
\begin{itemize}
\item the spherical coordinates are denoted $\{R,\Theta,\Phi\}$, 
\item the ellipsoidal coordinates are denoted $\{\sigma,\theta,\phi\}$,
\item the Greek indices $\alpha, \beta,...$ run from 0 to 3,
\item the Roman indices $i,j,...$ run from 1 to 3,
\item repeated Greek indices mean Einstein's summation from 0 to 3,
\item repeated Roman indices mean Einstein's summation from 1 to 3,
\item the unit matrix (also known as the Kronecker symbol) is denoted by $\delta_{ij}=\d^{ij}$,
\item the fully antisymmetric symbol Levi-Civita is denoted as $\varepsilon_{ijk}=\varepsilon^{ijk}$ with
$\varepsilon_{123}=+1$,
\item the bold letters ${\bm
a}=a^i, {\bm b}=b^i,$ etc.,  denote spatial
3-dimensional vectors,
\item a dot between two spatial vectors, for example
${\bm a}\cdot{\bm b}=a^1b^1+a^2b^2+a^3b^3=\d_{ij}a^i b^j$, means the Euclidean dot
product,
\item the cross between two vectors, for example $({\bm a}\times{\bm b})^i\equiv \varepsilon^{ijk}a^jb^k$,
means the Euclidean cross product,
\item we use a shorthand notation for partial derivatives $\partial_\alpha
=\partial/\partial x^\alpha$, 
\item covariant derivative with respect to a coordinate $x^\a$ is denoted as $\nabla_\a$;
\item  the Minkowski (flat) space-time metric $\eta_{\alpha\beta}={\rm diag}(-1,+1,+1,+1)$,
\item $g_{\a\b}$ is the physical spacetime metric,
\item the Greek indices are raised and lowered with the metric $\eta_{\alpha\beta}$,
\item the Roman indices are raised and lowered with the Kronecker symbol $\delta^i_j$,
\item $G$ is the universal gravitational constant,
\item $c$ is the fundamental speed of the Minkowski space,
\item $\o$ is a constant rotational velocity of rigidly rotating matter,
\item $\rho$ is a mass density distribution of matter,
\item $\r_{\rm c}$ is a constant central density of matter,
\item $a$ is a semi-major axis of the ellipsoid of revolution,
\item $b$ is a semi-minor axis of the ellipsoid of revolution,
\item $f$ is the geometric flattening: $f\equiv (a-b)/a$,
\item $\e$ is the {\it first} eccentricity of the Maclaurin ellipsoid: $\e\equiv\sqrt{a^2-b^2}/a=\sqrt{2f-f^2}$,
\item $\varkappa$ is the {\it second} eccentricity of the Maclaurin ellipsoid: $\varkappa\equiv\sqrt{a^2-b^2}/b=\e/(1-f)$,
\item $\alpha\equiv a\e=\sqrt{a^2-b^2}$,
\item $r\equiv R/\alpha$ is a dimensionless spherical radial coordinate,
\item $\kappa\equiv \pi G\r_{\rm c} a^2/c^2$ is a dimensionless parameter characterizing the strength of gravitational field on the surface of the field-generating body.
\end{itemize}
Other notations are explained in the text as they appear.

\section{Post-Newtonian Approximations}
\la{sec2}

\subsection{Harmonic coordinates and the metric tensor}
\la{hcmt3s}

Discussion of relativistic geodesy starts from the construction of the spacetime manifold for the case of a rigidly rotating fluid body having the same mass as the mass of the Earth. We shall employ Einstein's general relativity to build such a manifold though some other alternative theories of gravity discussed, for example in textbook \citep{will_1993book}, can be used as well. Einstein's gravitational field equations \eqref{eineq6} represent a system of ten non-linear differential equations in partial derivatives for ten components of the (symmetric) metric tensor, $g_{\a\b}$, which represents gravitational potentials generalizing the Newtonian gravitational potential $V$. Because the equations are difficult to solve exactly due to their non-linearity, we resort for their solution to the post-Newtonian approximations (PNA) \citep{will_1993book,kovlas_2004}.

The PNA are the most effective in case of slowly-moving matter having a weak gravitational field. This is exactly the situation in the solar system which makes PNA highly appropriate for constructing relativistic theory of reference frames \citep{Soffel_2003AJ}, and for relativistic celestial mechanics, astrometry and geodesy \citep{Soffel_1989_book,brumberg_1991_book,Kopeikin_2011_book}. The PNA are based on the assumption that solution of the Einstein equations for the metric tensor can be presented in the form of a Taylor expansion of the metric tensor with respect to the inverse powers of the fundamental speed, $c$, that is equal to the speed of light in vacuum and the speed of weak gravitational waves in general relativity. 

Exact mathematical formulation of a set of basic axioms required for doing the post-Newtonian expansion was given by Rendall \citep{rendall}. Practically, it requires having several small parameters characterizing the source of gravity which is often is an isolated astronomical system comprised of extended bodies. The parameters are:
$\varepsilon_i\sim v_i/c$, $\varepsilon_e\sim v_e/c$, and $\eta_i\sim U_i/c^2$, $\eta_e\sim U_e/c^2$, where $v_i$ is
a characteristic velocity of motion of matter inside the body, $v_e$ is a characteristic velocity of the relative
motion of the bodies with respect to each other, $U_i$ is the internal gravitational potential of each body,
and $U_e$ is the external gravitational potential between the bodies. If one denotes a characteristic radius of
a body as $\ell$ and a characteristic distance between the bodies as $R$, the estimates of the internal and external gravitational
potentials will be, $U_i\simeq GM/\ell$ and $U_e\simeq GM/R$, where $M$ is a characteristic mass of the body.
Due to the virial theorem of the Newtonian gravity theory \citep{Chandr_1965ApJ142_1513} the small parameters are not independent.
Specifically, one has
$\varepsilon_i^2\sim\eta_i$ and $\varepsilon_e^2\sim\eta_e$. Hence,
parameters, $\varepsilon_i$ and $\varepsilon_e$, characterizing the slow motion of matter, are sufficient in doing the iterative solution of the Einstein equations by the post-Newtonian approximations. Because within
the solar system these parameters do not significantly differ from each other, we shall not distinguish between them. Quite often we shall use notation, $\kappa\equiv \pi G\r_{\rm c} a^2/c^2\sim \eta_i$, to mark the powers of the fundamental speed $c$ in the post-Newtonian terms.

We assume that physical spacetime within the solar system has the metric tensor denoted $g_{\a\b}$. This spacetime is well-approximated in case of the slow-motion and weak-field post-Newtonian approximation, by a background manifold which is the Minkowski spacetime having the metric tensor denoted $ \eta_{\a\b}={\rm diag}(-1,1,1,1)$. Einstein's equations admit a gauge freedom associated with the arbitrariness in choosing coordinate charts covering the spacetime manifold. The gauge freedom is used to simplify the structure of Einstein's equations. The most convenient choice is associated with the harmonic coordinates $x^\a=(x^0,x^i)$, where
$x^0=ct$, and $t$ is the coordinate time. The class of the harmonic coordinates is used by the International Astronomical Union for description of the relativistic coordinates systems and for the data reduction \citep{Soffel_2003AJ,petit_2010} as well as in relativistic geodesy \citep{Mueller_2008JGeod,kopmaz_2016prd}
The harmonic coordinates are defined by imposing the de
Donder gauge condition on the metric tensor \citep{fock_1964book,weinberg_1972},
\be\la{har5}
\partial_\a\lt(\sqrt{- g} g^{\a\b}\rt)=0\;.
\ee
Imposing the harmonic gauge greatly simplifies the Einstein equations \eqref{eineq6} and allows us to solve them by the post-Newtonian iterations.

Because gravitational field of the solar system is weak and motion of matter is slow, we can solve Einstein's equations by post-Newtonian approximations. In fact, the first post-Newtonian approximation of general relativity is fully sufficient for the purposes of relativistic geodesy. We focus in this paper on calculation of the normal gravitational field of the Earth generated by uniformly rotating ideal (perfect) fluid.
Under these assumptions the spacetime interval has the following form \citep{Kopeikin_2011_book}
\be\la{me7g3v}
ds^2=g_{00}(t,{\bm x})c^2dt^2+2g_{0i}(t,{\bm x})cdtdx^i+g_{ij}(t,{\bm x})dx^idx^j\;,
\ee
where the post-Newtonian expressions for the metric tensor components read
\begin{subequations}\la{pnm39}
\ba\la{pnm1}
 g_{00}(t,{\bm x})&=&-1+\frac{2 V(t,{\bm x})}{c^2}-\frac{2V^2(t,{\bm x})}{c^4}+{\cal O}\lt(\frac1{c^6}\rt)\;,\\\la{ngetxb}
 g_{0i}(t,{\bm x})&=&-\frac{4 V^i(t,{\bm x})}{c^3}+{\cal O}\lt(\frac1{c^5}\rt)\;,\\\la{nrxvsyt}
 g_{ij}(t,{\bm x})&=&\delta_{ij}\lt[1+\frac{2 V(t,{\bm x})}{c^2}\rt]+{\cal O}\lt(\frac1{c^4}\rt)\;.
\ea
\end{subequations}
Herein, the scalar potential $V=V(t,{\bm x})$ and a (gravitomagnetic) vector potential $V^i=V^i(t,{\bm x})$ are functions of time and spatial coordinates satisfying the Poisson equations,
\ba\la{feq1}
\D  V&=&-4\pi G\r\lt[1+\frac1{c^2}\lt(2 v^2+2  V+\Pi+\frac{3 p}{\r}\rt)\rt]\;,\\\la{feq2}
\D  V^i&=&-4\pi G\r v^i\;,
\ea
with $\r=\r(t,{\bm x})$ being the mass density, $ p=p(t,{\bm x})$ and $v^i=v^i(t,{\bm x})$ -- pressure and velocity of matter respectively, and $\Pi=\Pi(t,{\bm x})$ is the specific internal energy of matter per unit mass. We emphasize that $\r$ is the local mass density of baryons per unit of invariant
(3-dimensional) volume element $d{\cal V}=\sqrt{- g}u^0d^3x$, where $u^0=dt/d\tau$ is the time component of the 4-velocity of
matter's particle, where $d\tau=\sqrt{-ds^2}/c$ is the proper time of the particle \footnote{The minus sign in definition of the proper time appears because $ds^2<0$ due to the choice of the metric signature shown in \eqref{pnm1}-\eqref{nrxvsyt}.}. The local mass density, $\rho$, relates in the post-Newtonian approximation to the invariant mass
density $\r^*=\sqrt{- g}u^0\rho$, which post-Newtonian expression is given by \citep{Kopeikin_2011_book}
\be
\r^*=\r+\frac{\r}{c^2}\lt(\frac12 v^2+3 V\rt)+{\cal O}\lt(\frac1{c^4}\rt)\;.
\ee
The internal energy, $\Pi$, is related to pressure, $ p$, and the local density, $\rho$, through the thermodynamic equation (the law of conservation of energy)
\be\la{trm2}
d\Pi+ pd\lt(\frac1{\r}\rt)=0\;,
\ee
and the equation of state, $ p= p(\r)$.

We shall further assume that the background matter rotates rigidly around fixed $z$ axis with a constant angular velocity $\omega$. This makes the background spacetime stationary with the background metric being independent of time. In the stationary spacetime, the mass density $\r^*$ obeys the {\it exact}, steady-state equation of continuity
\be\la{vel3}
\partial_i\lt(\r^*v^i\rt)=0\;.
\ee
The velocity of the rigidly rotating fluid is a linear function of spatial coordinates,
\be\la{vel4}
v^i=\varepsilon^{ijk} \omega^j x^k\;,
\ee
where $\omega^i=(0,0,\omega)$ is a constant angular velocity. Replacing velocity $v^i$ in \eqref{vel3} with \eqref{vel4},
and differentiating yield,
\be\la{ily7}
v^i\partial_i\r=0\;,
\ee
which is equivalent to $d\r/dt=0$, and means that the linear velocity $v^i$ of the fluid is tangent to the surfaces of constant density $\r$.

\subsection{Ellipsoidal and Spherical Coordinates}

Equipotential surfaces of gravitational field produced by a rigidly rotating fluid body are closely approximated by biaxial ellipsoids.  Therefore, it sounds reasonable to solve Einstein's equations in the oblate ellipsoidal coordinates. These coordinates are well known and widely used in geodesy \citep{Moritz_1967,Vanicek_1986_book,Torge_2012_book}. In order to introduce these coordinates let us consider a point ${\cal P}$ in space that is characterized by three Cartesian (harmonic) coordinates ${\bm x}=(x,y,z)$. We choose the origin of the coordinates at the center of mass of the rotating body with $z$-axis coinciding with the direction of the vector of the angular rotation, ${\bm\o}=(\o^i)$, and $x$ and $y$ axes lying in the equatorial plane. 

In the Newtonian theory the rotating, homogeneous fluid takes the form of an oblate ellipsoid of rotation (Maclaurin ellipsoid) with a semi-major, $a$, and a semi-minor axis, $b=a\sqrt{1-\e^2}$, where 
the constant parameter 
\be\la{fecc4}
\e=\frac{\sqrt{a^2-b^2}}{a}\;,
\ee
is called the {\it first eccentricity} \citep{Torge_2012_book}, and  $0\le\e\le1$. The oblate ellipsoidal coordinates associated with the ellipsoid of revolution, are defined by a set of surfaces of confocal ellipsoids and hyperboloids being orthogonal to each other (see [\textcolor{blue}{\url{https://en.wikipedia.org/wiki/Oblate_spheroidal_coordinates}}]). It means that the focal points of all the ellipsoids and hyperboloids coincide, and the distance of the focal points from the origin of the coordinates is given by the distance $\a=a\e$.

In order to connect the Cartesian coordinates, $(x,y,z)$, of the point ${\cal P}$ to the oblate ellipsoidal coordinates, $(\s,\th,\p)$, we pass through ${\cal P}$ the surface of the ellipsoid which is confocal with the Maclaurin ellipsoid formed by the rotating homogeneous fluid. Geodetic definition of the transformation from the Cartesian to the ellipsoidal coordinates used in geodesy, is given, for example, in \citep[eq. 1-103]{Moritz_1967}, and reads
\begin{subequations}
\la{bl3}
\ba
x&=&\a\sqrt{1+\s^2}\sin\th\cos\p\;,\\
y&=&\a\sqrt{1+\s^2}\sin\th\sin\p\;,\\
z&=&\a\s\cos\th\;,
\ea
\end{subequations}
where the radial coordinate $\s\ge 0$, $0\le\th\le\pi$, $0\le\p\le2\pi$, and the constant parameter $\a\equiv \sqrt{a^2-b^2}=a\e$.
%\ba\nonumber
%x&=&a\e\sqrt{1+\vk^2\up^2}\sin\xi\cos\psi\;,\\\la{hts3}
%y&=&a\e\sqrt{1+\vk^2\up^2}\sin\xi\sin\psi\;,\\\nonumber
%z&=&a\e\vk \up\cos\xi\;,
%\ea
%\be
%\vk\equiv\frac{\sqrt{1-\e^2}}{\e}\;.
%\ee
The interior domain, ${\cal V}$, of the ellipsoidal coordinate system is separated from the exterior domain, $({\cal V}_{\rm ext})$, by the surface ${\cal S}$ of the Maclaurin ellipsoid. The interior domain is determined by conditions $0\le \s\le 1/\vk$, and the exterior domain has $\s>1/\vk$ respectively where the constant  
\be\la{secc4}
\vk\equiv\frac{\e}{\sqrt{1-\e^2}}=\frac{\sqrt{a^2-b^2}}b\;,
\ee
is called the {\it second eccentricity} \citep{Torge_2012_book}, and we notice that $0\le\vk\le\infty$. In terms of the second eccentricity, the focal parameter $\a=b\vk $.

It is worth noticing that equation \eqref{bl3} looks similar to that used for definition of the Boyer–Lindquist coordinates which have been used in astrophysical studies of the Kerr black hole that is an exact axisymmetric solution of vacuum Einstein's equation \citep[chapter 17]{LPPT_1975_book}. Nonetheless, the oblate ellipsoidal coordinates, $\{\s,\theta,\phi\}$ in this paper don't coincide with the Boyer-Lindquist coordinates which are connected to the original, non-harmonic, Kerr coordinates.

The volume of integration in the ellipsoidal coordinates is
\be
d^3x=\a^3\lt(\s^2+\cos^2\th\rt)d\s d\O\;,
\ee
where $d\O=\sin\th d\th d\p$ is the infinitesimal element of the solid angle in the direction of the unit vector
\be
\hat{\bm x}=\sin\th\lt(\hat{\bm i}\cos\p+\hat{\bm j}\sin\p\rt)+\hat{\bm k}\cos\th\;,
\ee
where $(\hat{\bm i}, \hat{\bm j}, \hat{\bm k})$ are the unit vectors along the axes of the Cartesian coordinates $(x,y,z)$ respectively. 
Notice that the unit vector $\hat{\bm x}$ is different from the unit vector of the external normal $\hat{\bm n}$ to the surface ${\cal S}$ that is given by
\be\la{nor5c}
\hat{\bm n}=\frac{\sqrt{1-\e^2}\sin\th\lt(\hat{\bm i}\cos\p+\hat{\bm j}\sin\p\rt)+\hat{\bm k}\cos\th}{1-\e^2\sin^2\th}\;.
\ee

We also introduce the standard spherical coordinates, $\{R,\Theta,\Phi\}$ related to the harmonic coordinates, $x^\a=\{x,y,z\}$, by the relations
\ba\nonumber
x&=&R\sin\Theta\cos\Phi\;,\\\la{bl344}
y&=&R\sin\Theta\sin\Phi\;,\\\nonumber
z&=&R\cos\Theta\;.
\ea
In what follows, it will be more convenient to use a dimensionless radial coordinate $r$ by definition: $r\equiv R/\a$ so that $\a^2r^2=x^2+y^2+z^2$. The volume of integration in the spherical coordinates is
\be
d^3x=\a^3r^2dr d{\rm O}\;,
\ee
where $d{\rm O}=\sin\Theta d\Theta d\Phi$ is the infinitesimal element of the solid angle in the spherical coordinates in the direction of the unit vector
\be
\hat{\bm X}=\sin\Theta\lt(\hat{\bm i}\cos\Phi+\hat{\bm j}\sin\Phi\rt)+\hat{\bm k}\cos\Theta\;.
\ee

Comparing \eqref{bl3} and \eqref{bl344} we can find out a
transformation between the oblate elliptical coordinates, $(\s,\theta,\phi)$, and the spherical coordinates, $(r,\Theta,\Phi)$, given by relations,
\be\la{ytb5}
 \sqrt{1+\s^2}\sin\th =r\sin\Theta\;,\qquad\s\cos\th=r\cos\Theta\;,\qquad\p=\Phi\;.
\ee
The radial elliptical coordinate, $\s$ and the radial spherical coordinate, $r$, are interrelated
\be\la{nd5t}
r^2=\s^2+\sin^2\th\;.
\ee
%Equation \eqref{nd5t} can be also recast to the following, useful form
%\be\la{dfr3}
%\frac{z^2}{\s^2}-\a^2\s^2=\a^2-r^2\;.
%\ee
Solving \eqref{ytb5} and \eqref{nd5t} we get a direct transformation between the elliptical and spherical coordinates in explicit form \citep[equation (20.24)]{Vanicek_1986_book}
\be\la{mev8u}
\s=\sqrt{\frac{r^2-1}{2}\lt(1+\sqrt{1+\frac{4r^2\cos^2\Theta}{(r^2-1)^2}}  \rt)}\;,\qquad
%\sqrt{\frac{r^2-1+\sqrt{(r^2-1)^2+4r^2\cos^2{\Theta}}}{2}}\;,\\
\cos\th=\frac{r\cos\Theta}{\displaystyle\sqrt{\frac{r^2-1}{2}\lt(1+\sqrt{1+\frac{4r^2\cos^2\Theta}{(r^2-1)^2}}  \rt)}}\;.  
\ee
The approximate form of the relations \eqref{mev8u} for relatively large values of the radial coordinate, $r\gg 1$, reads
\be
\s\simeq r-\frac{\sin^2\Theta}{2r}+\ldots\;,\qquad \cos\th=\cos\Theta\lt(1+\frac{\sin^2\Theta}{2r}+\ldots\rt)\;.
\ee
\subsection{Green's Function of the Poisson equation in the Ellipsoidal Coordinates}

The Einstein equations \eqref{feq1}, \eqref{feq2} represent the Poisson equations with the known right-hand side. The most straightforward solution of these equations can be achieved with the technique of the Green function ${\cal G}({\bm x},{\bm x}')$ that satisfies the Poisson equation,
\be\la{poieq4}
\Delta {\cal G}({\bm x},{\bm x}')=-4\pi\d^{(3)}({\bm x}-{\bm x}')\;,
\ee
where, $\Delta=\partial^2_x+\partial^2_y+\partial^2_z$, is the Laplace operator, and $\d^{(3)}({\bm x}-{\bm x}')$ is the Dirac delta-function in the harmonic coordinates $\{x,y,z\}$. We need the Green function in the oblate ellipsoidal coordinates, $\{\s,\th,\phi\}$. In these coordinates the Laplace operator reads
\be\la{lapop2}
\Delta\equiv\frac{1}{\a^2\lt(\s^2+\cos^2\th\rt)}\lt[(1+\s^2)\frac{\partial^2}{\partial\s^2}+2\s\frac{\partial}{\partial\s}+\frac{\partial^2}{\partial\th^2}+\cot\th\frac{\partial}{\partial\th}+\frac{\s^2+\cos^2\th}{(1+\s^2)\sin^2\th}\frac{\partial^2}{\partial\phi^2}\rt]\;.
\ee
After substituting this form of the operator to the left side of \eqref{poieq4}, and applying a standard procedure of finding a Green function \citep{Arfken_2001book}, we get the Green function, ${\cal G}({\bm x},{\bm x}')$, in the ellipsoidal coordinates. It is represented in the form of expansion with respect to the ellipsoidal harmonics \citep{hobson_1931,Pohanka_2011CoGG}
\be\la{grfun1}
{\cal G}({\bm x},{\bm x}')=\frac1{|{\bm x}-{\bm x}'|}=\left\{ \begin{array}{ll}
&\displaystyle\frac1\a\sum_{\ell=0}^\infty\sum_{m=-\ell}^{m=\ell}\frac{(\ell-|m|)!}{(\ell+|m|)!}q_{\ell|m|}\lt(\s'\rt)p_{\ell|m|}\lt(\s\rt)Y^*_{\ell m}(\hat{\bm x}')Y_{\ell m}(\hat{\bm x})\;,\qquad\quad (\s\le\s')\;,\\
&\\
&\displaystyle\frac1\a\sum_{\ell=0}^\infty\sum_{m=-\ell}^{m=\ell}\frac{(\ell-|m|)!}{(\ell+|m|)!}p_{\ell|m|}\lt(\s'\rt)q_{\ell|m|}\lt(\s\rt)Y^*_{\ell m}(\hat{\bm x}')Y_{\ell m}(\hat{\bm x})\;,\qquad\quad(\s'\le\s)\;.\end{array}\right.
\ee
Here, $p_{\ell m}(u)$ and $q_{\ell m}(u)$ are the modified (real-valued) associated Legendre functions of a {\it real} argument $u$,  that are related to the associated Legendre functions of an {\it imaginary} argument,  $P_{\ell m}(iu)$ and $Q_{\ell m}(iu)$,  by the following definition \footnote{We remind that the associated Legendre functions of the imaginary argument, $z=x+iy$, are defined for all $z$ except at a cut line along the real axis, $-1\le x\le 1$. The associated Legendre functions of a real argument are defined only on the cut line, $-1\le x\le 1$ \citep[Section 12.10]{Arfken_2001book}.}
\be
P_{\ell m}(iu)=i^np_{\ell m}(u)\;,\qquad Q_{\ell m}(iu)=\frac{(-1)^m}{i^{\ell+1}}q_{\ell m}(u)\;,
\ee
where $i$ is the imaginary unit, $i^2=-1$. In case, when the index $m=0$ we shall use notations, $p_\ell(u)\equiv p_{\ell 0}(u)$, and, $q_\ell(u)\equiv q_{\ell 0}(u)$. 
We shall also use special notation for the associated Legendre functions taken on the surface of ellipsoid of rotation having a fixed radial coordinate $\sigma=1/\varkappa$. More specifically, we shall simply omit the argument of the surface functions, for example, we shall denote $p_\ell\equiv p_\ell(1/\varkappa)$ and $q_\ell\equiv q_\ell(1/\varkappa)$. Several modified associated Legendre functions which are ubiquitously used in the present paper, are shown in Table I.
\begin{table}
\la{tavbn6}
\caption{The modified associated Legendre functions used in the present paper.}
\[\begin{array}{|llll|llll|}
\hline
&&&&&&&\\
\displaystyle p_0(\s)=1&&&&&&&\displaystyle q_0(\s)=\phantom{-}\arccot\s\\
&&&&&&&\\
\displaystyle p_1(\s)=\s&&&&&&&\displaystyle q_1(\s)=-p_1(\s) q_0(\s)+1\\
&&&&&&&\\
\displaystyle p_2(\s)=\frac32\s^2+\frac12&&&&&&&\displaystyle q_2(\s)=\phantom{-}p_2(\s)q_0(\s)-\frac32\s\\
&&&&&&&\\
\displaystyle p_3(\s)=\frac{\s}2\lt(5\s^2+3\rt)&&&&&&&\displaystyle  q_3(\s)=-p_3(\s)q_0(\s)+\frac{5\s^2}2+\frac23\\
&&&&&&&\\
\displaystyle p_4(\s)=\frac{35}8\s^4+\frac{15}4\s^2+\frac38&&&&&&&\displaystyle q_4(\s)=\phantom{-}p_4(\s)q_0(\s)-\frac{35}8\s^3-\frac{55}{24}\s\\
&&&&&&&\\
\displaystyle p_{11}(\s)=\sqrt{1+\s^2}&&&&&&&\displaystyle q_{11}(\s)=p_{11}(\s)q_0(\s)-\frac{\s}{\sqrt{1+\s^2}}\\
&&&&&&&\\
\displaystyle p_{31}(\s)=\frac32\sqrt{1+\s^2}\lt(1+5\s^2\rt)&&&&&&&\displaystyle q_{31}(\s)=p_{31}(\s)q_0(\s)-\frac{\s}2\frac{13+15\s^2}{\sqrt{1+\s^2}}\\
&&&&&&&\\
\hline
\end{array}\]
\end{table}

Functions $Y_{\ell m}(\hat{\bm x})$ in \eqref{grfun1} are the standard spherical harmonics \footnote{Definition of the associated Legendre polynomials adopted in the present paper follows \citep[Sec. 8.81]{gradryzh}. It differs by a factor $(-1)^m$ from the definition of the associated Legendre polynomials adopted in the book \citep{Moritz_1967}. } 
\be\la{oin8}
Y_{\ell m}(\hat{\bm x})\equiv C_{\ell m}P_{\ell|m|}(\cos\th)e^{im\p}\;,
\ee
where $P_{\ell m}(\cos\th)$ are the associated Legendre polynomials, and  the normalization coefficient
\be
C_{\ell m}\equiv\sqrt{(2\ell+1)\frac{(\ell-|m|)!}{(\ell+|m|)!}}\;.
\ee
The spherical harmonics are complex, $Y^*_{\ell m}(\hat{\bm x})=Y_{\ell,-m}(\hat{\bm x})$, and form an orthonormal basis in the Hilbert space, that is for $|m|\le \ell$, $|m'|\le \ell'$ the integral over a unit sphere ${\cal S}^2$,
\be\la{inus52}
\oint\limits_{{\cal S}^2}Y^*_{\ell m}(\hat{\bm x})Y_{\ell'm'}(\hat{\bm x})d\O=4\pi\d_{\ell\ell'}\d_{mm'}\;,
\ee
where $\d_{\ell m}={\rm diag}(1,1,...,1)$ is a unit matrix (the Kronecker symbol).

In case when there is no dependence on the angle $\p$, the index $m=0$ in \eqref{oin8} and $C_{\ell 0}=\sqrt{2\ell+1}$. Then, Green's function \eqref{grfun1} takes on a more simple form,
\ba\la{grf1}
{\cal G}({\bm x},{\bm x}')=\frac1{|{\bm x}-{\bm x}'|}=\left\{ \begin{array}{ll}&\displaystyle\frac1\a\sum_{\ell=0}^\infty(2\ell+1)q_\ell\lt(\s'\rt)p_\ell\lt(\s\rt)P_{\ell}(\cos\th')P_{\ell}(\cos\th)\;,\qquad\quad(\s\le\s')\\&\\
&\displaystyle\frac1\a\sum_{\ell=0}^\infty(2\ell+1)p_\ell\lt(\s'\rt)q_\ell\lt(\s\rt)P_{\ell}(\cos\th')P_{\ell}(\cos\th)\;,\qquad\quad(\s'\le\s)\;,\end{array}\right.
\ea
where the Legendre polynomials $P_\ell(\cos\th)$ are normalized such that
\be\la{norm56}
\int\limits_{0}^\pi P_\ell(\cos\th)P_m(\cos\th)\sin\th d\th=\frac2{2\ell+1}\d_{\ell m}\;.
\ee

In what follows the following expressions are used for connecting different Legendre polynomials between themselves and with the trigonometric functions,
\ba\la{scl24}
\cos^2\th=\frac13\lt[1+2P_2\lt(\cos\th\rt)\rt]\;,\qquad\quad&&\qquad\quad\;\;\;
\sin^2\th=\frac23\lt[1-P_2\lt(\cos\th\rt)\rt]\;,\\
\la{scl242}
\sin\th P_{11}(\cos\th)=-\frac23\lt[1-P_2(\cos\th)\rt]\;,\qquad\quad&&\;\sin\th P_{31}(\cos\th)=-\frac{12}{7}\lt[P_2(\cos\th)-P_4(\cos\th)\rt]\;,\\
\la{scl241}
\sqrt{1+\s^2}q_{11}(\s)=\frac23\lt[q_0(\s)+q_2(\s)\rt]\;,\qquad\quad&&
\sqrt{1+\s^2}q_{31}(\s)=\frac{12}{7}\lt[q_2(\s)+q_4(\s)\rt]\;,
\ea
in order to make transformations of integrands in the process of calculation of gravitational potentials. 
%\be
%\int p_{\ell}(\s)d\s=\frac{p_{n+1}(\s)+p_{n-1}(\s)}{2n+1}\;,
%\ee

%\be
%p_2(\s)p_4(\s)=\frac1{77}\lt[22p_2(\s)-20p_4(\s)+35p_6(\s)\rt]\;,
%\ee

\section{Gravitational Field and Model of Matter Distribution}\la{sec3}

Gravitational field of a stationary-rotating matter is fully described by the particular solutions of the Einstein equations \eqref{feq1}, \eqref{feq2} for the metric tensor \eqref{pnm39} which include solution of the Poisson-type equation for scalar potential
\ba\la{inte1}
V({\bm x})&=&V_N({\bm x})+\frac1{c^2}V_{pN}({\bm x})\;,\\\la{nef7}
V_N({\bm x})&=&G\int\limits_{\cal V}\frac{\r({\bm x}')}{|{\bm x}-{\bm x}'|}d^3x'\;,\\\la{nvs8}
V_{pN}({\bm x})&=&G\int\limits_{\cal V}\frac{\r({\bm x}')}{|{\bm x}-{\bm x}'|}\lt[2 v^2({\bm x}')+2V({\bm x}')+\Pi({\bm x}')+\frac{3 p({\bm x}')}{\r({\bm x}')}\rt]d^3x'\;,
\ea
and that for a vector (often called gravitomagnetic \citep{Ciufolini_1995book,Kopeikin_2006IJMPD,Kopeikin_2010ASSL}) potential 
\be\la{hrt3}
V^i({\bm x})=G\int\limits_{\cal V}\frac{\r({\bm x}')v^i({\bm x}')}{|{\bm x}-{\bm x}'|}d^3x'\;,
\ee
where the field point has harmonic coordinates ${\bm x}$. In order to calculate the above integrals we have to know the distribution of mass density $\r$, pressure $p$, velocity $v^i$, and the internal energy density of the fluid $\Pi$, as well as the boundary of the volume ${\cal V}$ occupied by the fluid.

The real earth is near equilibrium shape, but not quite at it. Also, there are small measurable changes in shape from post-glacial viscous rebound, elastic adjustments to the shifting mass from melting glaciers, plate tectonics, and other long-wavelength variations in earth's geoid \citep{Hager_Richards_1989}. These factors are important in studying the problem of the dynamic earth. However, our goal in the present paper is more pragmatic and relates to the study of relativistic corrections in the earth's gravity field. Therefore, we shall neglect the dynamic changes in the distribution of masses and the earth's shape. According to previous studies \citep{Chandr_1967ApJ147_334,Kopejkin_1991}  the surfaces of the equal gravity potential, density, pressure, and the internal energy coincide both in the Newtonian and the post-Newtonian approximations so that in order to find out their shape it is enough to find out the surfaces of the equal potential. In what follows, we shall follow the model of the normal gravity field in classic geodesy and assume that the density of the fluid is almost uniform with a small post-Newtonian deviation from the homogeneity. 
\be\la{cond3}
\r({\bm x})=\r_{\rm c}\lt[1+\kappa F({\bm x})\rt]\;,
\ee
where $\r_{\rm c}$ is a constant density at the center of the spheroid, and $F({\bm x})$ is a homothetic function of ellipsoidal distribution with respect to its center,
\be\la{ncj7}
F({\bm x})={\cal A}\frac{q_1}{\vk^2}\lt(\frac{x^2+y^2}{a^2}+\frac{z^2}{b^2}\rt)\;,
\ee
$q_1\equiv q_1(1/\vk)$, and the constant parameter ${\cal A}={\cal A}(\e)$ is kept arbitrary in the course of the calculations that follow. We shall find the equations constraining the value of the parameter ${\cal A}$ later. Such type of the density distribution has been chosen because it is consistent with the distribution of pressure, at least in the post-Newtonian approximation (see below). The ratio $q_1/\vk^2$ was introduced to (\ref{ncj7}) explicitly to make the subsequent formulas look less cumbersome. We notice that the choice of the distribution \eqref{ncj7} allows us to handle calculations analytically in a closed form without series expansion while other assumptions on the mass distribution would lead to analytical results that are more complicated than the results given in this manuscript.

Distribution \eqref{ncj7} in the ellipsoidal coordinates takes on the following form  
  \be
\la{bvxr2}
F({\bm x})={\cal A}q_1(1-\e^2)R(\s,\th)\;,
\ee
where the function
\be
R(\s,\th)\equiv\lt(1+\s^2\rt)\sin^2\th+\lt(1+\vk^2\rt)\s^2\cos^2\th\;.
\ee
It is worth noticing that the ellipsoidal distribution of density \eqref{bvxr2} means that the surfaces of constant density are not the same as the surfaces of constant value of the radial coordinate $\sigma$. The density $\r$ remains dependent on the angular coordinate $\th$ everywhere inside the ellipsoid except at its surface, where $\s=1/\vk$ with the post-Newtonian accuracy, and $R({\vk^{-1}},\th)=\e^{-2}$. We also draw attention of the reader that in the limiting case of vanishing oblateness, $\vk\rightarrow 0$, the post-Newtonian correction to the density is not singular because $\lim\limits_{\vk\rightarrow 0}\vk^{-2}q_1= 1/3$.

Distribution of pressure inside the rotating homogeneous fluid is obtained by integrating the law of the hydrostatic equilibrium. Pressure enters calculations of the integrals characterizing the gravitational field, only at the post-Newtonian terms. Hence, it is sufficient to know its distribution to the Newtonian approximation which is easily obtained by solving the equation of hydrostatic equilibrium \citep{chandra_book,Tassoul_1978_book}
\be\la{ol4w}
p({\bm x})=2\pi G\r_{\rm c}^2 a^2\frac{q_1}{\vk^2}\lt[1-\e^2R(\s,\th)\rt]\;,
\ee
where we have denoted $q_1\equiv q_1(1/\vk)$ once again.
The internal energy is also required for calculation of the integrals only in terms of the post-Newtonian order of magnitude. In this approximation the internal energy can be considered as constant, 
\be
\Pi({\bm x})=\Pi_0\;,
\ee
in correspondence with the thermodynamic equation \eqref{trm2} solved for the constant density, $\r=\r_{\rm c}$. From now on we incorporate the constant thermodynamic energy to the central density and will not show $\Pi_0$ explicitly in our calculations. Because the fluid rotates uniformly in accordance with the law \eqref{vel4}, we have for the distribution of the velocity squared,
\be\label{skor}
v^2({\bm x})=\o^2\a^2\lt(1+\s^2\rt)\sin^2\th\;.\\
\ee

All integrals are calculated over a (yet unknown) volume occupied by the rotating fluid. The surface of the rotating, self-gravitating fluid is a surface of vanishing pressure that coincides with the surface of an equal gravitational potential \citep{Chandr_1965ApJ142_1513,Chandr_1967ApJ147_334,Kopejkin_1991}. In the Newtonian approximation it yields the Maclaurin ellipsoid. Deviations from the Maclaurin ellipsoid have a post-Newtonian order of magnitude \citep{Chandr_1967ApJ147_334,pyragas_1974,Petroff_2003PhRvD} that allows us to assume that the surface of the rotating fluid is given in the post-Newtonian approximation by the law,
%\be\la{ztw8}
%{\s_s}\equiv{\s_s}(\th)=\frac1{\vk}+\frac{4\pi }{c^2}{\cal B} G\r_{\rm c} b^2\frac{q_2}{\e^4}P_2(\cos\th)\;,
%\ee 
\be\la{ztw8qq}
{\s_s}=\frac1{\vk}\lt[1+{\cal B}\frac{\o^2a^2}{c^2\e^2}P_2(\cos\th)\rt]\;,
\ee 
where ${\cal B}={\cal B}(\e)$ is a constant arbitrary parameter which possible numerical value will be discussed below at the end of section \ref{mass6307} and in section \ref{nv2crh}. The reason for picking up the specific form of the numerical factor in \eqref{ztw8qq} is a matter of mathematical convenience in presentation of the equations which follow. 

It is worthwhile to make two remarks. First, we notice that 
appearance of $\e^2$, in the denominator in the right side of \eqref{ztw8qq} does not lead to divergence as the angular velocity of rotation $\o^2\sim\e^2$. Second, equation \eqref{ztw8qq} corresponds to the choice of the harmonic gauge \eqref{har5} being different from that adopted in our previous papers \citep{kopmaz_2016prd,kop_2016}. We have shown in those papers (see, e.g. \citep[Section 4]{kop_2016}) that the gauge freedom is described by four out of five parameters which parametrize the shape of the rotating post-Newtonian spheroid in the following form:
\be\la{ne6v8k}
\frac{x^2+y^2}{a^2}+\frac{z^2}{b^2}=1+\pi G\rho_{\rm c}a^2\left[{\cal K}_1\frac{x^2+y^2}{a^2}+{\cal K}_2\frac{z^2}{b^2}+{\cal B}_1\left(\frac{x^2+y^2}{a^2}\right)^2+{\cal B}_2\frac{z^4}{b^4}+{\cal B}_3\frac{x^2+y^2}{a^2}\frac{z^2}{b^2}\right]\;.
\ee 
The choice of the gauge made in papers \citep{kopmaz_2016prd,kop_2016} was ${\cal K}_1={\cal K}_2={\cal B}_1={\cal B}_2=0$ and the parameter ${\cal B}_3$ was linked to the solution of the equations of hydrostatic equilibrium of the rotating fluid (see \citep[Section 10]{kop_2016}). Equation \eqref{ztw8qq} of the present paper corresponds to different choice of the parametrization \eqref{ne6v8k}, for example, ${\cal B}_1={\cal B}_2=0$, and 
\be
{\cal K}_1=-16\frac{q_2}{\varkappa^3}{\cal B}\;,\qquad\qquad {\cal K}_2=8\frac{q_2}{\varkappa}{\cal B}\;,\qquad\qquad {\cal B}_3=-\frac{16}3\frac{q_2}{\varkappa}{\cal B}\;,
\ee
where parameter ${\cal B}$ is the same as in \eqref{ztw8qq}.

With the definition \eqref{ztw8qq} of the upper limit of the integration with respect to the radial coordinate $\s$, the volume integral from an arbitrary function 
\be
{\cal F}({\bm x},{\bm x}')=\left\{ \begin{array}{ll}
         {\cal F}_<(\s,\th,\s',\th') &\qquad \mbox{if $\s' \leq \s$}\;,\\
        {\cal F}_>(\s,\th,\s',\th') &\qquad \mbox{if $\s' \geq \s$}\;.\end{array}\right.  
\ee
can be calculated with sufficient accuracy as a sum of two terms:
\begin{itemize}
\item[1)] in case when the radial coordinate $\s$ of the field point ${\bm x}$ is taken inside the body
\ba\nonumber
\int\limits_{\cal V}{\cal F}({\bm x},{\bm x}')d^3x'&=&2\pi\a^3\int\limits_0^\s\int\limits_0^{\pi}{\cal F}_<(\s,\th,\s',\th')(\s'^2+\cos^2\th')d\s'd\th'\\\nonumber
&+&2\pi\a^3\int\limits_\s^{1/\vk}\int\limits_0^{\pi}{\cal F}_>(\s,\th,\s',\th')(\s'^2+\cos^2\th')d\s'd\th'\\\la{mryb7}
&+&2\pi{\cal B}\frac{\o^2a^4b}{c^2} \int\limits_0^\pi{\cal F}_>\lt(\s,\th,\vk^{-1},\th'\rt)\lt(\vk^{-2}+\cos^2\th'\rt)P_2(\cos\th')d\th'\;,\ea
\item[2)] in case when the radial coordinate $\s$ of the field point ${\bm x}$ is taken outside the body
\ba
\nonumber
\int\limits_{\cal V}{\cal F}({\bm x},{\bm x}')d^3x'&=&2\pi\a^3\int\limits_0^{1/\vk}\int\limits_0^{\pi}{\cal F}_<(\s,\th,\s',\th')(\s'^2+\cos^2\th')d\s'd\th'\\\la{mryb7z}
&+&2\pi{\cal B}\frac{\o^2a^4b}{c^2}\int\limits_0^\pi{\cal F}_<\lt(\s,\th,\vk^{-1},\th'\rt)\lt(\vk^{-2}+\cos^2\th'\rt)P_2(\cos\th')d\th'\;.
\ea
\end{itemize}
The very last integral in the right hand side of \eqref{mryb7} and \eqref{mryb7z} is of the post-Newtonian order of magnitude and will be treated as a post-Newtonian correction to the Newtonian gravitational potential.

\section{Newtonian Gravitational Potential}\la{sec4}

The Newtonian gravitational potential $V_N$ is given by \eqref{nef7}
where the density distribution, $\r({\bm x})$, is defined in \eqref{cond3} and the integration is performed over the volume bounded by the radial coordinate ${\s_s}$ in \eqref{ztw8qq}. The integral can be split in three parts:
\be\la{nxrv7}
V_N=V_N[\r_{\rm c}, {\cal S}]+V_N[\d\r, {\cal S}]+V_N[\r_{\rm c},\d {\cal S}]\;,
\ee
where $V_N[\r_{\rm c}, {\cal S}]$ is a contribution from the constant density $\r_{\rm c}$, $V_N[\d\r, {\cal S}]$ is a contribution from the variation $\d\r\equiv c^{-2}F({\bm x})$ of the density given by \eqref{bvxr2}, and $V_N[\r_{\rm c}, \d{\cal S}]$ is a contribution from the constant density $\r_{\rm c}$ enclosed in the volume between the real boundary and that of the Maclaurin ellipsoid.
The integrals $V_N[\r_{\rm c}, {\cal S}]$ and $V_N[\d\r, {\cal S}]$ are performed over the volume of the Maclaurin ellipsoid with the fixed value of the radial coordinate on its boundary, $\s=1/\vk$. The contribution $V_N[\r_{\rm c},\d {\cal S}]$ comes from the very last integrals in \eqref{mryb7} and \eqref{mryb7z}. Below we provide the specific details of calculations of the three constituents entering the right hand side of \eqref{nxrv7}. 

\subsection{Contribution from the constant density}
The contribution from the constant density $\r_{\rm c}$ to the Newtonian potential is given by the integral
\be\la{nw7h}
V_N[\r_{\rm c}, {\cal S}]=G\r_{\rm c}\int\limits_{\cal V}\frac{d^3x'}{|{\bm x}-{\bm x}'|}\;,
\ee
which can be calculated with making use of the Green function \eqref{grfun1}. Depending on the position of the field point ${\bm x}$ we distinguish the internal and external solutions.

\subsubsection{The internal solution}
The internal solution is valid for the field point ${\bm x}$ with the radial coordinate, $0\le\s\le1/\vk$. With the help of the Green function \eqref{grf1} it reads,
\ba\la{bevc6}
V_N[\r_{\rm c}, {\cal S}]&=&2\pi G\r_{\rm c}\a^2\sum_{\ell=0}^\infty(2\ell+1) q_\ell\lt(\s\rt)P_{\ell}(\cos\th)\int\limits_{0}^\s d\s'\int\limits_{0}^\pi\lt(\s'^2+\cos^2\th'\rt) p_\ell\lt(\s'\rt)P_{\ell}(\cos\th')\sin\th' d\th'\\\nonumber
&+&2\pi G\r_{\rm c}\a^2\sum_{\ell=0}^\infty(2\ell+1) p_\ell\lt(\s\rt)P_{\ell}(\cos\th)\int\limits_{\s}^{1/\vk} d\s'\int\limits_{0}^\pi\lt(\s'^2+\cos^2\th'\rt) q_\ell\lt(\s'\rt)P_{\ell}(\cos\th')\sin\th' d\th'\;.
\ea
We, first, integrate with respect to the angular variable $\th$ and, then, with respect to the radial coordinate $\s$. We also use the relation
\be\la{nd6t}
\s^2+\cos^2\th=\frac23\lt[p_2(\s)+P_{2}\lt(\cos\th\rt)\rt]\;,
\ee
that allows us to operate with the Legendre polynomials instead of the trigonometric functions, and use the condition of orthogonality \eqref{norm56}.
Then, after substituting \eqref{nd6t} to \eqref{bevc6} and making use of the normalization condition \eqref{norm56}, we obtain
\be\la{ndtb7}
V_N[\r_{\rm c}, {\cal S}]=V_0(\s)+V_2(\s)P_2(\cos\th)\;,
\ee
where
\ba
V_0(\s)&=&\frac{8\pi}3 G\r_{\rm c}\a^2\lt[q_0(\s)\int\limits_{0}^\s d\s'p_2(\s')+\int\limits_\s^{1/\vk} d\s'p_2(\s')q_0(\s')\rt]\;,\\
V_2(\s)&=&\frac{8\pi}3 G\r_{\rm c}\a^2\lt[q_2(\s)\int\limits_{0}^\s d\s'p_2(\s')
+p_2(\s)\int\limits_\s^{1/\vk} d\s'q_2(\s')\rt]\;.
\ea
%\ba
%V^{0,0}_N&=&\frac{4}{3}\pi G\r_{\rm c}\a^2\lt[\arccot{\s}\int\limits_{0}^\s d\s'\lt(3\s'^2+1\rt)+\int\limits_\s^{1/\vk} d\s'\lt(3\s'^2+1\rt)\arccot\s'\rt]\\
%V^{2,0}_N&=&\frac13\pi G\r_{\rm c}\a^2\lt(3\cos^2\th-1\rt)\lt[\lt(3\s^2+1\rt)\arccot\s-3\s\rt]\int\limits_{0}^\s d\s'\lt(3\s'^2+1\rt)\\\nonumber
%&+&\frac13\pi G\r_{\rm c}\a^2\lt(3\cos^2\th-1\rt)\lt(3\s^2+1\rt)\int\limits_\s^{1/\vk} d\s'\lt[\lt(3\s'^2+1\rt)\arccot\s'-3\s'\rt]\nonumber\;.
%\ea
Calculation of the integrals yields
\ba
V_0(\s)&=&\frac{2\pi G\r_{\rm c}\a^2}{3\vk^3}\lt[\vk\lt(1-\vk^2\s^2\rt)+2\lt(1+\vk^2\rt)\arctan\vk\rt]\;,\\
V_2(\s)&=&-\frac{2\pi G\r_{\rm c}\a^2}{3\vk^3}\lt[\vk+\vk\lt(3+2\vk^2\rt)\s^2-\lt(1+3\s^2\rt)\lt(1+\vk^2\rt)\arctan\vk\rt]\;.
\ea
Adding up the two expressions according to \eqref{ndtb7} and making the inverse transformation from the ellipsoidal to Cartesian coordinates, we get
%\be\la{ybv5}
%V_N[\r_{\rm c}]=\frac{\pi G\r_{\rm c}\a^2}{\vk^3}\lt\{\vk\lt[1+\s^2-\frac{z^2}{\a^2\s^2}-\frac{z^2}{\a^2}\lt(3+2\vk^2\rt)\rt]-\lt(1+\vk^2\rt)\lt(\s^2-\frac{z^2}{\a^2\s^2}-1-3\frac{z^2}{\a^2}\rt)\arctan\vk\rt\}
%\ee
%\be
%V_N=\frac{\pi G\r_{\rm c}}{\a}\lt[\frac{2\a b^2}3 + \frac{4a^2 b}3\arctan\lt(\frac{\a}{b}\rt)-2\a z^2+\lt(\frac{3\s^2}{\a^2}+1\rt)\lt(3\cos^2\th-1\rt)\lt(-\frac{\a b^2}3+ \frac{a^2 b}3 \arctan\lt(\frac{\a}{b}\rt)\rt)\rt]
%\ee
%Making use of equation \eqref{nd5t} we can make further simplifications of \eqref{ybv5}, and get it in the following form:
%\be
%\lt(3\s^2+1\rt)\lt(3\cos^2\th-1\rt)=\lt(3\s^2+1\rt)\lt(\frac{3z^2}{\a^2\s^2}-1\rt)=9\frac{z^2}{\a^2}+3\lt(\frac{z^2}{\a^2\s^2}-\s^2\rt)-1=2-3\frac{x^2+y^2}{\a^2}+6\frac{z^2}{\a^2}\;.
%\ee
%Then
\be\la{om4}
V_N[\r_{\rm c}, {\cal S}]=\pi G\r_{\rm c}\a^2\lt[2\frac{1+\vk^2}{\vk^3}\arctan\vk-2\frac{z^2}{\a^2}+\frac{x^2+y^2-2z^2}{\a^2\vk^2}\lt(1-\frac{1+\vk^2}{\vk}\arctan\vk\rt)\rt]\;.
\ee
This expression for the Newtonian potential $V_N[\r_{\rm c}]$ inside the Maclaurin ellipsoid is well-known from the classic theory of figures of rotating fluid bodies \citep{pizz_1913,chandra_book} (see also \citep[eq. 50]{kopmaz_2016prd}). It is straightforward to check by direct differentiation that \eqref{om4} satisfies the Poisson equation $\Delta V_N[\r_{\rm c}]=-4\pi G\r_{\rm c}$, in accordance with \eqref{nw7h}.

\subsubsection{The external solution}
Making use of the Green function \eqref{grf1} we get for the field point ${\bm x}$ with the radial ellipsoidal coordinate, $\s>\vk^{-1}$, the following external solution for the Newtonian potential of the homogeneous Maclaurin ellipsoid,
\ba\la{hhh4}
V_N[\r_{\rm c}, {\cal S}]&=&2\pi G\r_{\rm c}\a^2\sum_{\ell=0}^\infty(2\ell+1) q_{\ell}\lt(\s\rt)P_{\ell}(\cos\th)\int\limits_{0}^{1/\vk} d\s'\int\limits_{0}^\pi\lt(\s'^2+\cos^2\th'\rt) p_{\ell}\lt(\s'\rt)P_{\ell}(\cos\th')\sin\th' d\th'\;.
\ea
We again integrate over the angular variable $\th$ and, then, with respect to the radial variable $\s$. It yields,
\be\la{xcv67}
V_N[\r_{\rm c}, {\cal S}]=\frac{4\pi G\r_{\rm c}\a^2}{3\vk^3}\lt(1+\vk^2\rt)\lt[q_0(\s)+q_2(\s)P_2(\cos\th)\rt]\;.
%\frac{8\pi}3 G\r_{\rm c}\a^2\lt[q_0(\s)+q_2(\s)P_2(\cos\th)\rt]\int\limits_{0}^{1/\vk} d\s'p_2(\s')\\\nonumber
%&=&\frac{4\pi G\r_{\rm c}\a^2}{3\vk^3}\lt(1+\vk^2\rt)\lt\{\arccot\s+\frac12\lt[\lt(1+3\s^2\rt)\arccot\s-3\s\rt]P_2(\cos\th)\rt\}\;.
\ee
In the asymptotic regime at spatial infinity, when the radial coordinate $\s$ is very large, the Legendre functions have the following asymptotic behavior 
\be
q_0(\s)=\frac{1}{r}+{\cal O}\lt(\frac1{r^3}\rt)\;,\qquad q_2(\s)=\frac{2}{15r^3}+{\cal O}\lt(\frac1{r^5}\rt)\;,
\ee
so that the asymptotic expression for the Newtonian external gravitational potential at large distances from the body, is
\be
V_N[\r_{\rm c}, {\cal S}]=\frac{Gm_N}{r}+{\cal O}\lt(\frac1{r^3}\rt)\;,
\ee
where the notation $m_N\equiv M_N/\a$, and $M_N$ is the Newtonian mass of the Maclaurin ellipsoid
\be\la{M_N}
M_N=\frac{4\pi\r_{\rm c}\a^3}{3}\frac{1+\vk^2}{\vk^3}=\frac{4\pi\r_{\rm c}a^2}{3}\frac{\a}{\vk}=\frac{4\pi\r_{\rm c}a^2b}{3}\;.
\ee
Therefore, expression \eqref{xcv67} can be simplified to
\be\la{pw7n}
V_N[\r_{\rm c}, {\cal S}]=Gm_N\bigg[q_0(\s)+q_2(\s)P_2(\cos\th)\bigg]\;.
\ee
It is worth noticing that on the surface of the rotating body the two expressions for the internal and external gravitational potential, \eqref{ndtb7} and \eqref{pw7n} match smoothly for the gravitational potential is a continuous function. It is also useful to remark that for a fixed value of the fluid's density, $\r_{\rm c}$, the normalized mass, $m_N$, decreases inversely proportional to the eccentricity: $m_N\sim\vk^{-1}\sim\e^{-1}$.

The surface of the Maclaurin ellipsoid is equipotential, and it is defined by equation
\be
\frac12v^2+V_N[\r_{\rm c}, {\cal S}]=W_0={\rm const.}\;,
\ee
where $v^2$ is defined in \eqref{skor}. After taking into account \eqref{scl24} and \eqref{pw7n}, equation of the ellipsoid reads,
\be\la{opl87}
W_0=\frac{\o^2\a^2}{3\vk^2}\lt(1+\vk^2\rt)\lt[1-P_2(\cos\th)\rt]+Gm_N\bigl[q_0+q_2P_2(\cos\th)\bigr]\;,
%\frac{\o^2\a^2}{3\vk^2}\lt(1+\vk^2\rt)\lt[1-P_2(\cos\th)\rt]+\frac{4\pi G\r_{\rm c}\a^2}{3\vk^3}\lt(1+\vk^2\rt)\lt\{\arctan\vk+\frac1{2\vk^2}\lt[\lt(\vk^2+3\rt)\arctan\vk-3\vk\rt]P_2(\cos\th)\rt\}=W_0\;.
\ee
where we have introduced shorthand notations $q_0\equiv q_0(1/\vk)$ and $q_2\equiv q_2(1/\vk)$.
Since the left hand side of \eqref{opl87} is constant, the right hand side of it must be constant as well. It yields two relationships
\ba\la{nscbuwe}
\o^2&=&\frac{4\pi G\r_{\rm c}}{\vk}q_2\;,\\
\la{w0}
W_0&=&\frac{1}{3}\o^2a^2+Gm_Nq_0\;.
\ea
Equations \eqref{nscbuwe}, \eqref{w0} can be recast to yet another form,
\ba\la{n7d5}
\o^2&=&\frac{3Gm_N}{a^2}q_2\;,\\
\la{w1}
W_0&=&Gm_N\lt[q_0+q_2\rt]\;.
%\frac{2\pi G\r_{\rm c}\a^2}{\vk^5}\lt(1+\vk^2\rt)\lt[\lt(1+\vk^2\rt)\arctan\vk-\vk\rt]\;.
\ea

Equation \eqref{nscbuwe} yields  relation between the angular velocity of rotation of the homogeneous fluid and oblateness of the Maclaurin ellipsoid while \eqref{w0} or, equivalently, \eqref{w1} defines the gravity potential on its surface.  For small values of the second eccentricity $\vk\ll 1$, when the deviation of the ellipsoid from a sphere is very small, we can expand the Legendre function $q_2$ in the Taylor series, which yields the asymptotic expression for the angular velocity
\be
\o^2=\frac{8}{15}G\pi\r_{\rm c}\vk^2\lt(1-\frac67\vk^2\rt)+{\cal O}\lt(\vk^6\rt)\;,\qquad\qquad(\vk\ll 1)\;.
\ee
On the other hand, when the ellipsoid has a disk-like shape, we have $\vk\gg 1$, and the asymptotic expression of the angular velocity takes on another form,
 \be
\o^2=\frac{G\pi\r_{\rm c}}{\vk}\lt(\pi-\frac2{\vk}\rt)+{\cal O}\lt(\frac1{\vk^3}\rt)\;,\qquad\qquad(\vk\gg 1)\;.
\ee
Equation \eqref{nscbuwe} tells us that the angular velocity of rotation of the Maclaurin ellipsoid, $\o=\o(\vk)$, considered as a function of the eccentricity, $\vk$, has a maximum which is reached for $\vk\simeq 2.52931$ \citep{pizz_1913}. The maximal value of the angular velocity of the Maclaurin ellipsoid at this point is $\omega^2\simeq 0.45\pi G\r_{\rm c}$ \citep{Tassoul_1978_book} . 

\subsection{Contribution from the density inhomogeneity}\la{inhom3s}
The contribution from the non-homogeneous part of the mass density to the Newtonian potential is given by the integral
\be\la{nw7ha1}
V_N[\d\r_{\rm c}, {\cal S}]=\frac{G\r_{\rm c}}{c^2}\int\limits_{\cal V}\frac{F({\bm x}')d^3x'}{|{\bm x}-{\bm x}'|}\;,
\ee
where function $F({\bm x})$ is given in \eqref{cond3}--\eqref{bvxr2}. Making use of \eqref{bvxr2} in the integral \eqref{nw7ha1}, brings it to the form,
\be\la{nw4v6}
V_N[\d\r_{\rm c}, {\cal S}]={\cal A}\frac{\pi G^2\r_{\rm c}^2 \a^2b^2q_1}{c^2}I_1({\bm x})\;,
\ee
where we have introduced a notation
\be\la{qd6h}
I_1({\bm x})\equiv\frac1{\a^2}\int\limits_{\cal V}\frac{R(\s',\th')d^3x'}{|{\bm x}-{\bm x}'|}\;.
\ee
The integral \eqref{qd6h} is calculated with making use of the Green function \eqref{grf1}.
We consider the internal, $(\s\le 1/\vk)$, and external, $(\s\ge1/\vk)$, solutions separately.

\subsubsection{The internal solution}
Making use of the Green's functions \eqref{grf1} we get,
\ba
I_1(\s,\th)&=&2\pi\sum_{\ell=0}^\infty(2\ell+1) q_{\ell}\lt(\s\rt)P_{\ell}(\cos\th)\int\limits_{0}^\s d\s'\int\limits_{0}^{\pi}\lt(\s'^2+\cos^2\th'\rt)R(\s',\th') p_{\ell}\lt(\s'\rt)P_{\ell}(\cos\th')d\th'\\\nonumber
&+&2\pi\sum_{\ell=0}^\infty(2\ell+1)p_{\ell}\lt(\s\rt)P_{\ell}(\cos\th)\int\limits_{\s}^{1/\vk} d\s'\int\limits_{0}^\pi\lt(\s'^2+\cos^2\th'\rt)R(\s',\th') q_{\ell}\lt(\s'\rt)P_{\ell}(\cos\th')d\th'\;.
\ea
We, first, integrate with respect to the angular variable $\th$ and, then, with respect to the radial variable $\s$. The integrand of the above integral is 
\ba
\lt(\s^2+\cos^2\th\rt)R(\s,\th)&=&\frac2{105}\lt[ \lt(5 - 3 \vk^2\rt) p_2(\s) +
   4 \lt(3 + \vk^2\rt) p_4(\s)\rt]\\\nonumber
   & +& \frac2{105}\lt[5 - 3 \vk^2 + 8 \vk^2 p_4(\s) \rt] P_2(\cos\th)\\\nonumber
   & -& \frac8{105} \lt[ 3 + \vk^2 - 2 \vk^2 p_2(\s) \rt] P_4(\cos\th)\;.
\ea
After integration over the angular variable $\th$, the integral can be represented as a linear combination of several terms,
\be\la{cnb4}
I_1(\s,\th)=I_{10}(\s)+I_{12}(\s)P_2(\cos\th)+I_{14}(\s)P_4(\cos\th)\;,
\ee
where each part corresponds to its own Legendre polynomial, 
\ba
I_{10}(\s)&=&
\frac{8\pi}{105}\lt\{q_0(\s)\int\limits_{0}^\s d\s'\lt[ \lt(5 - 3 \vk^2\rt) p_2(\s') +
   4 \lt(3 + \vk^2\rt) p_4(\s')\rt]\rt.\\\nonumber
   &&\phantom{+++++++++++++++++++}+\lt.\int\limits_{\s}^{1/\vk} d\s'\lt[ \lt(5 - 3 \vk^2\rt) p_2(\s') +
      4 \lt(3 + \vk^2\rt) p_4(\s')\rt]q_0(\s') \rt\}\;,
\\
I_{12}(\s)&=&\frac{8\pi}{105}\lt\{q_2(\s)\int\limits_{0}^\s d\s'\lt[5 - 3 \vk^2 + 8 \vk^2 p_4(\s') \rt]p_2(\s')
   +p_2(\s)\int\limits_{\s}^{1/\vk} d\s'\lt[5 - 3 \vk^2 + 8 \vk^2 p_4(\s') \rt] q_2(\s') \rt\}\;,\\
I_{14}(\s)&=&-\frac{32\pi}{105}\lt\{q_4(\s)\int\limits_{0}^\s d\s'\lt[ 3 + \vk^2 - 2 \vk^2 p_2(\s') \rt]p_4(\s')
   +p_4(\s)\int\limits_{\s}^{1/\vk} d\s'\lt[ 3 + \vk^2 - 2 \vk^2 p_2(\s') \rt] q_4(\s') \rt\}\;.
\ea
Calculation of the integrals reveals
\ba\la{cnb5}
I_{10}(\s)&=& \frac{\pi}{15 \vk ^5} \left[\vk  \left(1-\vk ^2 \sigma ^2\right) \left(3+5 \vk ^2+3 \vk ^2 \sigma ^2+\vk ^4 \sigma ^2\right)+12 \left(1+\vk ^2\right)^2 \arctan\vk\right]\;,  \\\la{cnb6}
I_{12}(\s)&=& \frac{2 \pi}{21 \vk ^5} \left[-\vk  \left(3+5 \vk ^2+9 \sigma ^2+15 \vk ^2 \sigma ^2+2 \vk ^4 \sigma ^2+2 \vk ^6 \sigma ^4\right)+3 \left(1+\vk ^2\right)^2 \left(1+3 \sigma ^2\right) \arctan\vk \right]\;, \\\la{cnb7}
I_{14}(\s)&=& \frac{\pi}{105 \vk ^5} \biggl[\vk  \left(9+15 \vk ^2+90 \sigma ^2+150 \vk ^2 \sigma ^2+48 \vk ^4 \sigma ^2+105 \sigma ^4+175 \vk ^2 \sigma ^4+56 \vk ^4 \sigma ^4-8 \vk ^6 \sigma ^4\right)\\\nonumber
&&\phantom{---}-3 \left(1+\vk ^2\right)^2 \left(3+30 \sigma ^2+35 \sigma ^4\right) \arctan\vk \biggr]\;.
\ea

\subsubsection{The external solution}
The external solution is obtained by making use of the Green function \eqref{grf1}
\ba
I_1(\s,\th)&=&2\pi\sum_{\ell=0}^\infty(2\ell+1) q_{\ell}(\s)P_{\ell}(\cos\th)\int\limits_{0}^{1/\vk} d\s'\int\limits_{0}^\pi \lt(\s'^2+\cos^2\th'\rt)R(\s',\th') p_{\ell}\lt(\s'\rt)P_{\ell}(\cos\th')d\th'\;.
\ea
The result is
\ba
I_1(\s,\th)&=&\frac{8\pi}{105}q_0(\s)\int\limits_{0}^{1/\vk}d\s'\lt[ \lt(5 - 3 \vk^2\rt) p_2(\s') + 4 \lt(3 + \vk^2\rt) p_4(\s')\rt]\\\nonumber
&+&\frac{8\pi}{105}q_2(\s)P_2(\cos\th)\int\limits_{0}^{1/\vk}d\s'\lt[5 - 3 \vk^2 + 8 \vk^2 p_4(\s') \rt]p_2(\s')\\\nonumber
&-&\frac{32\pi}{105}q_4(\s)P_4(\cos\th)\int\limits_{0}^{1/\vk}d\s'\lt[ 3 + \vk^2 - 2 \vk^2 p_2(\s') \rt]p_4(\s')\;.
\ea
After performing the integrals it results in
\be
I_1(\s,\th)=\frac{4\pi}{35}\frac{\lt(1+\vk^2\rt)^2}{\vk^5}\lt[7q_0(\s)+5q_2(\s)P_2(\cos\th)-2q_4(\s)P_4(\cos\th)\rt]\;.
\ee
\subsection{Contribution from the spheroidal deviation from the Maclaurin ellipsoid}

Contribution $V_N[\r_{\rm c}, \d{\cal S}]$ from the spheroidal deviation of the shape of the rotating fluid from the Maclaurin ellipsoid is given by equations \eqref{mryb7}, \eqref{mryb7z}, where function ${\cal F}$ is proportional to the Green function \eqref{grf1}. More specifically,  
\be
V_N[\r_{\rm c}, \d{\cal S}]=\frac{4\pi}{c^2}{\cal B} G\r_{\rm c}\o^2 \a^4I_2(\s,\th)\;,
\ee
where
\be\la{br9m7}
I_2(\s,\th)=\left\{\begin{array}{ll}
&\displaystyle\frac12\sum_{\ell=0}^\infty(2\ell+1)q_{\ell} p_{\ell}(\s)P_{\ell}(\cos\th)\int\limits_0^\pi\lt(\vk^{-2}+\cos^2\th'\rt)P_{2}(\cos\th')P_{\ell}(\cos\th')d\th'\;,\qquad (\s\le1/\vk)\\
&\displaystyle\frac12\sum_{\ell=0}^\infty(2\ell+1)p_{\ell} q_{\ell}(\s)P_{\ell}(\cos\th)\int\limits_0^\pi\lt(\vk^{-2}+\cos^2\th'\rt)P_2(\cos\th')P_{\ell}(\cos\th')d\th'\;,\qquad(\s\ge1/\vk)
\end{array}\right.
\ee
Calculation of the integral in \eqref{br9m7} is performed with the help of \eqref{norm56} 
%and the expansion 
%\be
%\cos^2\th P_2(\cos\th)=\frac2{15} +\frac{11}{21}  P_2(\cos\th) + \frac{12}{35} P_4(\cos\th)\;.
%\ee
yielding
\be
\int\limits_0^\pi\lt(\vk^{-2}+\cos^2\th'\rt)P_{2}(\cos\th')P_{\ell}(\cos\th')d\th'=\frac{2}{2\ell+1}\lt[\frac2{15}\d_{0\ell}+\lt(\frac1{\vk^2} +\frac{11}{21} \rt)\d_{2\ell}+\frac{12}{35}\d_{4\ell}\rt]\;.
\ee
It yields,
\be\la{br9qx}
I_2(\s,\th)=\left\{\begin{array}{ll}
&\displaystyle\frac{2q_0}{15}+\lt(\frac1{\vk^2} +\frac{11}{21} \rt)q_2p_2(\s)P_2(\cos\th)+\frac{12q_4}{35}p_4(\s)P_4(\cos\th)\;,\qquad\qquad (\s\le1/\vk)\\
&\displaystyle\frac2{15}q_0(\s)+\lt(\frac1{\vk^2} +\frac{11}{21} \rt)p_2q_2(\s)P_2(\cos\th)+\frac{12p_4}{35} q_4(\s)P_4(\cos\th)\;,\qquad(\s\ge1/\vk)
\end{array}\right.
\ee
where $q_0\equiv q_0(1/\vk)$, $q_2\equiv q_2(1/\vk)$, $q_4\equiv q_4(1/\vk)$, $p_2\equiv p_2(1/\vk)$, $p_4\equiv p_4(1/\vk)$.

\section{Post-Newtonian Potentials}\la{sec5}
\subsection{Scalar potential $V_{pN}$}\la{scap8}

The post-Newtonian correction \eqref{nvs8} to the Newtonian gravity potential obeys the Poisson equation
\be\la{puk9d}
\D  V_{pN}({\bm x})=-4\pi G\r_{pN}({\bm x})\;,
\ee
where
\be\la{vr7d}
\r_{pN}({\bm x})\equiv \r_{\rm c}\lt[2 v^2({\bm x})+2  V_N({\bm x})+\Pi({\bm x})\rt]+3p({\bm x})\;,
\ee
and the functions entering the right hand side of \eqref{vr7d} are defined by equations \eqref{ol4w}--\eqref{skor} and \eqref{ndtb7}.
Fock had proved (see \citep[Eq. 73.26]{fock_1964book}) that for any (including a homogeneous) distribution of mass density the following equality holds
\be
\r({\bm x})\lt[\frac12v^2({\bm x})+V_N({\bm x})\rt]=p({\bm x})+\r({\bm x})\Pi({\bm x})\;.
\ee
It can be used in order to re-write \eqref{vr7d} as follows
\be\la{vr8d}
\r_{pN}({\bm x})\equiv \r_{\rm c}\lt[v^2({\bm x})+3\Pi({\bm x})\rt]+5p({\bm x})\;.
\ee
This allows to eliminate the Newtonian gravitational potential $V_N$ from the calculation of the post-Newtonian gravitational potential $V_{pN}$ by solving \eqref{puk9d}.

After making use of expressions \eqref{ol4w}--\eqref{skor} and including the constant term $3\Pi=3\Pi_0$, to the constant density (re-normalizing the central density $\r_{\rm c}$) we get in the elliptical coordinates,
\be
\r_{pN}({\bm x})=\r_{\rm c}\o^2\a^2\lt(1+\s^2\rt)\sin^2\th-10\pi G\r_{\rm c}^2b^2q_1R(\s,\th)\;.
\ee
Integrating \eqref{puk9d} directly with the help of the Green function \eqref{grf1}, yields
\be
V_{pN}({\bm x})=-10\pi G^2\r^2_0\a^4\frac{q_1}{\vk^2}I_1(\s,\th)+G\r_{\rm c}\o^2\a^4I_3(\s,\th)\;,
\ee
where the integral $I_1(\s,\th)$ has been calculated in section \ref{inhom3s}, and $I_3(\s,\th)$ is the integral from function $(1+\s^2)\sin^2\th$ to the post-Newtonian gravitational potential $V_{pN}$. The integral $I_3(\s,\th)$ is performed as follows.

\subsection{Integral from the source $(1+\s^2)\sin^2\th$}\la{seccs}
The explicit form of the integral $I_3(\s,\th)$ is as follows,
\be\la{inc0}
I_3(\s,\th)=\frac1{\a^2}\int\limits_{\cal V}\frac{(1+\s'^2)\sin^2\th' d^3x'}{|{\bm x}-{\bm x}'|}\;,
\ee
where
\be\la{inc1}
\lt(\s^2+\cos^2\th\rt)(1+\s^2)\sin^2\th=\frac{16}{105}\Biggl\{p_2(\s) + p_4(\s)
    +\lt[1-p_4(\s)\rt] P_2(\cos\th)
    - \lt[ 1 + p_2(\s) \rt] P_4(\cos\th)\Biggr\}\;.
\ee
Substituting \eqref{inc1} to \eqref{inc0} and integrating with respect to the angular variables we get the internal and external solutions.

\subsubsection{The internal solution}
Making use of the Green's functions \eqref{grf1} the internal solution of \eqref{inc0} takes on the following form
\ba
I_3(\s,\th)&=&2\pi\sum_{\ell=0}^\infty(2\ell+1) q_{\ell}(\s)P_{\ell}(\cos\th)\int\limits_{0}^\s d\s'\int\limits_{0}^\pi\lt(\s'^2+\cos^2\th'\rt)(1+\s'^2)\sin^2\th' p_{\ell}\lt(\s'\rt)P_{\ell}(\cos\th')d\th'\\\nonumber
&+&2\pi\sum_{\ell=0}^\infty(2\ell+1)p_{\ell}(\s)P_{\ell}(\cos\th)\int\limits_{\s}^{1/\vk} d\s'\int\limits_{0}^\pi\lt(\s'^2+\cos^2\th'\rt)(1+\s'^2)\sin^2\th' q_{\ell}\lt(\s'\rt)P_{\ell}(\cos\th')d\th'\;,
\ea
which can be represented as a linear combination of the Legendre polynomials,
\be\la{inc2}
I_3(\s,\th)=I_{20}(\s)+I_{22}(\s)P_2(\cos\th)+I_{24}(\s)P_4(\cos\th)\;,
\ee
where the coefficients are functions of the radial coordinate $\sigma$,
\begin{subequations}\la{net7b4}
\ba
I_{20}(\s)&=&\phantom{-}
\frac{64\pi}{105}\lt\{q_0(\s)\int\limits_{0}^\s d\s'\lt[ p_2(\s')+p_4(\s')\rt]
   +\int\limits_{\s}^{1/\vk} d\s'\lt[ p_2(\s')+p_4(\s')\rt]q_0(\s') \rt\}
   \;,\\
I_{22}(\s)&=&\phantom{-}\frac{64\pi}{105}\lt\{q_2(\s)\int\limits_{0}^\s d\s'\lt[1-p_4(\s')\rt]p_2(\s')
   +p_2(\s)\int\limits_{\s}^{1/\vk} d\s'\lt[1-p_4(\s')\rt]q_2(\s') \rt\}
   \;,\\
I_{24}(\s)&=&-\frac{64\pi}{105}\lt\{q_4(\s)\int\limits_{0}^\s d\s'\lt[1+p_2(\s')\rt]p_4(\s')
   +p_4(\s)\int\limits_{\s}^{1/\vk} d\s'\lt[1+p_2(\s')\rt]q_4(\s') \rt\}
   \;.
\ea
\end{subequations}
Calculation of the integrals in \eqref{net7b4} yields
\begin{subequations}
\ba
I_{20}(\s)&=& \frac{2\pi}{15 \vk ^5} \Biggl\{\vk  \left[1 + 2 \vk^2 - 2 \vk^4 \s^2 - \vk^4 \s^4\right]
+4 \left(1+\vk ^2\right)^2\arctan\vk\Biggr\}
\;,\\
I_{22}(\s)&=& \frac{2 \pi}{21 \vk ^7}  \Biggl\{\vk  \left[3+4 \vk ^2-\vk ^4+\lt(9 +12 \vk ^2-3 \vk ^4-2 \vk ^6\rt) \sigma ^2+2\vk^6\sigma ^4\right]- \left(1+\vk ^2\right)^2 \left(3-\vk ^2\right)\left(1+3 \sigma ^2\right)\arctan\vk\Biggr\}
\;,\\
I_{24}(\s)&=& \frac{\pi}{70 \vk ^7} \Biggl\{\vk  \left[15+34 \vk ^2+23 \vk ^4+\lt(150 +340 \vk ^2 +230 \vk ^4 +32 \vk ^6 \rt)\sigma ^2+\lt(175 +\frac{1190}{3} \vk ^2 +\frac{805}{3} \vk ^4+\frac{128}{3} \vk ^6\rt)\sigma ^4\right]\\\nonumber
&&\phantom{---}-\left(1+\vk ^2\right)^2 \left(5+3 \vk ^2\right) \left(3+30 \sigma ^2+35 \sigma ^4\right)\arctan\vk \Biggr\}
\;.
\ea
\end{subequations}
\subsubsection{The external solution}
The external solution of \eqref{inc0} is obtained by making use of the Green function \eqref{grf1}
\ba
I_3(\s,\th)&=&2\pi\sum_{\ell=0}^\infty(2\ell+1) q_{\ell}(\s)P_{\ell}(\cos\th)\int\limits_{0}^{1/\vk} d\s'\int\limits_{0}^\pi\lt(\s'^2+\cos^2\th'\rt)(1+\s'^2)\sin^2\th' p_{\ell}\lt(\s'\rt)P_{\ell}(\cos\th')d\th'\;,
\ea
which is reduced after implementing \eqref{inc1} and integrating over the angular variable $\th$ to
\ba
I_3(\s,\th)&=&\frac{64\pi}{105}q_0(\s)\int\limits_{0}^{1/\vk}d\s'\lt[p_2(\s')+ p_4(\s')\rt]
+\frac{64\pi}{105}q_2(\s)P_2(\cos\th)\int\limits_{0}^{1/\vk}d\s'\lt[1 - p_4(\s') \rt]p_2(\s')\\\nonumber
&&-\frac{64\pi}{105}q_4(\s)P_4(\cos\th)\int\limits_{0}^{1/\vk}d\s'\lt[ 1 + p_2(\s') \rt]p_4(\s')\;.
\ea
Performing the integrals over the radial variable yields the external solution
\be
I_3(\s,\th)=\frac{4\pi\lt(1+\vk^2\rt)^2}{\vk^5}\lt[\frac2{15}q_0(\s)-\frac{3-\vk^2}{21\vk^2}q_2(\s)P_2(\cos\th)-\frac{5+3\vk^2}{35\vk^2}q_4(\s)P_4(\cos\th)\rt]\;.
\ee

\subsection{Vector potential $V^i$}\la{sec6.1}

Vector potential $V^i$ is defined above by equation \eqref{hrt3}. As we need it only in the post-Newtonian approximation, the density of the fluid entering \eqref{hrt3} can be treated as constant $\r_{\rm c}$.
%obeys the Poisson equation
%\be\la{pom7}
%\D V^i({\bm x})=-4\pi G\r({\bm x}) v^i({\bm x})\;,
%\ee
%which has a particular solution
%\be\la{potui}
%V^i({\bm x})=G\int\limits_{\cal V}\frac{\r({\bm x}') v^i(x')}{|{\bm x}-{\bm x}'|}d^3x'\;.
%\ee
Each element of a rigidly rotating fluid has velocity, $v^i(x)=\varepsilon^{ijk}\omega^j x^k$, where $\varepsilon^{ijk}$ is the Levi-Civita symbol, so that \eqref{hrt3} can be written as follows
\be\la{kui8}
V^i({\bm x})=\varepsilon^{ijk}\hat k^j {\cal D}^k({\bm x})\;,
\ee
where $\hat k^i=\omega^i/\omega$ is the unit vector along $z$-axis which coincides with the direction of the angular velocity vector, $\omega^i$, and the Cartesian  vector ${\cal D}^k=\{{\cal D}^x, {\cal D}^y, {\cal D}^z\}$ is given by
\be\la{om7}
{\cal D}^k({\bm x})=G\omega\r_{\rm c}\int\limits_{\cal V}\frac{x'^k d^3x'}{|{\bm x}-{\bm x}'|}\;.
\ee
We denote, ${\cal D}^+\equiv{\cal D}^x+ i{\cal D}^y$. In the ellipsoidal coordinates \eqref{om7} one has \footnote{Notice that ${\cal D}^z\not=0$ but we don't need this component for calculating $V^+$.}
\be\la{d+}
{\cal D}^+({\bm x})=G\omega\r_{\rm c}\a\int\limits_{\cal V}\frac{ \sqrt{1+\s'}^2\sin\th' e^{i\p'}d^3x'}{|{\bm x}-{\bm x}'|}\;.
%\qquad {\cal D}^z=G\r_{\rm c}\a\int\frac{\s'\cos\th'd^3x'}{|{\bm x}-{\bm x}'|}\;.
\ee
Because the angular velocity, $\o^i=(0,0,\o)$, the vector potential $V^i=\lt(V^x,V^y,V^z\rt)$ has $V^z=0$. The remaining two components of the vector potential can be combined together
\be\la{vep3m}
V^+=V^x+iV^y=i{\cal D}^+\;.
\ee
Equation \eqref{vep3m} reveals that calculation of the vector potential is reduced to calculation of the integral in the right hand side of \eqref{d+} which depends on the point of integration and is separated into the internal and external solutions. We discuss these solutions below.

\subsubsection{The internal solution}

The internal solution is obtained for the field points located inside the volume occupied by the rotating fluid. Making use of the Green function \eqref{grfun1} we have
\ba\nonumber
{\cal D}^+&=&G\omega\r_{\rm c}\a^3\sum_{\ell=0}^\infty\sum_{m=-\ell}^{m=+\ell}\frac{(\ell-|m|)!}{(\ell+|m|)!} q_{\ell|m|}\lt(\s\rt)Y_{\ell m}(\hat{\bm x})\int\limits_{0}^\s d\s'\oint\limits_{S^2}d\O'\lt(\s'^2+\cos^2\th'\rt)\sqrt{1+\s'^2}\sin\th' e^{i\p'} p_{\ell|m|}\lt(\s'\rt)Y^*_{\ell m}(\hat{\bm x}')\\\la{nrt69}
&+&G\omega\r_{\rm c}\a^3\sum_{\ell=0}^\infty\sum_{m=-\ell}^{m=+\ell}\frac{(\ell-|m|)!}{(\ell+|m|)!} p_{\ell|m|}\lt(\s\rt)Y_{\ell m}(\hat{\bm x})\int\limits_{\s}^{1/\vk} d\s'\oint\limits_{S^2}d\O'\lt(\s'^2+\cos^2\th'\rt)\sqrt{1+\s'^2}\sin\th' e^{i\p'} q_{\ell|m|}\lt(\s'\rt)Y^*_{\ell m}(\hat{\bm x}')\;,
\ea
where, $d\O'=\sin\th' d\th' d\phi'$, is the element of the solid angle, and we integrate in \eqref{nrt69} over the unit sphere.
%\ba
%{\cal D}^z&=&G\r_{\rm c}\a^3\sum_{\ell=0}^\infty\sum_{m=-n}^{m=+n}\frac{(n-|m|)!}{(n+|m|)!} q_{n|m|}\lt(\s\rt)Y_{\ell m}(\hat{\bm x})\int\limits_{0}^\s d\s'\oint\limits_{S^2}d\O'\lt(\s'^2+\cos^2\th'\rt) \s'\cos\th'  p_{n|m|}\lt(\s'\rt)Y^*_{\ell m}(\hat{\bm x}')\\\nonumber
%&+&G\r_{\rm c}\a^3\sum_{\ell=0}^\infty\sum_{m=-n}^{m=+n}\frac{(n-|m|)!}{(n+|m|)!} p_{n|m|}\lt(\s\rt)Y_{\ell m}(\hat{\bm x})\int\limits_{\s}^{1/\vk} d\s'\oint\limits_{S^2}d\O'\lt(\s'^2+\cos^2\th'\rt) \s'\cos\th' q_{n|m|}\lt(\s'\rt)Y^*_{\ell m}(\hat{\bm x}')\;,
%\ea
We expand functions under the sign of integrals in terms of the spherical harmonics
\ba\la{vv6g}
\sqrt{1+\s^2}\lt(\s^2+\cos^2\th\rt)\sin\th e^{i\p}&=&-\sqrt{1+\s^2}\lt[\frac{2}{15}\frac{Y_{31}(\hat{\bm x})}{C_{31}}+\lt(\s^2+\frac{1}{5}\rt)\frac{Y_{11}(\hat{\bm x})}{C_{11}}\rt]   
\;,
%\\
%\lt(\s^2+\cos^2\th\rt) \s\cos\th&=& \s\lt[\frac{2}{5}\frac{Y_{30}(\hat{\bm x})}{C_{30}}+\lt(\s^2+\frac{3}{5}\rt)\frac{Y_{10}(\hat{\bm x})}{C_{10}}\rt]\;.
\ea
and perform calculations of \eqref{nrt69} with the help of the orthogonality relation \eqref{inus52}.
Finally, we obtain the internal solution of the potential \eqref{vep3m} in the following form
\ba
{\cal D}^+&=&\frac{\pi G\omega\r_{\rm c}\a^3\sqrt{1+\s^2}}{30\vk^5}\biggl\{ 12\vk(1+\s^2\vk^4)P_{11}(\cos\th)
+\vk\Bigl[8\s^2\vk^4+(1+5\s^2)(3+5\vk^2)\Bigr]P_{31}(\cos\th)\\\nonumber
&-& 3(1+\vk^2)^2\Bigl[4P_{11}(\cos\th)+(1+5\s^2)P_{31}(\cos\th)\Bigr]\arctan\vk\biggr\}e^{i\p}\;.
%\\
%{\cal D}^z&=&\frac{2\pi G\r_{\rm c}\a^3\s}{15\vk^5}\biggl[ 3\vk(2+3\vk^2-\s^2\vk^4)P_{1}(\cos\th)
%+\vk\Bigl(9+6\vk^2+15\s^2+2\s^2\vk^2(5-\vk^2)\Bigr)P_{3}(\cos\th)\\\nonumber
%&-& 3(1+\vk^2)\Bigl(2P_{1}(\cos\th)+(3+5\s^2)P_{3}(\cos\th)\Bigr)\arctan\vk\biggr]\;.
\ea

\subsubsection{The external solution}
The external solution of \eqref{vep3m} is obtained for the field points lying outside the volume occupied by the rotating fluid. Making use of the Green functions \eqref{grfun1}, we obtain
\ba\la{vse4}
{\cal D}^+&=&G\omega\r_{\rm c}\a^3\sum_{\ell=0}^\infty\sum_{m=-\ell}^{m=+\ell}\frac{(\ell-|m|)!}{(\ell+|m|)!} q_{\ell|m|}\lt(\s\rt)Y_{\ell m}(\hat{\bm x})\int\limits_{0}^{1/\vk} d\s'\oint\limits_{S^2}d\O'\lt(\s'^2+\cos^2\th'\rt)\sqrt{1+\s'^2}\sin\th' e^{i\p'} p_{\ell|m|}\lt(\s'\rt)Y^*_{\ell m}(\hat{\bm x}')\;.
%\\\la{vse5}
%{\cal D}^z&=&G\r_{\rm c}\a^3\sum_{\ell=0}^\infty\sum_{m=-n}^{m=+n}\frac{(n-|m|)!}{(n+|m|)!} q_{n|m|}\lt(\s\rt)Y_{\ell m}(\hat{\bm x})\int\limits_{0}^{1/\vk} d\s'\oint\limits_{S^2}d\O'\lt(\s'^2+\cos^2\th'\rt) \s'\cos\th'  p_{n|m|}\lt(\s'\rt)Y^*_{\ell m}(\hat{\bm x}')\;,
\ea
After making use of \eqref{vv6g} and integrating over the angular variables, we get
\ba
{\cal D}^+&=&-2\pi G\omega\r_{\rm c}\a^3\lt[ \frac1{45}q_{31}\lt(\s\rt)P_{31}(\cos\th)\int\limits_{0}^{1/\vk}\sqrt{1+\s'^2}p_{31}(\s') d\s'\rt.\\\nonumber
&&+\lt.q_{11}\lt(\s\rt)P_{11}(\cos\th)\int\limits_{0}^{1/\vk}\sqrt{1+\s'^2}\lt(\s'^2+\frac{1}5\rt)p_{11}(\s') d\s'\rt]e^{i\p}\;.
%\\
%{\cal D}^z&=&4\pi G\r_{\rm c}\a^3\lt[ \frac2{5}q_3\lt(\s\rt)P_3(\cos\th)\int\limits_{0}^{1/\vk}\s'p_3(\s') d\s'+q_1\lt(\s\rt)P_1(\cos\th)\int\limits_{0}^{1/\vk}\lt(\s'^3+\frac{3\s'}5\rt)p_1(\s') d\s'\rt]\;.
\ea
Integration with respect to the radial coordinate $\s$ yields
\be\la{ppp3m}
{\cal D}^+={\cal D}e^{i\p}\;,
\ee
where
\be\la{nrt7n}
{\cal D}=-2\pi G\omega\r_{\rm c}\a^3\frac{(1+\vk^2)^2}{5\vk^5}\lt[q_{11}\lt(\s\rt)P_{11}(\cos\th)+ \frac1{6}q_{31}\lt(\s\rt)P_{31}(\cos\th)
\rt]\;.\ee
%and
%\be
%{\cal D}^z=4\pi G\r_{\rm c}\a^3\frac{1+\vk^2}{5\vk^5}\lt[q_3\lt(\s\rt)P_3(\cos\th)+q_1\lt(\s\rt)P_1(\cos\th)\rt]\;.
%\ee
Similar result has been obtained in \citep{tessand_1978PhRvD,cheng_2007ChAA}.

We notice that \eqref{nrt7n} can be transformed to yet another (differential) form
\be
{\cal D}=-\frac{3S}{4\a^2}\sqrt{1+\s^2}\sin\th\frac{d}{d\s}\lt[q_1(\s)+\frac16 q_3(\s)\frac{dP_3(\cos\th)}{d\cos\th}\rt]\;,
\ee
where $S=2Ma^2\o/5$ is the angular momentum of the rotating spheroid.

\section{Relativistic Multipole Moments of the Spinning Spheroid}\la{sec6}

Expansion of gravitational field of an extended massive body to multipoles is ubiquitously used in celestial mechanics and geodesy in order to study the distribution of the matter inside the Earth and other planets of the solar system \citep{Moritz_1967,Vanicek_1986_book,Torge_2012_book}. General relativity brings about several complications making the multipolar decomposition of gravitational fields more difficult. First, all multipole moments of the gravitational field should include the relativistic corrections to their definition. Second, besides the multipole moments of a single gravitational potential $V$ of the Newtonian theory, one has to include the multipole moments of all components of the metric tensor. There is a vast literature devoted to clarification of various aspects of the multipolar decomposition of relativistic gravitational fields but it goes beyond the scope of the present paper (see , for example,  \citep{hansen_1974JMP,Thorne_1980,1986RSPTA.320..379B,quevedo_1990,damiyer_1991PhRvD,blanchet_1998CQG,Kopeikin_2011_book}). The present paper needs two types of the post-Newtonian multipole moments which appear in the decomposition of the scalar potential $V$, and the vector potential $V^i$. The scalar and vector multipoles are more commonly known as mass and spin multipole moments \citep{1986RSPTA.320..379B} following the names of the leading terms in the multipolar decompositions of the potentials $V$ and $V^i$ respectively. 

\subsection{Mass Multipole Moments}\la{mass6307}

Mass multipole moments of the external gravitational field of the rotating spheroid are defined by expanding  the scalar potential $V$ entering $g_{00}$ component of the metric tensor \eqref{pnm1}, in the asymptotic series for a large radial distance $r$ in the spherical coordinates. The scalar potential $V$ takes into account the post-Newtonian contributions from the internal energy $\Pi$, pressure $p$, the kinetic energy of rotation and the internal gravitational energy as well as the spheroidal shape of matter distribution and its inhomogeneity,
\be\la{sc7n}
V=V_N[\r_{\rm c}, {\cal S}]+V_N[\d\r_{\rm c}, {\cal S}]+V_N[\r_{\rm c}, \d{\cal S}]+\frac1{c^2}V_{pN}\;,
\ee
where the terms standing in the right hand side of this formula have been provided above in sections \ref{sec4} and \ref{sec5}. As we consider the multipolar expansion of $V$ outside the body, we need only the external solutions which are 
\ba\la{u4v8na}
V_N[\r_{\rm c}, {\cal S}]&=&Gm_N\lt[q_0(\s)+q_2(\s)P_2(\cos\th)\rt]\;,\\
V_N[\d\r_{\rm c}, {\cal S}]&=&\frac{9{\cal A}}{20}\frac{G^2m_N^2}{c^2}\frac{q_1}{\vk}\lt[q_0(\s)+\frac57q_2(\s)P_2(\cos\th)-\frac27q_4(\s)P_4(\cos\th)\rt]\;,\\
V_N[\r_{\rm c}, \d{\cal S}]&=&\frac{2{\cal B}}{5}\frac{Gm_N}{c^2}\omega^2a^2\lt[q_0(\s)+\frac{15}2\lt(\frac{11}{42}+\frac9{7\vk^2}+\frac{3}{2\vk^4}\rt)q_2(\s)P_2(\cos\th)+\frac9{28}\lt(3+\frac{30}{\vk^2}+\frac{35}{\vk^4}\rt)q_4(\s)P_4(\cos\th)\rt],\\\la{cjusgt4}
V_{pN}&=&\frac{2}{5}Gm_N\omega^2a^2\lt[q_0(\s)-\frac{5}2\frac{3-\vk^2}{7\vk^2}q_2(\s)P_2(\cos\th)-\frac32\frac{5+3\vk^2}{7\vk^2}q_4(\s)P_4(\cos\th)\rt]\\\nonumber
&&-\frac{9}{2}\frac{q_1}{\vk}Gm_N\lt[q_0(\s)+\frac57q_2(\s)P_2(\cos\th)-\frac27q_4(\s)P_4(\cos\th)\rt]\;,
\ea
where $m_N=M_N/\a$, and the constant Newtonian mass, $M_N$, is given by \eqref{M_N}. 

After replacing \eqref{u4v8na} - \eqref{cjusgt4} in \eqref{sc7n} and reducing similar terms, the scalar potential takes on the following form,
\be\la{brt8}
V={\cal E}_0q_0(\s)+{\cal E}_2q_2(\s)P_2(\cos\th)+\frac1{c^2}{\cal E}_4q_4(\s)P_4(\cos\th)\;,
\ee
where the constant numerical coefficients are
\ba\la{pou67}
{\cal E}_0&\equiv&Gm_N\lt\{1+\frac1{10c^2}\lt[({\cal A}-10)\frac{9q_1}{2\vk}Gm_N+4({\cal B}+1)\o^2a^2\rt]\rt\}\;,\\
{\cal E}_2&\equiv&  Gm_N\lt\{1+\frac1{c^2}\lt[({\cal A}-10)\frac{9q_1}{28\vk}Gm_N+3{\cal B}\lt(\frac{11}{42}+\frac1{7\vk^2}+\frac{3}{2\vk^4}\rt)\o^2a^2+\frac{1}7\lt(1-\frac{3}{\vk^2}\rt)\o^2 a^2\rt]\rt\}\;, \\\la{nr8v}
{\cal E}_4&\equiv&\frac{9Gm_N}{70}\lt\{(10-{\cal A})\frac{q_1}{\vk}Gm_N+{\cal B}\lt(3+\frac{30}{\vk^2}+\frac{35}{\vk^4}\rt)\o^2a^2-\frac{2}3\lt(3+\frac{5}{\vk^2}\rt)\o^2 a^2\rt\}\;.
\ea
It is remarkable that the three constants, ${\cal E}_0$, ${\cal E}_2$, ${\cal E}_4$ are interrelated. Indeed, by direct inspection of \eqref{pou67}--\eqref{nr8v}, we obtain 
\be\la{ztv4f}
{\cal E}_2={\cal E}_0+\frac{{\cal E}_4}{c^2}\;.
\ee
Therefore, equation \eqref{brt8} takes on a more simple form,
\be\la{brt8a}
V={\cal E}_0\lt[q_0(\s)+q_2(\s)P_2(\cos\th)\rt]+\frac1{c^2}{\cal E}_4\lt[q_2(\s)P_2(\cos\th)+q_4(\s)P_4(\cos\th)\rt]\;.
\ee
The scalar potential \eqref{brt8a} is given in the ellipsoidal coordinates in terms of the ellipsoidal harmonics which are the modified Legendre functions $q_0(\s)$, $q_2(\s)$ and $q_4(\s)$. The advantage in using the ellipsoidal harmonics is that it allows us to represent the post-Newtonian scalar potential $V$ with a finite number of a few terms only. The residual terms in \eqref{brt8a} are of the post-post-Newtonian order of magnitude ($\sim 1/c^4$) which we systematically neglect in the present paper.

In spite of the finite form of the expansion \eqref{brt8a} in terms of the ellipsoidal functions it is a more common practice to discuss the multipolar structure of external gravitational field of an isolated body in terms of spherical coordinates \eqref{bl344}. Mass multipole moments of the gravitational field are defined in general relativity similarly to the Newtonian gravity as coefficients in the expansion of scalar potential $V$ with respect to the spherical harmonics \citep{Thorne_1980}. For axially-symmetric body the spherical multipolar expansion of the scalar potential reads as follows \citep{Moritz_1967,Thorne_1980},
\be\la{m8s7b}
V=
%V_N+\frac1{c^2}V_{pN}=
\frac{GM}R\lt[1-\sum\limits_{\ell=1}^\infty J_{2\ell}\lt(\frac{a}{R}\rt)^{2\ell}P_{2\ell}(\cos\Theta)\rt]\;,
\ee
where $M$ is the relativistic mass, and $J_{2\ell}$ are the relativistic multipole moments of the gravitational field that are defined (in terms of the spherical coordinates) by integrals over the body's volume
\ba\la{ndf53v}
M&=&\int\limits_{\cal V}\lt[\r({\bm x})+\frac1{c^2}\r_{pN}({\bm x})\rt]R^{2}dRd{\rm O}\;,\\
J_{2\ell}&=&-\frac{2}{Ma^{2\ell}}\int\limits_{\cal V}\lt[\r({\bm x})+\frac1{c^2}\r_{pN}({\bm x})\rt]R^{2\ell+2}P_{2\ell}(\cos\Theta)dRd{\rm O}\;,\qquad (n\ge 1)\;,
\ea
with $d{\rm O}\equiv\sin\Theta d\Theta d\Phi$ is the infinitesimal element of the solid angle in the spherical coordinates.

In order to read the multipole moments of the potential $V$ out of \eqref{brt8} we have to transform \eqref{brt8} to spherical coordinates. This can be achieved with the help of the auxiliary formulas representing expansions of the ellipsoidal harmonics in series with respect to the spherical harmonics. Exact transformations between ellipsoidal and spherical harmonic expansions have been derived by Jekeli \citep{jekeli_1988} for numerical computations. However, Jekeli's transformation lacks a convenient analytic form and are not suitable for our purposes. Therefore, below we present a general idea of calculation of the series expansion of the ellipsoidal harmonics in terms of the spherical harmonics \footnote{Our method is partially overlapping with a similar development given in \citep[Section 2.9]{Moritz_1967}}. 

The ellipsoidal harmonics are solutions of the Laplace equation and are represented by the products of the modified Legendre functions $q_{\ell m}(\s)$ or $p_{\ell m}(\s)$ with the associated Legendre polynomials $P_{\ell m}(\cos\th)$. We are interested in the expansion of the $\ell$-th ellipsoidal harmonic $q_\ell(\s)P_\ell(\cos\th)$ in series of the spherical harmonics which are also solutions of the Laplace equation. The most general expansion of this type reads
\be\la{n3c6f}
q_\ell(\s)P_\ell(\cos\th)=\sum_{n=\ell}^{\infty}\frac{A_nP_{n}(\cos\Theta)}{r^{n+1}}\;,
\ee
where $A_n$ are the numerical coefficients depending on $n$. As both sides of \eqref{n3c6f} are analytic harmonic functions, they are identical at any value of the coordinates. In order to calculate the numerical coefficients $A_n$, it is instructive to take the point with $\th=0$. At this point, we also have $\Theta=0$, while $\s=r$, so that the expansion \eqref{n3c6f} is reduced to 
\be\la{n3c6fw3}
\Bigl.q_\ell (\s)\Bigr|_{\s=r}=\sum_{n=\ell}^{\infty}\frac{A_n}{r^{n+1}}\;,
\ee
which means that the coefficients $A_n$ are simply the coefficients of the asymptotic expansion of the modified Legendre function $q_\ell(r)$ for large values of its argument. These coefficients are found by writing down the modified Legendre function $q_\ell(\s)$ in the left side of \eqref{n3c6fw3} in terms of the hypergeometric function ${}_2F_1$ (see \citep[Eq. VI-56b ]{snow_1952}) 
\be\la{hypleg49}
\Bigl.q_\ell(\s)\Bigr|_{\s=r}=\frac{\sqrt{\pi}}{2^{\ell+1}}\frac{\Gamma(\ell+1)}{\Gamma\lt(\ell+\displaystyle{\frac32}\rt)}\frac{\displaystyle{}_2F_1\lt(1+\frac{\ell}2;\frac{\ell+1}2;\ell+\frac32;-\frac1{r^2}\rt)}{r^{\ell+1}}\;,
\ee
and equating the coefficients of the expansion to $A_n$ in the right side of \eqref{n3c6fw3}.  

After applying the above procedure, we get for the first several elliptic harmonics the following series,
\ba\la{ex28}
q_0(\s)&=&+\sum_{\ell=0}^{\infty}\frac{(-1)^\ell}{2\ell+1}\frac{P_{2\ell}(\cos\Theta)}{r^{2\ell+1}}\;,\\
q_2(\s)P_2(\cos\th)&=&-\sum_{\ell=1}^{\infty}\frac{2\ell(-1)^\ell}{(2\ell+1)(2\ell+3)}\frac{P_{2\ell}(\cos\Theta)}{r^{2\ell+1}}\;,\\\la{ex30}
q_4(\s)P_4(\cos\th)&=&+\sum_{\ell=2}^{\infty}\frac{4\ell(\ell-1)(-1)^\ell}{(2\ell+1)(2\ell+3)(2\ell+5)}\frac{P_{2\ell}(\cos\Theta)}{r^{2\ell+1}}\;.
\ea
Replacing expansions \eqref{ex28}--\eqref{ex30} in \eqref{brt8a}, reducing terms of the same power in $1/r^{2\ell+1}$, and comparing the terms of the expansion obtained with similar terms in \eqref{m8s7b}, we conclude that
\ba\la{ct3c}
GM&=&{\cal E}_0\;,\\\la{on3c5}
%J_2&=&\frac13\lt(1-\frac25\frac{{\cal E}_2}{{\cal E}_0}\rt)\;,\\
%J_{2n}&=&\frac{(-1)^{n+1}}{2n+1}\lt[1-\frac{2n}{2n+3}\frac{{\cal E}_2}{{\cal E}_0}+\frac{4n(n-1)}{(2n+3)(2n+5)}\frac{{\cal E}_4}{{\cal E}_0}\rt]\;,\qquad (n\ge 1)\;,
J_{2\ell}&=&\frac{3(-1)^{\ell+1}\e^{2\ell}}{(2\ell+1)(2\ell+3)}\lt[1-\frac{14}{3c^2}\frac{\ell}{2\ell+5}\frac{{\cal E}_4}{{\cal E}_0}\rt]\;,\qquad (n\ge 1)
\ea
where the second term in the square brackets yields the post-Newtonian correction to the Newtonian multipole moments of the Maclaurin ellipsoid which are defined as the coefficient standing in front of the square brackets in \eqref{on3c5}. 

It is convenient from practical point of view to express the relativistic multipole moments \eqref{on3c5} in terms of the dynamical form factor $J^{\rm dyn}_{2}$ of an extended body with an arbitrary internal distribution of mass density. The dynamical form factor is expressed in terms of the difference between the polar, $C$ and equatorial, $A$, moments of inertia,
\be
J^{\rm dyn}_{2}=\frac{C-A}{Ma^2}\;.
\ee  
We follow the technique developed by Heiskanen and Moritz \citep[Section 2.9]{Moritz_1967} according to which the quadrupole moment of the homogeneous ellipsoid must be exactly equal to the dynamical form factor
\be 
J_2=J^{\rm dyn}_{2}\;.
\ee
It is rather straightforward to prove that in terms of the dynamical form factor equation \eqref{on3c5} reads  
\be\la{mv4yi}
J_{2\ell}=\frac{3(-1)^{\ell+1}\e^{2\ell}}{(2\ell+1)(2\ell+3)}\lt[1-\ell+5\ell\frac{J^{\rm dyn}_{2}}{\e^2}+\frac{4\ell}{3c^2}\frac{\ell-1}{2\ell+5}\frac{{\cal E}_4}{{\cal E}_0}\rt]\;,
\ee
where $C$ and $A$ are the principal moments of inertia. Equation \eqref{mv4yi} looks quite different from \eqref{on3c5} but the difference is illusory since the coefficient in the square brackets of \eqref{mv4yi} is identically equal to the corresponding term in \eqref{on3c5} because for the model of the (almost) homogeneous spheroid accepted in the present paper, the ratio 
\be
\frac{C-A}{Ma^2}=\frac{\e^2}5\left(1-\frac{2}{3c^2}\frac{{\cal E}_4}{{\cal E}_0}\right)\;,
\ee
that can be easily checked by direct calculation of the integrals defining the moments of inertia. Equation \eqref{mv4yi} is a relativistic generalization of the result obtained previously by Heiskanen and Moritz \citep[Equation 2-92]{Moritz_1967}. 

Post-Newtonian equation \eqref{mv4yi} allows to calculate the multipole moments of the normal gravity field at any order as soon as the other parameters of the spheroid are defined. As a particular example we adopt the model of GRS80 international ellipsoid that is characterized by the following parameters (see \citep[Section 4.3]{Torge_2012_book} and \citep[Table 1.2]{petit_2010}):
\ba\nonumber
GM&=&398600.5\times 10^9\;{\rm m}^3{\rm s}^{-2}\;,\\\nonumber
a&=&6378137\;{\rm m}\;,\\\nonumber
J^{\rm dyn}_{2}&=&1082.63\times 10^{-6}\;,\\\nonumber
\omega&=&7.292115\times 10^{-5}\;{\rm rad}\,{\rm s}^{-1}\;,\\\nonumber
1/f&=&298.257222101\;.
\ea
Corresponding (derived) values for the first and second eccentricities of GRS80 are 
\ba\nonumber
\e&=&0.08181919104282\;,\\\nonumber
\varkappa&=&0.08209443815192\;.
\ea
The value of the ratio ${\cal E}_4/{\cal E}_0$ can be calculated on the basis of equation \eqref{on3v4} which is derived below in Section \ref{nv2crh} from the condition of the hydrostatic equilibrium.
Making use of the numerical values of the parameters of GRS80 model, we get
\be
\frac{{\cal E}_4}{c^2{\cal E}_0}=-5.58671\times 10^{-7}\left({\cal B}-0.0023445\right)\;,
\ee
where ${\cal B}$ is the parameter defining in our spheroidal model a small deviation from the shape of the Maclaurin ellipsoid. The value of this parameter can be chosen arbitrary in the range being compatible with the limitations imposed by the post-Newtonian approximation, say, $-10\lesssim{\cal B}\lesssim 10$. 
Calculating the scalar multipole moments in our spheroidal model by means of \eqref{mv4yi} yields for the case of pure ellipsoid,
\be
J_4=-2.37091222014\times 10^{-6}\;,\quad J_6=+6.08347064201\times 10^{-9}\;,\quad J_8=-1.42681406953\times 10^{-11}\;,\qquad({\cal B}=0)
\ee
and for spheroid 
\be
J_4=-2.37091158429\times 10^{-6}\;,\quad J_6=+6.08346483750\times 10^{-9}\;,\quad J_8=-1.42680988487\times 10^{-11}\;,\qquad({\cal B}=1)\;.
\ee
These values can be compared with the corresponding values of the GSR80 geodetic model \citep[Equation 4.77c]{Torge_2012_book} which does not take into account relativistic corrections,
\be
J^{\rm GRS80}_4=-2.37091221865\times 10^{-6}\qquad J^{\rm GRS80}_6=+6.08347062840\times 10^{-9}\qquad J^{\rm GRS80}_8=-1.42681405972\times 10^{-11}
\ee
One can see that accounting for relativistic corrections gives slightly different numerical values of the multipole moments for different models of the normal gravity field generated by rotating spheroid. Significance of these deviations for practical applications in geodynamics is a matter of future theoretical and experimental studies. Nonetheless, already now we can state that general relativity changes the classic model of the normal gravity field. Hence, the separation of the observed value of the field into the normal gravity and its perturbation differs from the Newtonian theory and has certain consequences for interpretation of the gravity field anomalies.

\subsection{Spin Multipole Moments} \la{spin1643}

The spin multipole moments are defined as coefficients in the expansion of vector potential $V^i$ with respect to vector spherical harmonics \citep{Thorne_1980}
\be\la{mul3c}
V^i(r,\Theta,\Phi)=\sum_{\ell=0}^\infty\sum_{m=-\ell}^{m=+\ell}\lt[E^{\ell m}(r)Y_{E,\ell m}^i(\Theta,\Phi)+B^{\ell m}(r)Y_{B,\ell m}^i(\Theta,\Phi) +R^{\ell m}(r)Y_{R,\ell m}^i(\Theta,\Phi)\rt]\;,
\ee
where $E^{\ell m}, B^{\ell m}, R^{\ell m}$ are the spin multipole moments depending on the radial coordinate $r$, and $Y_{E,\ell m}^i, Y_{B,\ell m}^i, Y_{R,\ell m}^i$ are the Cartesian components of the three vector spherical harmonics, ${\bm Y}_{E,\ell m}, {\bm Y}_{B,\ell m}, {\bm Y}_{R,\ell m}$. The harmonics ${\bm Y}_{E,\ell m}$ and ${\bm Y}_{R,\ell m}$ are of ``electric-type" parity $(-1)^\ell$, while ${\bm Y}_{B,\ell m}$ have ``magnetic-type" parity $(-1)^{\ell+1}$ \citep{Thorne_1980}.  Only the ``magnetic-type" harmonics present in the expansion of the vector potential in case of an axially-symmetric gravitational field \citep{tessand_1978PhRvD}, hence, we don't consider the ``electric-type" harmonics below.

The ``magnetic-type" harmonics are defined as follows \citep{Arfken_2001book}
\be
{\bm Y}_{B,\ell m}(\Theta,\Phi)=i\frac{{\bm L}Y_{\ell m}(\Theta,\Phi)}{\sqrt{\ell(\ell+1)}}\;,
\ee
where ${\bm L}=-i{\bm x}\times{\bm\nabla}$ is the operator of the angular momentum, the cross `${\bm\times}$' denotes the Euclidean product of vectors, and ${\bm\nabla}$ is the gradient operator.  
The Cartesian components $(L_x, L_y, L_z)$ of the vectorial operator of the angular momentum ${\bm L}$ expressed in terms of the spherical coordinates, are \citep[Exercise 2.5.14]{Arfken_2001book}
\begin{subequations}
\ba
L_x&=&i\lt(\cos\Phi\cot\Theta\frac{\partial}{\partial\Phi}+\sin\Phi\frac{\partial}{\partial\Theta}\rt)\;,\\
L_y&=&i\lt(\sin\Phi\cot\Theta\frac{\partial}{\partial\Phi}-\cos\Phi\frac{\partial}{\partial\Theta}\rt)\;,\\
L_z&=&-i\frac{\partial}{\partial\Phi}\;.
\ea
\end{subequations}

We have found in section \ref{sec6.1} that all of the non-vanishing components of the vector potential $V^i$ are included to the potential $V^+$ defined in \eqref{vep3m}. This potential is proportional to the components of the vector spherical harmonics, ${\bm Y}_{+,\ell m}\sim L_+Y_{\ell m}$ where the action of the operator $L_+$ on the standard spherical harmonics is as follows \citep[Exercise 12.6.7]{Arfken_2001book}
\be
L_+Y_{\ell m}(\Theta,\Phi)=\sqrt{(\ell-m)(\ell+m+1)}Y_{\ell,m+1}(\Theta,\Phi)\;,
%L_-Y_{\ell m}(\Theta,\Phi)&=&\sqrt{(n+m)(n-m+1)}Y_{n,m-1}(\Theta,\Phi)\;,
\ee
which tells us that $V^+\sim Y_{\ell,m+1}$. On the other hand, due to the fact that the angular coordinates of the ellipsoidal and spherical coordinates coincide, $\Phi=\phi$, and $V^+=i{\cal D}e^{i\phi}$ as follows from \eqref{vep3m} and \eqref{ppp3m}, we conclude that the multipolar expansion \eqref{mul3c} of $V^+$ with respect to the spherical harmonics contains only the spherical harmonics with $m=1$, that is $V^+=i{\cal D}^+\sim Y_{\ell 1}\sim P_{\ell 1}e^{i\phi}$. This can be seen directly after applying the Green function in spherical coordinates and taking into account the rotational symmetry with respect to the angle $\Phi$ which yields equation  \eqref{ppp3m} with function ${\cal D}$ having the following form 
\be\la{muk8}
{\cal D}=\frac{GS}{2R^2}\lt[\sin\Theta+\sum_{\ell=1}^\infty \frac{S_{2\ell+1}}{2\ell+1}\lt(\frac{a}{R}\rt)^{2\ell}P_{2\ell+1,1}(\cos\Theta)\rt]\;,
\ee
where 
\ba\la{cyt58n}
S\equiv S_1&=&\phantom{-}\o\int\limits_{\cal V}\r({\bm x})R^4\sin^2\Theta dRd{\rm O}\;,\\
S_{2\ell+1}&=&\frac1{\ell+1}\frac{\o}{Sa^{2\ell}}\int\limits_{\cal V}\r({\bm x})R^{2\ell+4}\sin\Theta P_{2\ell+1,1}(\Theta) dRd{\rm O}\;,\qquad\qquad (\ell>1)
\ea
are the absolute value of the angular momentum (spin) of the rotating spheroid and the spin multipole moments of the higher order, and ${\cal V}$ is the volume bounded by the surface of the Maclaurin ellipsoid. It is sufficient to perform calculation of the integral in \eqref{cyt58n} with the constant value of density, $\r({\bm x})=\r_{\rm c}$. It yields
\be\la{sdef8}
S=\frac{2}{5}Ma^2\o\;,
\ee
which coincides with the result obtained in textbooks on classic mechanics. 

Calculation of the spin multipoles $S_{2\ell+1}$ can be also performed directly but it will be more instructive to find them out from the Taylor expansion of function ${\cal D}$ given in the ellipsoidal coordinates by equation \eqref{nrt7n}. It is convenient to write down this equation by replacing the central density $\r_{\rm c}$ with the total mass, $M$, as follows
\be\la{nx3a}
{\cal D}=-\frac3{10}GM\omega\frac{(1+\vk^2)}{\vk^2}\lt[q_{11}\lt(\s\rt)P_{11}(\cos\th)+ \frac1{6}q_{31}\lt(\s\rt)P_{31}(\cos\th)\rt]\;.\ee
In order to calculate the spin multipole moments, we have to transform \eqref{nx3a} from the ellipsoidal to spherical harmonics. For we have in \eqref{nx3a}  the ellipsoidal harmonics $q_{\ell m}(\s)P_{\ell m}(\cos\th)$ with the index $m=1$, and the odd index $\ell=2k+1$, we have to apply a slightly different approach to get the transformation formula as compared with that employed in the previous section \ref{mass6307}. More specifically, because both the ellipsoidal and spherical harmonics are solutions of the Laplace equation, we have
\be\la{n3c6fqq}
q_{2\ell+1,1}(\s)P_{2\ell+1,1}(\cos\th)=\sum_{n=\ell}^{\infty}\frac{B_nP_{2n+1,1}(\cos\Theta)}{r^{2n+2}}\;,
\ee
where $B_n$ are the numerical coefficients depending on $n$. As both sides of \eqref{n3c6fqq} are analytic harmonic functions, they are identical at any value of the coordinates. In order to calculate the numerical coefficients $B_n$, we take the point with $\th=\pi/2$. At this point we also have $\Theta=\pi/2$ and $\cos\Theta=0$, while the radial coordinate, $\s=\sqrt{r^2-1}$. The Legendre polynomials 
\be\la{x7b2k5}
P_{2n+1,1}(0)=(-1)^{n+1}\frac2{\sqrt{\pi}}\frac{\Gamma(n+\displaystyle\frac32)}{\Gamma(n+1)}=(-1)^{n+1}\frac{(2n+1)!!}{2^n n!}
\ee
so that the expansion \eqref{n3c6fqq} is reduced to 
\be\la{udr35v}
\Biggl.(-1)^{\ell+1}\frac{(2\ell+1)!!}{2^\ell \ell!}q_{2\ell+1,1}(\s)\Biggr|_{\s=\sqrt{r^2-1}}=\sum_{n=\ell}^{\infty}(-1)^{n+1}\frac{(2n+1)!!}{2^n n!}\frac{B_n}{r^{2n+2}}\;,
\ee
which means that the coefficients $B_n$ are simply the coefficients of the asymptotic expansion of the modified Legendre function $q_{\ell1}(\sqrt{r^2-1})$ for large values of its argument, $r\gg 1$. These coefficients are found by writing down the modified Legendre function in the left side of \eqref{udr35v} in terms of the hypergeometric function (see \citep[Eq. VI-57b ]{snow_1952}) 
\be\la{n3c6b3}
\Biggl.q_{2\ell+1,1}(\s)\Biggr|_{\s=\sqrt{r^2-1}}=\frac{\sqrt{\pi}}{2^{2\ell+2}}\frac{\Gamma(2\ell+3)}{\Gamma\lt(2\ell+\displaystyle{\frac52}\rt)}\frac{\displaystyle{}_2F_1\lt(\ell+\frac{3}2;\ell+\frac{1}2;2\ell+\frac52;\frac1{r^2}\rt)}{r^{2\ell+2}}\;.
\ee
Comparing the coefficients of the expansion of the right side of \eqref{n3c6b3} with the numerical coefficients in the right side of \eqref{udr35v}, we can find coefficients $B_n$.  

After applying this procedure, we get for the first two ellipsoidal harmonics the following series expansions,
\ba
q_{11}(\s)P_{11}(\cos\th)&=&\phantom{-}2\sum_{\ell=0}^{\infty}\frac{(-1)^\ell}{(2\ell+1)(2\ell+3)}\frac{P_{2\ell+1,1}(\cos\Theta)}{r^{2\ell+2}}\;,\\
q_{31}(\s)P_{31}(\cos\th)&=&-24\sum_{\ell=1}^{\infty}\frac{\ell(-1)^\ell}{(2\ell+1)(2\ell+3)(2\ell+5)}\frac{P_{2\ell+1,1}(\cos\Theta)}{r^{2\ell+2}}\;.
\ea
Replacing these expansions to \eqref{nx3a} and reducing similar terms, we obtain
\be\la{pm4f}
{\cal D}=\frac{GMa^2\omega}{5R^2}\lt[\sin\Theta-15\sum_{\ell=1}^{\infty}\frac{(-1)^\ell\e^{2\ell}}{(2\ell+1)(2\ell+3)(2\ell+5)}\lt(\frac{a}{R}\rt)^{2\ell}P_{2\ell+1,1}(\cos\Theta)\rt]\;.
\ee
Comparing expansion \eqref{pm4f} with \eqref{muk8}, we conclude that the coefficients of the multipolar expansion of the vector potential in \eqref{muk8} are
\ba\la{z5s0m}
S_{2\ell+1}&=&\frac{15(-1)^{\ell+1}\e^{2\ell}}{(2\ell+3)(2\ell+5)}\;,\qquad (\ell\ge 1).
\ea
The spin multipole moments, $S_{2\ell+1}$, are uniquely related to the mass multipole moments, $J_{2\ell}$, of a homogeneous and uniformly rotating Maclaurin ellipsoid as follows
\be\la{n4v8n}
S_{2\ell+1}=5\,\frac{2\ell+1}{2\ell+5}J_{2\ell}\;.
\ee
Relations \eqref{z5s0m} and \eqref{n4v8n} have been also derived by Teyssandier \citep{tessand_1978PhRvD} by making use of a different mathematical technique.

It is convenient to give, yet another form of the expansion \eqref{pm4f} in terms of the derivatives of the Legendre polynomials. To this end, we employ the relation \citep[Eq. 8.752-1]{gradryzh}
\be
P_{\ell m}(u)=(-1)^m\lt(1-u^2\rt)^{m/2}\frac{d^mP_\ell(u)}{du^m}\;,
\ee 
which allows us to recast \eqref{pm4f} to the following form
\be\la{pm4faa}
{\cal D}=\frac{GMa^2\omega\sin\Theta}{5R^2}\lt[1+15\sum_{\ell=1}^{\infty}\frac{(-1)^\ell\e^{2\ell}}{(2\ell+1)(2\ell+3)(2\ell+5)}\lt(\frac{a}{R}\rt)^{2\ell}\frac{dP_{2\ell+1}(\cos\Theta)}{d\cos\Theta}\rt]\;.
\ee
After accounting for relations \eqref{vep3m}, \eqref{ppp3m}, the vector potential $V^i$ can be written down explicitly in a vector form
\be\la{x6b4}
V^i=\frac{G}2\frac{\lt({\bm S}\times{\bm x}\rt)^i}{R^3}\lt[1+15\sum_{\ell=1}^{\infty}\frac{(-1)^\ell\e^{2\ell}}{(2\ell+1)(2\ell+3)(2\ell+5)}\lt(\frac{a}{R}\rt)^{2\ell}\frac{dP_{2\ell+1}(\cos\Theta)}{d\cos\Theta}\rt]\;,
\ee
where the angular momentum vector ${\bm S}=\{0,0,S\}$, and $S$ is defined in \eqref{sdef8}.  We notice that a similar expansion formula given by Soffel and Frutos \citep[Eq. 23]{Soffel_2016_JGeod} for the vector potential has a typo and should be corrected in accordance with \eqref{x6b4}. 

\section{Relativistic Normal Gravity Field}
\la{sec7}

\subsection{Equipotential Surface}
\label{tdfe}

Harmonic coordinates introduced in section \ref{hcmt3s} represent an inertial reference frame in space which is used to describe the motion of probe masses (satellites) and light (radio) signals in metric \eqref{pnm1}--\eqref{nrxvsyt}. Let us consider a continuous ensemble of observers rotating rigidly in space with respect to $z$ axis of the inertial reference frame with the angular velocity of the rotating spheroid, $\omega^i$.  Each observer moves with respect to the inertial reference frame along a world line $x^i\equiv \{x(t),y(t),z(t)\}$ such that $z(t)$ and $x^2(t)+y^2(t)$ remain constant. We assume that each observer carries out a clock measuring its own proper time $\tau=\tau(t)$ where $t=x^0/c$ is the coordinate time of the harmonic coordinates $x^\a$ introduced in section \ref{hcmt3s}. 

The proper time of the clock is defined by equation $-c^2d\tau^2=ds^2$ where the interval $ds$ is calculated along the world line of the clock \citep{brumberg_1991_book,Soffel_2013_book}. In terms of the metric tensor \eqref{me7g3v} the interval $d\tau$ of the proper time reads,
\be\la{kop100}
d\tau=\lt(-g_{00}(t,{\bm x})-\frac{2}{c}g_{0i}(t,{\bm x})v^i-\frac1{c^2}g_{ij}(t,{\bm x})v^iv^j\rt)^{1/2}dt\;,
\ee
where, ${\bm x}=\{x^i(t)\}$ is taken on the world line of the clock, and $v^i=dx^i/dt=\lt({\bm\omega}\times{\bm x}\rt)^i$ is a constant linear velocity of the clock with respect to the inertial reference frame.  The ensemble of the observers is static with respect to the rotating spheroid and represents a realization of a rigidly rotating reference frame extending to the outer space outside the spheroid. It should be understood that the rigidly rotating observers are generally not in a free fall except of those which are at the radial distance corresponding to the orbit of geostationary satellites. The rotating reference frame is local - it does not go to a spatial infinity and is limited by the radial distance at which the linear velocity equates to the speed of light, $v\le c$, that is $|{\bm x}|\le c/\omega$.  For the Earth this distance does not exceed 27.5 AU - a bit less than the radius of Neptune's orbit.  

After replacing the metric \eqref{pnm1}--\eqref{nrxvsyt} in \eqref{kop100} and extracting the root square, we get the fundamental time delay equation in the post-Newtonian approximation \citep{issi_2017arXiv}
\be\label{kop4}
\frac{d\tau}{dt}=1-\frac{W}{c^2}+{\cal O}\left(c^{-6}\right)\;,
\ee
where the time-independent function, $W$ is given by 
\be\la{kop5}
W=\frac12v^2+V+\frac1{c^2}\left(\frac18 v^4+\frac32 v^2 V-4v^i V^i-\frac12 V^2\right)\;.
\ee
Function $W$ is the post-Newtonian potential of the {\it normal gravity field} taken at the point of localization of the clock \citep{Kopejkin_1991,Kopeikin_2011_book}. 

The equipotential surface is defined by the condition of the constant rate of clock's proper time with respect to the coordinate time, that is \citep{issi_2017arXiv,kop_2015PhLA,Philipp_2017PhRvD}
\be
\frac{d\tau}{dt}=W={\rm const}\;.
\ee
In case of a stationary spacetime generated by a rigidly rotating body through Einstein's equations, the equipotential surface is orthogonal at each point to the direction of the gravity force (the plumb line) \citep{Kopejkin_1991,kop_2015PhLA,Philipp_2017PhRvD,Oltean_2016}. Inside the rotating fluid the equipotential surface also coincides with the levels of equal density - $\r$, pressure - $p$, and thermodynamic energy - $\Pi$ \citep{Kopejkin_1991,Kopeikin_2011_book}. 

\subsection{Normal Gravity Field Potential}

The post-Newtonian potential, $W$, of the normal gravity field inside the rigidly rotating fluid body has been derived in detail in our previous publications \citep{kopmaz_2016prd,kop_2016}, and its derivation corresponds to the internal solution for the potentials worked out in the previous sections. The present paper focuses on the structure of the normal gravity field outside the rotating spheroid which is described by the external solutions of the metric tensor coefficients discussed above. 

The external solution of the scalar potential $V$ entering \eqref{kop5} has been given in \eqref{brt8}. 
Vector (gravitomagnetic) potential outside the body is given by \eqref{kui8} or, more explicitly, 
\be\la{ub4f7}
V^x=-\sin\phi{\cal D}\;,\qquad V^y=\cos\phi{\cal D}\;,\qquad V^z=0\;,
\ee
where ${\cal D}$ is given in \eqref{nx3a}.
It is straightforward to show that the Euclidean dot product, $v^iV^i=v^xV^x+v^yV^y$, entering the post-Newtonian part of $W$, is
\be
v^iV^i=\o\a\sqrt{1+\s^2}{\cal D}\sin\th\;,
\ee
or, by making use of \eqref{scl241}, \eqref{scl242}
\be
v^iV^i=\frac2{15} Gm_N\o^2a^2
\lt\{q_{0}(\s)+q_2(\s)-\lt[q_0(\s)-\frac5{49}q_2(\s)-\frac{54}{49}q_4(\s)\rt]P_2(\cos\th)-\frac{54}{49}\Biggl[q_2(\s)+q_4(\s)\Biggr]P_4(\cos\th)\rt\}\;.
\ee
The rest of the terms entering expression \eqref{kop5} for the normal gravity potential $W$ are
\ba
v^2&=&\o^2\a^2\lt(1+\s^2\rt)\sin^2\th=\frac23\o^2\a^2\lt(1+\s^2\rt)\lt[1-P_2(\cos\th)\rt]\;,\\
v^4&=&\o^4\a^4\lt(1+\s^2\rt)^2\sin^4\th=8\o^4\a^4\lt(1+\s^2\rt)^2\lt[\frac{1}{15}-\frac{2}{21}P_2(\cos\th)+\frac1{35}P_4(\cos\th)\rt]\;,\\
%V_{pN}&=&\frac9{140}\lt(10-{\cal A}\rt)\frac{G^2M_N^2}{\a^2}\frac{\arctan\vk-\vk}{\vk^2}\lt[7q_0(\s)+5q_2(\s)P_2(\cos\th)-2q_4(\s)P_4(\cos\th)\rt]\\\nonumber
%&+&\frac{GM_N\o^2\a}{35}\frac{1+\vk^2}{\vk^4}\lt[14\vk^2q_0(\s)-5\lt(3-\vk^2\rt)q_2(\s)P_2(\cos\th)-3\lt(5+3\vk^2\rt)q_4(\s)P_4(\cos\th)\rt]\;,\\
v^2V_N&=&\frac23Gm_N\o^2\a^2\lt(1+\s^2\rt)
\lt\{q_0(\s)-\frac15q_2(\s) -\lt[q_0(\s)-\frac{5}{7}q_2(\s)\rt]P_2(\cos\th)-\frac{18}{35}q_2(\s)P_4(\cos\th)\rt\}\;,\\
%v^iV^i&=&\o^2\a\sqrt{1+\s^2}{\cal D}\sin\th\\\nonumber
%&=&\frac15 GM_N\o^2\a\frac{1+\vk^2}{\vk^2}\sqrt{1+\s^2}
%\lt\{q_{11}\lt(\s\rt)-\lt[q_{11}\lt(\s\rt)-\frac{3}7q_{3|1}\lt(\s\rt)\rt]P_2(\cos\th)-\frac{3}7q_{3|1}\lt(\s\rt)P_4(\cos\th)\rt\}\;,\\
V_N^2&=&G^2m_N^2\lt\{q_0^2(\s)+\frac{1}{5}q_2^2(\s)+2q_2(\s)\lt[q_0(\s)+\frac{1}{7}q_2(\s)\rt]P_2(\cos\th)+\frac{18}{35}q_2^2(\s)P_4(\cos\th)\rt\}
\ea

%\ba
%v^2&=&\o^2\a^2\lt(1+\s^2\rt)\sin^2\th\;,\\
%v^4&=&\o^4\a^4\lt(1+\s^2\rt)^2\sin^4\th\;,\\
%V_{pN}&=&\frac9{140}\lt(10-{\cal A}\rt)\frac{G^2M_N^2}{\a^2}\frac{\arctan\vk-\vk}{\vk^2}\lt[7q_0(\s)+5q_2(\s)P_2(\cos\th)-2q_4(\s)P_4(\cos\th)\rt]\\\nonumber
%&+&\frac{GM_N\o^2\a}{35}\frac{1+\vk^2}{\vk^4}\lt[14\vk^2q_0(\s)-5\lt(3-\vk^2\rt)q_2(\s)P_2(\cos\th)-3\lt(5+3\vk^2\rt)q_4(\s)P_4(\cos\th)\rt]\;,\\
%v^2V_N&=&GM_N\o^2\a\lt(1+\s^2\rt)\lt[q_0(\s)+q_2(\s)P_2(\cos\th)\rt]\sin^2\th\;,\\
%v^iV^i&=&\o^2\a\sqrt{1+\s^2}{\cal D}\sin\th\\\nonumber
%&=&\frac1{35} GM_N\o^2\a\frac{1+\vk^2}{\vk^2}
%\lt\{7q_{0}(\s)+10q_2(\s)+3q_4(\s)+15[q_2(\s)+q_4(\s)]P_2(\cos\th)\rt\}\sin^2\th\;,\\
%V_N^2&=&\frac{G^2M_N^2}{\a^2}\lt\{q_0^2(\s)+\frac{1}{5}q_2^2(\s)+2q_2(\s)\lt[q_0(\s)+\frac{1}{7}q_2(\s)\rt]P_2(\cos\th)+\frac{18}{35}q_2^2(\s)P_4(\cos\th)\rt\}
%\ea
Summing up all terms in \eqref{brt8} we can reduce it to a polynomial 
\be\la{berv7}
W(\s,\th)={\cal W}_0(\s)+{\cal W}_2(\s)P_2(\cos\th)+\frac1{c^2}{\cal W}_4(\s)P_4(\cos\th)\;,
\ee
which coefficients are functions of the radial coordinate $\s$,
\ba
{\cal W}_0(\s)&=& \frac13\o^2\a^2(1+\s^2)+Gmq_0(\s)\\\nonumber
&&+\frac1{c^2}\o^2\a^2(1+\s^2)\lt\{ \frac1{15}\o^2\a^2(1+\s^2)+Gm\lt[q_0(\s)-\frac15q_2(\s)\rt]\rt\} \\\nonumber
&&-\frac{Gm}{c^2}\lt\{\frac{8}{15}\o^2a^2\Biggl[q_0(\s)+q_2(\s)\Biggr]
+\frac12Gm\lt[q_0^2(\s)+\frac15q_2^2(\s)\rt]\rt\}\;,\\
{\cal W}_2(\s)&=&  -\frac13\o^2\a^2(1+\s^2)+Gmq_2(\s)+\\\nonumber
&&+\frac1{c^2}\lt\{{\cal E}_4q_2(\s)-\o^2\a^2(1+\s^2)\lt[\frac2{21}\o^2\a^2(1+\s^2)+Gm\lt(q_0(\s)-\frac57q_2(\s)\rt)\rt]\rt\} \\\nonumber
&&+\frac{Gm}{c^2}\lt\{\frac{8}{15}\o^2a^2\lt[q_0(\s)-\frac5{49}q_2(\s)-\frac{54}{49}q_4(\s)\rt]
-Gmq_2(\s)\lt[q_0(\s)+\frac17q_2(\s)\rt]\rt\}\;, \\
{\cal W}_4(\s)&=& {\cal E}_4q_4(\s)+  \frac1{35}\o^2\a^2(1+\s^2)\lt[ \o^2\a^2(1+\s^2)-18Gmq_2(\s)\rt] \\\nonumber
&&-\frac{9}{35}Gm\lt\{Gmq_2^2(\s)-\frac{16}{7}\o^2a^2\Biggl[q_2(\s)+q_4(\s)\Biggr]\rt\}\;,
\ea
and we have denoted, $m\equiv M/\alpha$, where $M$ is the relativistic mass \eqref{ndf53v} that is related to the Newtonian mass $m_N$ through equations \eqref{ct3c} and \eqref{pou67}.

\subsection{The Figure of Equilibrium}\la{nv2crh}

The surface of a rotating fluid body is defined by the boundary condition of vanishing pressure, $p=0$. This surface coincides with the level of the constant gravitational potential \citep{Kopeikin_2011_book,kop_2015PhLA} that is defined by the condition,
\be\la{x5f7}
{\cal W}({\s_s},\th)=W_0={\rm const.},
\ee
for the value of the radial coordinate ${\s_s}={\s_s}(\th)$ defined above in \eqref{ztw8qq}. After expanding the left hand side of \eqref{x5f7} around the constant value of the radial coordinate $1/\vk$, the equation of the level surface takes on the following form
\be\la{x5f7a1}
\bar{\cal W}_0-\frac{\g_2\o^2a^3}{5\vk\e c^2}{\cal B}+\lt[\bar{\cal W}_2-\lt(\g_0+\frac27\g_2\rt)\frac{\o^2a^3}{\vk\e c^2}{\cal B}\rt]P_2(\cos\th)+\frac1{c^2}\lt[\bar{\cal W}_4-\frac{18}{35}\frac{\g_2\o^2a^3}{\vk\e}{\cal B}\rt]P_4(\cos\th)=W_0,
\ee
where $\g_0$ and $\g_2$ are the components of the gravity force defined below in \eqref{grfco}, and $\bar{\cal W}_0\equiv{\cal W}_0(\vk^{-1})$, $\bar{\cal W}_2\equiv{\cal W}_2(\vk^{-1})$, $\bar{\cal W}_4\equiv{\cal W}_4(\vk^{-1})$.

Because the potential is constant on the level surface the left hand side of \eqref{x5f7a1} cannot depend on the angle $\th$ which means that the coefficients in front of the Legendre polynomials, $P_2(\cos\th)$ and $P_4(\cos\th)$, must vanish. Equating, 
\be
\bar{\cal W}_4-\frac{18}{35}\frac{\g_2\o^2a^3}{\vk\e}{\cal B}=0\;,
\ee
yields 
\be\la{on3v4}
q_4{\cal E}_4+\frac{108}{35}\frac{G^2m^2}{\vk}q_1q_2{\cal B}=-\frac{54}{245}G^2m^2q_2\lt( q_2+8 q_4\rt)\;,
\ee
%\be
%{\cal E}_4=-\frac{6}{245}\o^4 a^4\lt[\frac1{q_4}+\frac{8}{q_2}\rt]+\frac{18}{35}\frac{\g_2\o^2a^3}{\vk\e c^2}{\cal B}\;,
%\ee
where ${\cal E}_4$ has been defined in \eqref{nr8v}, $q_1=q_1(1/\vk)$, $q_2\equiv q_2(1/\vk)$, $q_4\equiv q_4(1/\vk)$, and we have made use of \eqref{n7d5}. Equation \eqref{on3v4} determines the coefficient ${\cal B}$ in the equation of the spheroidal surface \eqref{ztw8qq} of the rotating fluid body as a function of the coefficient ${\cal A}$ defining the deviation of the internal density of the fluid, $\r$, from the uniform distribution by equations \eqref{cond3}, \eqref{ncj7}. 

Equating, 
\be
\bar{\cal W}_2-\lt(\g_0+\frac27\g_2\rt)\frac{\o^2a^3}{\vk\e c^2}{\cal B}=0\;,
\ee
yields a relationship generalizing the Maclaurin equation \eqref{nscbuwe} to the post-Newtonian approximation, 
\be\la{qv5f3}
\o^2a^2=3Gm q_2\lt[1-\frac4{245}\frac{Gm}{c^2}\Bigl(20q_2+49q_0+36q_4\Bigr)\rt]+\frac{q_2}{c^2}\lt[{\cal E}_4-\frac{3G^2m^2}{7\vk^3}\lt(21+11\vk^2\rt)q_1{\cal B}\rt]\;,
\ee
where $q_0$, $q_1$, $q_2$ and $q_4$ have the same meaning as in \eqref{on3v4} above, and the constant coefficient ${\cal B}$ relates to ${\cal E}_4$ by means of \eqref{on3v4}. In its own turn the coefficient ${\cal E}_4$ is given by equation \eqref{nr8v}.

% In the small eccentricity approximation, when $\vk\ll 1$, equation \eqref{qv5f3} reads
%\be
%\o^2=\frac{3Gm}{a^2}q_2\lt[1-\frac{4}{5}\frac{GM}{ac^2}\lt(1+\frac{2{\cal B}}{35}\rt)+{\cal O}\lt(\vk^2\rt)\rt]\;.
%\ee

%Equating $W=W_0={\rm const.}$  and considering $\s=\vk^{-1}$ we get 3 equations by grouping together free terms and terms proportional to $P_2(\cos\th)$ and $P_4(\cos\th)$. Due to orthogonality of the Legendre polynomials this allows us to set each equation to zero and to solve them for $\o^2,\:{\cal A}$ and $W_0$. As a result we get
%\ba\la{eccom7}
%\o^2&=&\frac{3GM_N\vk^2q_2}{\a^3(1+\vk^2)}\biggl[1-\frac{GM_N}{70\a\vk^2c^2q_4}\Bigl(5\vk^2q_2(54q_2+5q_4)+35q_4(9q_2+8\vk^2q_0)\\\nonumber
%&-&12(15q_2q_{3|1}+14q_4q_{11}-6q_4q_{3|1})\vk\sqrt{1+\vk^2}\Bigr)\biggr]\;,
%\ea
%\be
%{\cal A}=10-\frac{2q_2}{q_4(\vk-\arctan\vk)}\lt(6\vk^2q_2+q_4(5+3\vk^2)-4\vk q_{3|1}\sqrt{1+\vk^2}\rt)\;,
%\ee

Finally, the constant value of the gravity potential on the surface of the rotating ellipsoid is, 
\be
W_0\equiv \bar{\cal W}_0-\frac{\g_2\o^2a^3}{5\vk\e c^2}{\cal B}\;,
\ee
or, more explicitly, 
\be\la{uniq73}
W_0=Gm(q_0+q_2)-\frac{G^2m^2}{2c^2}\lt(q_0^2+\frac{7}5q_0q_2+\frac{17}{5}q_2^2+\frac{12}{\vk}q_1q_2{\cal B}\rt)\;.
\ee
In the small eccentricity approximation the value of the gravity potential on the level surface is
\be\la{oc6gd}
W_0=Gm(q_0+q_2)-\frac{G^2M^2}{2a^2c^2}\lt[1+\lt(\frac{13}{25}+\frac{8{\cal B}}{15}\rt)\vk^2+{\cal O}\lt(\vk^4\rt)\rt]\;.
\ee

The condition of the hydrostatic equilibrium \eqref{x5f7} imposes a constraint on the linear combination of the constant parameters ${\cal E}_4$ and ${\cal B}$ through \eqref{on3v4} which establishes the correspondence between the constants ${\cal A}$ and ${\cal B}$ defining the distribution of mass density \eqref{ncj7} inside the rotating body and the shape of its surface \eqref{ztw8qq}. There are not any other limitations on these parameters. Hence, one of them can be chosen arbitrary.  One choice is to accept ${\cal A}=0$ that is to admit that the density of the fluid is homogeneous at any order of the post-Newtonian approximations. This makes the figure of the equilibrium of the rotating fluid deviate from the ellipsoid of revolution. This choice was made, for example, in papers \citep{Chandr_1967ApJ147_334,Chandr_1974MNRAS167,Bardeen_1971ApJ,Petroff_2003PhRvD, Meinel_2008,kopmaz_2016prd} that consider the corresponding figures of the post-Newtonian rotating homogeneous spheroids with emphasis on astrophysical applications. On the other hand, one can postulate the equipotential surface to be exactly the Maclaurin ellipsoid at any post-Newtonian approximation which is achieved by choosing the parameter ${\cal B}=0$. Such an ellipsoidal figure of equilibrium of a rotating fluid body has a non-homogeneous distribution of mass density so that the parameter  ${\cal A}\not=0$. This case had been considered in our paper \cite{kop_2016}. There is also a possibility to choose the constant parameter ${\cal B}$ in such a way that the post-Newtonian formula \eqref{qv5f3} connecting the angular velocity of rotation, $\o$, with the geometric parameters of the figure of equilibrium, will formally coincide with the classic Maclaurin relationship \eqref{n7d5}. This case deserves a special attention but we shall not dwell upon it over here as it relates to the problem of consistency of a set of astronomical constants which is a prerogative of the International Astronomical Union  [\textcolor{blue}{\url{http://asa.usno.navy.mil/SecK/Constants.html}}].

The gravity potential $W_0$ is defined by formula \eqref{uniq73}. The value of $W_0$ is the gravity potential on the surface of geoid, that is currently chosen as a defining constant \footnote{It is equivalent to a constant $L_G=W_0/c^2=6.969290134\times 10^{-10}$ that determines the difference between TT and TCG time scales (see \citep{issi_2017arXiv,Soffel_2003AJ} or \citep[Appendix C.2, Resolution B1.9.]{Kopeikin_2011_book})} {\it without taking into account} the post-Newtonian contribution as follows, $W_0=62636856.0\pm 0.5$ m$^2$s$^{-2}$  \citep[Table 1.1]{petit_2010}. The fractional uncertainty in the IERS Convention 2010 value of $W_0$ is $\delta W_0/W_0\simeq 8\times 10^{-9}$. This uncertainty is significantly less than that of the global geometric reference system and is no longer accurate enough to match the operational precision of VLBI and SLR measurements as well as satellite laser altimetry of ocean's surface which have reached the level of one millimeter. This accuracy is at the level of the post-Newtonian correction to the Newtonian value of $W_0$ (the first term in a right hand side of \eqref{uniq73}) that can be easily evaluated on the basis of the approximate formula \eqref{oc6gd}, and is about $GM_\oplus/2a_\oplus c^2\simeq 3.5\times 10^{-10}$ or, in terms of height, $2.2$ mm on the Earth surface. This is within the operational precision of current geodetic techniques. We suggest that the significance of the post-Newtonian correction to the defining constant $W_0$ should be thoroughly discussed by corresponding IAU/IAGG working groups developing a new generation of the system of the geopotential-based geodetic constants (see discussion in \citep[Section 4.3]{Torge_2012_book}). 

\subsection{The Somigliana Formula}

The Somigliana formula in classic geodesy gives the value of the normal gravity $\g^i$ on the reference ellipsoid \citep{Vanicek_1986_book,Moritz_1967,Torge_2012_book}. Vector of the normal gravity is perpendicular to the equipotential surface, and is calculated in the post-Newtonian approximation in accordance with equation \citep{Kopeikin_2011_book,kop_2015PhLA}
\be\la{n4v8m}
\g^i=-c^2\Lambda^{ij}\frac{\partial}{\partial x^j} \log\lt(1-\frac1{c^2}{\cal W}\rt) \;,
\ee
where the matrix operator
\be
\Lambda^{ij}=\d^{ij}-\frac1{c^2}\lt(\frac12 v^iv^j+\d^{ij}V_N\rt)+{\cal O}\lt(\frac1{c^4}\rt)\;,
\ee
defines transformation to the inertial frame of a local observer being at rest with respect to the rotating frame of reference.

We are looking for the normal component, $\g_n=\hat n^i\g^i$, of the vector $\g^i$ in the direction of the plumb line that is given by the unit vector $\hat{\bm n}$ defined in \eqref{nor5c}. A particular interest represents the value of $\g_n$ taken on the surface of the ellipsoid which corresponds to the classical derivation of the formula of {\it Somigliana} \citep{Moritz_1967,Torge_2012_book}.  After taking the partial derivative in \eqref{n4v8m}, and making use of the ellipsoidal coordinates, it reads
\be\la{be7b}
\g_n=-\frac1\a\lt[\lt(1+\frac1{2c^2}\o^2\a^2(1+\s^2)\sin^2\th\rt)\lt(\frac{1+\s^2}{\s^2+\cos^2\th}\rt)^{1/2}\frac{\partial{\cal W}}{\partial\s}\rt]_{\s={\s_s}}\;,
\ee
or, more explicitly
\be\la{s5v8l}
\g_n=\lt(1+\frac1{2c^2}\o^2a^2\sin^2\th\rt)\lt(\frac{1+{\s_s}^2}{{\s_s}^2+\cos^2\th}\rt)^{1/2}\lt[\g_0(\s)+\g_2(\s) P_2(\cos\th)+\frac1{c^2}\g_4(\s) P_4(\cos\th) \rt]_{\s={\s_s}}\;,
\ee
where we have introduced the following notations for the partial derivatives of the components of the normal gravity potential
\be\la{grfco}
\g_0(\s)\equiv-\frac1\a \frac{\partial{\cal W}_0}{\partial\s}\;,\qquad
\g_2(\s)\equiv -\frac1\a\frac{\partial{\cal W}_2}{\partial\s}\;,\qquad
\g_4(\s)\equiv -\frac1\a\frac{\partial{\cal W}_4}{\partial\s}\;,
\ee
and the radial coordinate, ${\s_s}={\s_s}(\th)$, in accordance with \eqref{ztw8qq}. It means that $\g_0({\s_s})$, $\g_2({\s_s})$, and $\g_4({\s_s})$ depend on the angular coordinate $\th$.

In what follows, it is more convenient to expand all functions entering the right hand side of \eqref{s5v8l} into the Taylor series expanded around the value of the radial coordinate $\s=\vk^{-1}$. This brings \eqref{s5v8l} to the following form,
\be\la{h4tc}
\g_n=\lt(\frac{1+\vk^2}{1+\vk^2\cos^2\th}\rt)^{1/2}\lt[1+\frac1{2c^2}\o^2a^2\sin^2\th\lt(1+{\cal B}\frac{1-3\cos^2\th}{1+\vk^2\cos^2\th}\rt)\rt]\lt[\bar\g_0+\bar\g_2 P_2(\cos\th)+\frac1{c^2}\bar\g_4 P_4(\cos\th) \rt]\;,
\ee
where now we have
\ba\la{1q7n}
\bar\g_0&=&\lt[\g_0+\frac15\frac{\o^2a^2}{5\e^2c^2}\frac{\partial\g_2}{\partial\s}{\cal B}\rt]_{\s=\vk^{-1}}\;,\\
\bar\g_2&=&\lt[\g_2+\frac{\o^2a^2}{\e^2c^2}\lt(\frac{\partial\g_0}{\partial\s}+\frac27\frac{\partial\g_2}{\partial\s}\rt){\cal B}\rt]_{\s=\vk^{-1}}\;,\\
\bar\g_4&=&\lt[\g_4+\frac{18}{35}\frac{\o^2a^2}{\e^2c^2}\frac{\partial\g_2}{\partial\s}{\cal B}\rt]_{\s=\vk^{-1}}\;.
\ea

According to {\it Somigliana} \citep{Moritz_1967} it is more convenient to write down the normal gravity force $\g_n$ in terms of two constants which are the values of the normal gravity force taken at two particular positions on the surface: 1) a point $a$ on the equator with $\th=\pi/2$, and 2) a point $b$ at the pole with $\th=0$. At these points \eqref{s5v8l} takes on the following forms,
\ba\la{a1bm}
\g_a&=&\sqrt{1+\vk^2}\lt[1+\lt({\cal B}+1\rt)\frac{\o^2a^2}{2c^2}\rt]\lt(\bar\g_{0}-\frac{\bar\g_{2}}2+\frac{3\bar\g_4}{8c^2}\rt)\;,\\\la{a2bm}
\g_b&=&\bar\g_{0}+\bar\g_{2}+\frac{\bar\g_4}{c^2}\;.
\ea
Solving these equations with respect to $\bar\g_0$ and $\bar\g_2$ we get  the post-Newtonian generalization of the theorem of {\it Pizzetti} \citep{pizz_1913,Torge_2012_book},
\be\la{piz1j}
\bar\g_0=\frac{2b}{3a}\lt(1-\frac{\o^2a^2}{2c^2}\rt)\g_a+\frac13\g_b-\frac{7}{12c^2}\bar\g_4\;,
\ee
and the theorem of {\it Clairaut} \citep{Moritz_1967},
\be\la{cl7v}
\bar\g_2=-\frac{2b}{3a}\lt(1-\frac{\o^2a^2}{2c^2}\rt)\g_a+\frac23\g_b-\frac{5}{12c^2}\bar\g_4\;.
\ee
%\ba\la{a1bm}
%\g_a&=&\sqrt{1+\vk^2}\lt[1+\lt({\cal B}+1\rt)\frac{\o^2a^2}{2c^2}\rt]\lt[\bar\g_0-\frac{\bar\g_2}2+\frac{3\bar\g_4}{8c^2}-\frac{\o^2a^2}{2\e^2c^2}\lt(\frac{\partial\bar\g_0}{\partial{\s_s}}-\frac12\frac{\partial\bar\g_2}{\partial{\s_s}}\rt){\cal B}\rt]\;,\\\la{a2bm}
%\g_b&=&\bar\g_0+\bar\g_2+\frac{\bar\g_4}{c^2}+\frac{\o^2a^2}{\e^2c^2}\lt(\frac{\partial\bar\g_0}{\partial{\s_s}}+\frac{\partial\bar\g_2}{\partial{\s_s}}\rt){\cal B}\;,
%\ea
%where
%\be\la{g8n6}
%\bar\g_0\equiv-\frac1\a \frac{\partial{\cal W}_0}{\partial\s}\Biggr|_{\s=\vk^{-1}}\;,\qquad
%\bar\g_2\equiv -\frac1\a\frac{\partial{\cal W}_2}{\partial\s}\Biggr|_{\s=\vk^{-1}}\;,\qquad
%\bar\g_4\equiv -\frac1\a\frac{\partial{\cal W}_4}{\partial\s}\Biggr|_{\s=\vk^{-1}}\;,
%\ee
%do not depend on the angular coordinate $\th$.
Equations \eqref{piz1j} and \eqref{cl7v} coincide with the corresponding post-Newtonian formulations of {\it Pizzetti} and {\it Clairaut} theorems given in our paper \citep[Sections 11 and 12]{kop_2016} after making transformation of the parameters to the new gauge defined by the parametrization  \eqref{ztw8qq} of the shape of the rotating spheroid \footnote{For more details about the gauge transformations of the post-Newtonian spheroid the reader is referred to \citep[Section 4]{kop_2016}.}.

Theorems {\it Pizzetti} and {\it Clairaut} are used to derive the formula of {\it Somigliana} describing the magnitude of the normal gravity field vector in terms of the forces of gravity measured at equator and at pole \citep{Moritz_1967}. This is achieved by replacing \eqref{piz1j} and \eqref{cl7v} back into \eqref{h4tc} and expanding it with respect to $1/c^2$. It results in the post-Newtonian generalization of the formula of {\it Somigliana},
\ba\la{x5v2i}
\g_n&=&\frac{a\g_b\cos^2\th+b\g_a\sin^2\th}{\sqrt{a^2\cos^2\th+b^2\sin^2\th}}+\frac1{32c^2}\frac{4\o^2a^2\lt(a\g_b-b\g_a\rt)-35a\g_4}{\sqrt{a^2\cos^2\th+b^2\sin^2\th}}\sin^22\th\\\nonumber\\
&+&{\cal B}\frac{\o^2b}{8c^2}\frac{3b(a\g_b-b\g_a)\sin^2\th-a(a\g_a+2b\g_b)}{\lt(a^2\cos^2\th+b^2\sin^2\th\rt)^{3/2}}\sin^22\th\;.\nonumber
\ea
The first term in the right side of \eqref{x5v2i} is the canonical formula of {\it Somigliana} used ubiquitously in classic geodesy \footnote{One should notice that in classic geodesy the {\it Somigliana} formula is usually expressed in terms of the geographic latitude $\Phi$ on ellipsoid that is related to the ellipsoidal angle $\th$ by $\th=\beta-\pi/2$, and, $a\tan\beta=b\tan\Phi$, \citep[Eq. 2-77]{Moritz_1967}}, and the second and third terms being proportional to $1/c^2$, are the explicit post-Newtonian corrections. It is worth noticing that the post-Newtonian corrections to the {\it Somigliana} formula \eqref{x5v2i} are also included implicitly to the canonical (first) term through the values of the normal gravity force at equator, $\g_a$, and at the pole, $\g_b$, as follows from \eqref{1q7n}--\eqref{a2bm}. 

It is instructive to compare the results of the post-Newtonian formalism of the previous sections with the post-Newtonian approximations of axially-symmetric {\it exact} solutions of the Einstein equations. There are plenty of the known solutions (see, e.g., \citep{Stephani_2009_book}) and some of their aspects have been analyzed in the application to relativistic geodesy in \citep{Soffel_2016_JGeod,Philipp_2017PhRvD}. Below, we focus on the post-Newtonian approximation of the Kerr metric and comment on its practical usefulness in geodesy.

\section{Normal Gravity Field of the Kerr metric}\la{sec8}

The Kerr metric is an exact, axisymmetric, stationary solution of the Einstein equations found by Roy Kerr \citep{Stephani_2009_book}. The Kerr metric is a vacuum solution representing rotating black hole. It is often assumed in relativistic mechanics that the Kerr metric can be used to describe the external gravitational field of rotating extended body as well. However, the exact internal solution generating the external Kerr metric has not yet been found, and may not exist \citep{Hernandez_1967PhRv}. The internal Kerr solution having been recently found by Hernandez-Pastora \& Herrera \citep{herpas_2017PhRvD} requires further study for verification. Nonetheless, it is instructive to investigate the geometric properties of the Kerr metric from the point of view of its application in relativistic geodesy. Preliminary study of this problem has been performed in \citep{Soffel_2016_JGeod,Philipp_2017PhRvD}. The present paper focuses on the comparison of the multipolar structure of gravitational field of the Kerr metric with that of gravitational field of rotating spheroid. Because we employed the harmonic coordinates for description of the normal gravitational field, we will need the Kerr metric expressed in the harmonic coordinates defined by the condition \eqref{har5}.

%Taking the results of this paper, one can compare the metric \eqref{pnm1}-\eqref{nrxvsyt} with the post-Newtonian expansion of the Kerr metric. Though the two metrics do not coincide
%precisely, in the case of a weak field they show rather good agreement. For instance, if the Earth is taken as the body under consideration and its
%numerical parameters are used to estimate the metric components, then the relative error in the (most interesting) component $g_{0\phi}$ turns out to be less than $0.1\%$ everywhere outside the surface. This discrepancy is maximal at the surface and tends to zero as $r^{-2}$.
\subsection{Harmonic and ellipsoidal coordinates}\la{el8c30}

The Kerr metric in harmonic coordinates, $x^\a=\{x,y,z\}$, has been derived in \citep{jiang_2014,jiang_2014PRD}.  We introduce the Kerr ellipsoidal coordinates, $\{\varsigma,\vartheta,\varphi\}$, connected to the harmonic coordinates $x^\a$ by equations
\begin{subequations} \la{bl3az}
\ba
x&=&\a_{\rm K}\sqrt{1+\varsigma^2}\sin\vartheta\cos\varphi\;,\\
y&=&\a_{\rm K}\sqrt{1+\varsigma^2}\sin\vartheta\sin\varphi\;,\\
z&=&\a_{\rm K}\varsigma\cos\vartheta\;,
\ea
\end{subequations}
which look similar but not equal to the ellipsoidal coordinates  \eqref{bl3} for uniformly rotating fluid body. The difference between the two types of the ellipsoidal coordinates is due to the fact that the geometric meaning of the parameter $\a$ in \eqref{bl3} is different from that of the Kerr parameter $\a_{\rm K}$ which is equal (by definition of the Kerr geometry) to the ratio of the angular momentum, $S$, to mass, $M$, of the rotating body: $\a_{\rm K}=S/Mc$. Dimension of the Kerr parameter $\a_{\rm K}$ is the same (length) as one of the parameter $\a=a\e$ which is used in definition \eqref{bl3} of the ellipsoidal coordinates in classic geodesy. Nonetheless, the two parameters, $\a$ and $\a_{\rm K}$, have different numerical values in the most general case making the two types of the ellipsoidal coordinates related to each other by transformation
\be
\a_{\rm K}\sqrt{1+\varsigma^2}\sin\vartheta=\a\sqrt{1+\sigma^2}\sin\theta\;,\qquad  \a_{\rm K}\varsigma\cos\vartheta=\a\sigma\cos\theta\;,\qquad \varphi=\phi\;.
\ee

We can establish a connection between two parameters, $\a_{\rm K}$ and $\a$, by making use of relationship, $S=I\o$, that expresses the angular momentum $S$ of rotating extended body with its rotational moment of inertia, $I$, and the angular velocity of rotation, $\o$. The moment of inertia can be expressed in terms of mass of the body and its equatorial radius, $I=\lambda Ma^2$, where $\lambda$ is a dimensionless integral parameter that depends on the distribution of matter inside the rotating body and is determined by an equation of state. For example, in case of a homogeneous ellipsoid, $\lambda=2/5$ \citep{Tassoul_1978_book,essen_2014}. 

Let us now assume that the Kerr metric is generated by an extended, rigidly rotating ellipsoid with some non-homogeneous distribution of mass density. Homogeneous distribution of density is excluded as the multipolar expansion of gravitational field of the Kerr metric does not coincide with that of the homogeneous ellipsoid of rotation as we show below. 
%Physically-consistent internal solution for the Kerr metric is not yet known (see, e.g., \citep{1981JMP....22.1696S,wolf_1992CQGra,Islam_1985_book,herpas_2017PhRvD}). 
Then, in accordance with definition of the Kerr parameter, we should accept $\a_{\rm K}=S/Mc=a\e_{\rm K}$, where $\e_{\rm K}=\lambda\o a/c$ is the effective oblateness of the extended body that supposedly generates the Kerr metric \eqref{kmetr}. Thus,  parameter $\a_{\rm K}$ can be equal to $\a=a\e$, if and only if, $\e=\e_{\rm K}$. This condition imposes a certain limitation on the oblateness $\e$ of rotating extended body which, on the other hand, is also a function of the rotational angular velocity and the average density, $\r$, of the body, $\e=\e(\r,\o)$, as follows from the condition of hydrostatic equilibrium of the body's matter in the rotating frame \footnote{For example, in case of a rigidly rotating homogeneous perfect fluid the relation, $\e=\e(\r,\o)$,  is simply given by the Maclaurin formula \eqref{nscbuwe}}. 
It means that for a given total mass $M$ and oblateness $\e$ of extended rotating body, its angular velocity cannot be adjusted independently of the other parameters to make $\a_{\rm K}=\a$. This is because the other parameters of the body like density, $\r$, or a semi-major axis, $a$, must be appropriately chosen to maintain the condition of hydrostatic equilibrium in the rotating frame. We proceed by assuming that $\a_{\rm K}\not=\a$. At the same time, we postulate that the total mass $M$ and angular momentum $S$ of the Kerr metric exactly coincide with the total mass $M$ and angular momentum $S$ of rotating body. This can be always done as these parameters are defined in terms of conserved integrals given at the spatial infinity of  asymptotically-flat spacetime \citep{avr}.  

\subsection{Post-Newtonian approximation of the Kerr metric}

In the ellipsoidal coordinates \eqref{bl3az} the exterior solution of Einstein's equations for the Kerr metric reads \citep{jiang_2014,jiang_2014PRD}
\ba\la{kmetr}
ds^2&=&-c^2dt^2+\a_{\rm K}^2\lt[(\varsigma+\mu_{\rm K})^2+\cos^2\vartheta\rt]\lt(\frac{d\varsigma^2}{1+\varsigma^2-\mu_{\rm K}^2}+d\vartheta^2\rt)\\\nonumber
&&+\frac{2\mu_{\rm K}(\varsigma+\mu_{\rm K})}{(\varsigma+\mu_{\rm K})^2+\cos^2\vartheta}\lt[\frac{\a_{\rm K} \mu_{\rm K}^2 \sin^2\vartheta d\varsigma}{(1+\varsigma^2-\mu_{\rm K}^2)(1+\varsigma^2)}-\a_{\rm K}\sin^2\vartheta d\phi+cdt \rt]^2\\\nonumber
&&+\a_{\rm K}^2\sin^2\vartheta\lt[(\varsigma+\mu_{\rm K})^2+1\rt]\lt[\frac{ \mu_{\rm K}^2 d\varsigma}{(1+\varsigma^2-\mu_{\rm K}^2)(1+\varsigma^2)}- d\phi \rt]^2\;,
\ea 
where the Kerr mass parameter $\mu_{\rm K}\equiv Gm_{\rm K}/c^2$, and $m_{\rm K}\equiv M/\a_{\rm K}$.   The interior solution for the Kerr metric is still unknown despite of numerous attempts to find it out \citep{krasin_1978AnPhy,quevedo_1990,wolf_1992CQGra,Islam_1985_book,Stephani_2009_book}. Recently, a certain progress has been made by Hernandez-Pastora and Herrera \citep{herpas_2017PhRvD} who used a model of a viscous, anisotropic distribution of mass density inside rotating body.

The post-Newtonian approximation of the Kerr metric \eqref{kmetr} in the ellipsoidal coordinates is
\ba\la{mdvdt9}
ds^2&=&\lt[-1+\frac{2\mu_{\rm K}\varsigma}{\varsigma^2+\cos^2\vartheta}-2\mu_{\rm K}^2\frac{\varsigma^2-\cos^2\vartheta}{(\varsigma^2+\cos^2\vartheta)^2}\rt]c^2dt^2-\frac{4\a_{\rm K}\mu_{\rm K}\varsigma\sin^2\vartheta}{\varsigma^2+\cos^2\vartheta}cdt d\phi\\\nonumber
&+&\a_{\rm K}^2\lt[\frac{\varsigma^2+\cos^2\vartheta+2\mu_{\rm K}\varsigma}{1+\varsigma^2}d\varsigma^2+\lt(1+\varsigma^2\rt)\sin^2\vartheta\lt(1+\frac{2\mu_{\rm K}\varsigma}{\varsigma^2+\cos^2\vartheta}\rt)d\phi^2
+\lt(\varsigma^2+\cos^2\vartheta+2\mu_{\rm K}\varsigma\rt)d\vartheta^2\rt]\;.
\ea 
Harmonic coordinates are asymptotically Cartesian with the Euclidean metric $\delta_{ij}$ at spatial infinity. Components of the Euclidean metric in the ellipsoidal coordinates can be obtained directly from the coordinate transformation \eqref{bl3az}, and the result reads 
\ba\la{rgy6}
\delta_{ij}dx^idx^j&=&\a_{\rm K}^2\lt[\frac{\varsigma^2+\cos^2\vartheta }{1+\varsigma^2}d\varsigma^2+\lt(1+\varsigma^2\rt)\sin^2\vartheta d\phi^2+\lt(\varsigma^2+\cos^2\vartheta\rt)d\vartheta^2\rt]\;.
\ea 
Comparing \eqref{rgy6} with \eqref{mdvdt9} allows us to recast the spacetime metric \eqref{mdvdt9} to the following form 
\be\la{bsdr4}
ds^2=\lt[-1+\frac{2V_{\rm K}}{c^2}-\frac{2V_{\rm K}^2}{c^4}+\frac{1}{c^4}\frac{2G^2m_{\rm K}^2\cos^2\vartheta}{(\varsigma^2+\cos^2\vartheta)^2}\rt]c^2dt^2-\frac{8{V}_{\rm K}^i}{c^2}dt dx^i+\lt(1+\frac{2{V}_{\rm K}}{c^2}\rt)\delta_{ij}dx^idx^j\;,
\ee
where 
\ba
{V}_{\rm K}&=&\frac{Gm_{\rm K}\varsigma}{\varsigma^2+\cos^2\vartheta}\;,\\\la{vp4b5}
{V}_{\rm K}^i&=&\frac{1}2\frac{\lt({\bm S}\times{\bm x}\rt)^i}{1+\varsigma^2}\frac{{V}_{\rm K}}{\a^3_{\rm K}m_{\rm K}}\;,
\ea
and ${\bm S}=\{0,0,S\}$ is a vector of the total angular momentum (spin) of the body directed along the $z$ axis of the harmonic coordinates which coincides with the direction of the rotational axis, $S=\a_{\rm K}Mc$.
Potentials ${V}_{\rm K}$ and ${V}_{\rm K}^i$ are harmonic functions and satisfy the Laplace equation
\be
\Delta{V}_{\rm K}=0\;,\qquad\qquad\Delta{V}_{\rm K}^i=0\;,
\ee
where the Laplace operator in the ellipsoidal coordinates is defined in \eqref{lapop2} after a corresponding replacement $\s\rightarrow \varsigma$. 

The very last term in time-time component of the metric \eqref{bsdr4} is, actually, of the post-post-Newtonian order of magnitude, and can be dropped out in the post-Newtonian approximation when comparing with the post-Newtonian metric \eqref{pnm39} of extended body. The thing is that the metric \eqref{bsdr4} is written in dimensionless coordinates, while the parameter $\a_{\rm K}=S/Mc$, that has been used to make them dimensionless, includes a relativistic factor of $1/c$ which is the main parameter of the post-Newtonian expansions. When we go back to the dimensional coordinates and also use the angular momentum, $S$ and mass $M$, instead of $m=M/\a_{\rm K}$, we have
\be\la{n3c5}
\frac{1}{c^4}\frac{2G^2m_{\rm K}^2\cos^2\vartheta}{(\varsigma^2+\cos^2\vartheta)^2}=\frac{1}{c^6}\frac{2G^2S^2\cos^2\vartheta}{\lt(R^2+\a^2_{\rm K}\cos 2\vartheta\rt)^2}\;,
\ee 
where $R^2=x^2+y^2+z^2=\a_{\rm K}^2\lt(\varsigma^2+\sin^2\vartheta\rt)$, and $S=\a_{\rm K}Mc$ is the angular momentum of the rotating body. The right hand side of \eqref{n3c5} is apparently of the order of $1/c^6$ which is the post-post-Newtonian term not entering the first post-Newtonian approximation \citep{Kopeikin_2011_book}.

\subsection{Normal gravity field potential of the Kerr metric}

The normal gravity field represented by the Kerr metric can be expressed in a closed form similarly to the normal gravity field of uniformly rotating perfect fluid. Potential $W$ of the normal field in case of the Kerr metric is defined by formula \eqref{kop5} where we have to use $V_{\rm K}$ and $V_{\rm K}^i$ for scalar and vector gravitational potentials. We have
\ba
W&=&\frac12\o^2\a_{\rm K}^2\lt(1+\varsigma^2\rt)\sin^2\vartheta+\frac{Gm_{\rm K}\varsigma}{\varsigma^2+\cos^2\vartheta}\\\nonumber
&+&\frac1{c^2}\lt\{\frac18\o^4\a_{\rm K}^4\lt(1+\varsigma^2\rt)^2\sin^4\vartheta+\lt[\frac32(1+\varsigma^2)-2\lambda\lt(1+\frac1{\vk^2}\rt)\rt]\frac{Gm_{\rm K}\varsigma}{\varsigma^2+\cos^2\vartheta}\o^2\a^2_{\rm K}\sin^2\vartheta -\frac12\frac{G^2m^2_{\rm K}\varsigma^2}{\lt(\varsigma^2+\cos^2\vartheta\rt)^2} \rt\}\;,
\ea
where $\lambda$ is the parameter introduced above in section \ref{el8c30}, to connect the moment of inertia of a rotating body with its mass and the equatorial radius.

\subsection{Multipolar expansion of scalar potential.}

Analytic comparison of the Kerr metric with an external solution of uniformly rotating ellipsoid is achieved by comparing the multipolar expansions of the corresponding gravitational potentials. The multipolar expansion of the Kerr metric is obtained by applying the partial fraction decomposition, 
\be
\frac{\varsigma}{\varsigma^2+\cos^2\vartheta}=-\frac{1}{2i}\lt(\frac{1}{i\varsigma+\cos\vartheta}+\frac{1}{i\varsigma-\cos\vartheta}\rt)\;,\ee
and a generating function for the Legendre functions of the 2-nd type \citep[formula 8.791]{gradryzh}
\be
\frac{1}{u-v}=\sum_{\ell=0}^{\infty}(2\ell+1)P_\ell(v)Q_\ell(u)\;.
\ee
It allows us to represent the Newtonian potential ${V}_{\rm K}$ of the Kerr metric in the following form
\be\la{ex3u}
{V}_{\rm K}=Gm_{\rm K}\sum_{\ell=0}^{\infty}(-1)^\ell(4\ell+1)q_{2\ell}(\varsigma)P_{2\ell}(\cos\vartheta)=m_{\rm K}\Bigl[q_0(\varsigma)-5q_2(\varsigma)P_2(\cos\vartheta)+9q_4(\varsigma)P_4(\cos\vartheta)+\ldots\Bigr]\;.
\ee

The post-Newtonian terms in time-time component of the metric \eqref{bsdr4} can be written as a partial derivative of the Newtonian potential
\be
-V_{\rm K}^2+\frac{G^2m_{\rm K}^2\cos^2\vartheta}{(\varsigma^2+\cos^2\vartheta)^2}=m_{\rm K}\frac{\partial V_{\rm K}}{\partial\varsigma}\;.
\ee
Derivative of the Legendre function $q_{\ell}(\varsigma)$ is \citep[formula 8.832]{gradryzh}, \citep[formula (78)]{Pohanka_2011CoGG}
\be
\frac{dq_\ell(\varsigma)}{d\varsigma}=-\frac{1+\ell}{1+\varsigma^2}\lt[q_{\ell+1}(\varsigma)+\varsigma q_\ell(\varsigma)\rt]\;.
\ee
It yields for the partial derivative of the Newtonian potential
\be
\frac{\partial{V}_{\rm K}}{\partial\varsigma}=-\frac{Gm}{1+\varsigma^2}\sum_{\ell=0}^{\infty}(-1)^\ell(4\ell+1)(2\ell+1)\lt[q_{2\ell+1}(\varsigma)+\varsigma q_{2\ell}(\varsigma)\rt]P_{2\ell}(\cos\vartheta)\;.
\ee

For analytical comparison of the Kerr metric with the post-Newtonian metric generated by extended rotating body it is sufficient to compare the multipolar expansion of potential $V$ of the metric of the rotating body and that of $V_{\rm K}$ of the Kerr metric, and to deduce the correspondence between the parameters of the expansions \citep{quevedo_1990}. Any kind of a multipolar expansion performed in either ellipsoidal or spherical coordinates can be used. Nonetheless, in theoretical practice, the comparison of multipolar expansions of gravitational fields is usually done in spherical coordinates. We follow this practice and introduce the spherical coordinates $\{\varrho,\Theta,\Phi\}$
\begin{subequations}\la{bl95}
\ba
x&=&\a_{\rm K}\varrho\sin\Theta\cos\Phi\;,\\
y&=&\a_{\rm K}\varrho\sin\Theta\sin\Phi\;,\\
z&=&\a_{\rm K}\varrho\cos\Theta\;,
\ea
\end{subequations}
where the angular coordinates $\Theta, \Phi$ are the same as in \eqref{bl3} and the radial coordinate $\varrho$ is related to $R$ by a scale transformation, $R=\a_{\rm K}\varrho$ so that, $\varrho=(\a/\a_{\rm K})r$. 

To get the multipolar expansion of potential $V_{\rm K}$ we use the following expansion
\ba\la{ne7b2c}
\frac{\varsigma}{\varsigma^2+\cos^2\vartheta}&=&\sum_{\ell=0}^{\infty}(-1)^\ell\frac{P_{2\ell}(\cos\Theta)}{\varrho^{2\ell+1}} \;,
\ea
that can be easily confirmed by direct inspection for the point with angular coordinate, $\th=\Theta=0$, when $\varsigma=\varrho$, and applying expansion of the elementary function
\be
\frac{\varrho}{\varrho^2+1}=\sum_{\ell=0}^{\infty}\frac{(-1)^\ell}{\varrho^{2\ell+1}}\;.
\ee
The expansion \eqref{ne7b2c} can be re-written in the form of a multipolar expansion of a scalar potential,
\be\la{1x2}
{V}_{\rm K}=\frac{GM}{R}\lt[1-\sum_{\ell=1}^{\infty}J^{\rm K}_{2\ell}\lt(\frac{a}{R}\rt)^{2\ell}P_{2\ell}(\cos\Theta)\rt]\;,
%\frac{Gm_{\rm K}}{\varrho}\lt[1+ \sum_{n=1}^{\infty}(-1)^n\frac{P_{2n}(\cos\Theta)}{\varrho^{2n}}\rt]\;,
\ee
where the mass multipole moments of the Kerr metric
\be\la{x6v29n}
J^{\rm K}_{2\ell}=(-1)^{\ell+1}\e_{\rm K}^{2\ell}=(-1)^{\ell+1}\lt(\frac{\lambda\o a}{c}\rt)^{2\ell}\;,\qquad (\ell\ge 1)
\ee
It should be compared with the multipolar expansion \eqref{m8s7b} of the potential $V$ of rotating homogeneous spheroid where the multipole moments $J_{2n}$ are defined in \eqref{on3c5}. 
%\be\la{n3x}
%V=\frac{Gm}{r}\lt\{1+3\sum_{n=1}^{\infty}\frac{(-1)^n}{(2\ell+1)(2n+3)}\lt[1-\frac23\frac{{\cal E}_4}{{\cal E}_0}\rt]\frac{P_{2n}(\cos\Theta)}{r^{2n}}\rt\}\;,
%\ee
%where ${\cal E}_0$, ${\cal E}_4$ are given by \eqref{ct3c} and \eqref{on3v4} correspondingly.
Comparing $V_{\rm K}$ and $V$ and assuming that each of the potentials is generated by a corresponding axially-symmetric body, one can see that monopole terms ($\sim 1/R$) in \eqref{1x2} and \eqref{m8s7b} match perfectly so that the total mass of the Kerr metric can, indeed, be equated to the post-Newtonian mass of rotating spheroid, as it has been postulated above. The second order (quadrupole) moments can be matched as well. Indeed, we are allowed to equate $J_2^{\rm K}=J_2$, under condition
\be\la{pm4b8}
\e_{\rm K}=\frac{\e}{\sqrt{5}}\;,\qquad\qquad (\ell=1)\;.
\ee
This result means that the Kerr metric can imitate gravitational field of a uniformly rotating Maclaurin ellipsoid in the quadrupole approximation. However, as soon as the condition \eqref{pm4b8} is satisfied, the multipole moments of the higher order (octupole, decapole, etc.) of multipole expansions of two potentials, $V$ and $V_{\rm K}$, cannot be matched as follows from comparison of two expressions -- \eqref{on3c5} and \eqref{x6v29n}. It means that the mass multipole moments of the Kerr metric, $J^{\rm K}_{2\ell}$, are, in general, different from the multipole moments $J_{2\ell}$ of rotating ellipsoid for $\ell\ge 2$. 

\subsection{Multipolar expansion of vector potential.}

Gravitomagnetic vector-potential, $V_{\rm K}^i$ of the Kerr metric is given in \eqref{vp4b5}. It has only two non-vanishing components, $V_{\rm K}^i=\lt(V_{\rm K}^x,V_{\rm K}^y,0\rt)$, which can be combined together in the complex potential, $V_{\rm K}^+\equiv V_{\rm K}^x+iV_{\rm K}^y$, where $i$ is the imaginary unit, c.f. \eqref{vep3m}. The explicit form of the potential is
\be
V^+_{\rm K}=i{\cal D}_{\rm K}e^{i\phi}\;,
\ee
where
\be\la{n5v3}
{\cal D}_{\rm K}=\frac{c}2\frac{\sin\vartheta}{\sqrt{1+\varsigma^2}}V_{\rm K}=\frac{c}2\frac{\sin\vartheta}{\sqrt{1+\varsigma^2}}\frac{Gm_{\rm K}\varsigma}{\varsigma^2+\cos^2\vartheta}\;.
\ee
We make use of the expansion \eqref{ex3u} for $V_{\rm K}$, and equation 
\be
\sin\vartheta P_{\ell}(\cos\vartheta)=\frac1{4\ell+1}\lt[P_{2\ell-1,1}(\cos\vartheta)-P_{2\ell+1,1}(\cos\vartheta)\rt]\;,
\ee
that is given in \citep[formula 8.733-4]{gradryzh}, to bring \eqref{n5v3} to the following form
\be\la{b4v3p}
{\cal D}_{\rm K}=-\frac{Gm_{\rm K}c}2\sum_{\ell=0}^{\infty}(-1)^\ell\frac{q_{2\ell+2}(\varsigma)+q_{2\ell}(\varsigma)}{\sqrt{1+\varsigma^2}}P_{2\ell+1,1}(\cos\vartheta)\;.
\ee
We are now use \citep[formula 8.734-5]{gradryzh}
\be
\sqrt{1+\varsigma^2}q_{2\ell+1,1}(\varsigma)=\frac{(2\ell+1)(2\ell+2)}{4\ell+3}\Bigl[q_{2\ell}(\varsigma)+q_{2\ell+2}(\varsigma)\Bigr]\;,
\ee
to obtain the expansion of ${\cal D}_{\rm K}$ in the final form
\ba
{\cal D}_{\rm K}&=&-\frac{Gm_{\rm K}c}4\sum_{\ell=0}^{\infty}\frac{(-1)^\ell(4\ell+3)}{(\ell+1)(2\ell+1)}q_{2\ell+1,1}(\varsigma)P_{2\ell+1,1}(\cos\vartheta)\\\nonumber
&=&-\frac{3m_{\rm K}c}4\lt[q_{11}(\varsigma)P_{11}(\cos\vartheta)-\frac7{18}q_{31}(\varsigma)P_{31}(\cos\vartheta)+\frac{11}{45}q_{51}(\varsigma)P_{51}(\cos\vartheta)+\ldots\rt)\;.
\ea

Expansion of function ${\cal D}_{\rm K}$ in spherical harmonics reads
\be\la{hu8m4}
{\cal D}_{\rm K}=-\frac{Gm_{\rm K}c}{2}\sum_{\ell=0}^{\infty}\frac{(-1)^\ell}{2\ell+1}\frac{P_{2\ell+1,1}(\cos\Theta)}{\varrho^{2\ell+2}}\;.
\ee
This formula can be checked by taking a point with coordinate, $\vartheta=\Theta=\pi/2$, where $\varsigma=\sqrt{\varrho^2-1}$ in accordance with \eqref{bl3az} and \eqref{bl95}. At this value of the angular coordinate expression \eqref{n5v3} yields, 
\be
{\cal D}_{\rm K}=\frac{Gm_{\rm K}c}{2}\frac1{\varrho\sqrt{\varrho^2-1}}=\frac{Gm_{\rm K}c}{2}\sum_{\ell=0}^{\infty}\frac{(2\ell-1)!!}{2^\ell \ell !}\frac1{\varrho^{2\ell+2}}\;,
\ee
but this is exactly the same formula as \eqref{hu8m4} for $\Theta=\pi/2$ with a special value of the polynomial $P_{2\ell+1,1}(0)$ taken from \eqref{x7b2k5}.

Coming back to a physical domain of the dimensional coordinates, and expressing the Legendre polynomial, $P_{2\ell+1,1}$ in terms of the first derivative with respect to its argument, we obtain
\be\la{v5x7v}
{\cal D}_{\rm K}=\frac{GS}2\frac{\sin\Theta}{R^2}\lt[1+\sum_{\ell=1}^{\infty}\frac{(-1)^\ell\e_{\rm K}^{2\ell}}{2\ell+1}\lt(\frac{a}{R}\rt)^{2\ell}\frac{dP_{2\ell+1}(\cos\Theta)}{d\cos\Theta}\rt]\;,
\ee
where $S=\a_{\rm K}Mc$ is a spin of the body generating the Kerr metric. Comparison of the multipole expansion \eqref{v5x7v} with \eqref{muk8} allows us to read out spin multipole moments, $S^{\rm K}_{2\ell+1}$, of the Kerr metric,
\be\la{u46v9}
S^{\rm K}_{2\ell+1}= (-1)^{\ell+1}\e_{\rm K}^{2\ell}\;,\qquad\qquad (\ell\ge 1)\;.
\ee
As one can see, the spin multipole moments, $S^{\rm K}_{2\ell+1}$,  of the Kerr metric given in \eqref{u46v9} are significantly different from those, $S_{2n+1}$, of the homogeneous rotating ellipsoid given in \eqref{z5s0m}. Relation between spin and mass multipole moments of the Kerr metric, 
\be\la{s7b2l8n}
S^{\rm K}_{2\ell+1}=J^{\rm K}_{2\ell}\;,
\ee
does not coincide with similar relation \eqref{n4v8n} between spin and mass multipole moments of a rotating spheroid made of homogeneous fluid. We also notice that relation \eqref{s7b2l8n} corresponds to a well-known relation between Geroch-Hansen mass and spin moments of the Kerr black hole \citep[Chapter 14]{bambi_2017}.

Replacing \eqref{v5x7v} to vector potential of the Kerr metric yields
\be\la{ty5c5}
V^i_{\rm K}=\frac{G}2\frac{({\bm S}\times{\bm x})^i}{R^3}\lt[1+\sum_{\ell=1}^{\infty}\frac{(-1)^\ell\e_{\rm K}^{2\ell}}{2\ell+1}\lt(\frac{a}{R}\rt)^{2\ell}\frac{dP_{2\ell+1}(\cos\Theta)}{d\cos\Theta}\rt]\;,
\ee
We can compare the multipole expansion \eqref{ty5c5} of vector potential of the Kerr metric with similar expansion \eqref{x6b4} given for rotating spheroid. We notice that the first terms of these expansions match exactly each other so that the angular momentum, ${\bm S}$, of the body generating the Kerr geometry can be equated to the angular momentum of the rotating spheroid. As we have already equated the mass of the body generating gravitational field of the Kerr metric to that of the spheroid, we have to conclude that equating the angular momenta of the two bodies also requires imposing a limitation, $\a_{\rm K}=\a$, or, in other words, $\e_{\rm K}=\e$. However, this equation is not compatible with the matching condition \eqref{pm4b8} for the quadrupoles. It means that the Kerr metric is compatible with the normal gravity field of rotating ellipsoid merely in the mass-monopole spin-dipole approximation. This approximation is not suitable for geodetic applications. We conclude that the Kerr metric should not be used for the purposes of relativistic geodesy. Our conclusion complement the arguments set forth by Soffel and Frutos in \citep{Soffel_2016_JGeod}.
\newline
\newline

{\bf Acknowledgements}
\newline
\newline
\noindent
We thank the anonymous referees for valuable comments and suggestions that helped us to improve the manuscript.
S. M. Kopeikin is grateful to the Shanghai Astronomical Observatory of the Chinese Academy of Sciences for hospitality and financial support of travel and lodging expenses. W.-B. Han has been supported by the Youth Innovation Promotion Association of the Chinese Academy of Sciences. The work of I. Yu. Vlasov has been sponsored by the grant \textnumero 14-27-00068 of the Russian Science Foundation. This paper contributes to the research project ``Spacetime Metrology, Clocks and Relativistic Geodesy'' [\textcolor{blue}{\url{http://www.issibern.ch/teams/spacetimemetrology/}}] funded by the International Space Science Institute (ISSI) in Bern, Switzerland.   
\newline
\newline
\bibliographystyle{unsrt}
\bibliography{PN_spheroid_bib}

%\appendix

\end{document}